\documentclass[apj]{emulateapj}
\usepackage{lscape}
\usepackage{apjfonts}
\usepackage{graphicx}

\slugcomment{Accepted for publication in ApJS}
\shorttitle{Galactic Star Clusters in SDSS I.}
\shortauthors{An \& Johnson, et~al.}

\begin{document}
\title{Galactic Globular and Open Clusters in the Sloan Digital Sky Survey. I.\\
Crowded Field Photometry and Cluster Fiducial Sequences in \MakeLowercase{\it ugriz}}

\author{Deokkeun An\altaffilmark{1}, Jennifer A.\ Johnson\altaffilmark{1},
James L.\ Clem\altaffilmark{2},
Brian Yanny\altaffilmark{3},
Constance M.\ Rockosi\altaffilmark{4},\\
Heather L.\ Morrison\altaffilmark{5},
Paul Harding\altaffilmark{5},
James E.\ Gunn\altaffilmark{6},
Carlos Allende Prieto\altaffilmark{7},\\
Timothy C.\ Beers\altaffilmark{8},
Kyle M.\ Cudworth\altaffilmark{9},
Inese I.\ Ivans\altaffilmark{6,10},
\v{Z}eljko Ivezi\'{c}\altaffilmark{11},\\
Young Sun Lee\altaffilmark{8},
Robert H.\ Lupton\altaffilmark{6},
Dmitry Bizyaev\altaffilmark{12},
Howard Brewington\altaffilmark{12},\\
Elena Malanushenko\altaffilmark{12},
Viktor Malanushenko\altaffilmark{12},
Dan Oravetz\altaffilmark{12},\\
Kaike Pan\altaffilmark{12},
Audrey Simmons\altaffilmark{12},
Stephanie Snedden\altaffilmark{12},\\
Shannon Watters\altaffilmark{12}, and
Donald G.\ York\altaffilmark{13,14}
}
\altaffiltext{1}{Department of Astronomy, Ohio State University,
140 West 18th Avenue, Columbus, OH 43210;
deokkeun,jaj@astronomy.ohio-state.edu.}
\altaffiltext{2}{Department of Physics \& Astronomy, Louisiana State
University, 202 Nicholson Hall, Baton Rouge, LA 70803.}
\altaffiltext{3}{Fermi National Accelerator Laboratory,
P.O. Box 500, Batavia, IL 60510.}
\altaffiltext{4}{UCO/Lick Observatory, University of California,
Santa Cruz, CA 95064.}
\altaffiltext{5}{Department of Astronomy, Case Western Reserve University,
Cleveland, OH 44106.}
\altaffiltext{6}{Department of Astrophysical Sciences, Princeton University,
Princeton, NJ 08544.}
\altaffiltext{7}{McDonald Observatory and Department of Astronomy, The University of
Texas, 1 University Station, C1400,  Austin, TX 78712-0259.}
\altaffiltext{8}{Department of Physics \& Astrophysics, CSCE:
Center for the Study of Cosmic Evolution, and JINA: Joint Institute for
Nuclear Astrophysics, Michigan State University, E. Lansing, MI  48824.}
\altaffiltext{9}{Yerkes Observatory, University of Chicago,
373 West Geneva Street, Williams Bay, WI 53191.}
\altaffiltext{10}{The Observatories of the Carnegie Institution of
Washington, 813 Santa Barbara St., Pasadena, CA 91101.}
\altaffiltext{11}{Department of Astronomy, University of Washington,
Box 351580, Seattle, WA 98195.}
\altaffiltext{12}{Apache Point Observatory, P.O. Box 59, Sunspot, NM 88349.}
\altaffiltext{13}{Department of Astronomy and Astrophysics,
University of Chicago, 5640 South Ellis Avenue, Chicago, IL 60637.}
\altaffiltext{14}{Enrico Fermi Institute, University of Chicago,
5640 South Ellis Avenue, Chicago, IL 60637.}

\begin{abstract}
We present photometry for globular and open cluster stars observed with
the Sloan Digital Sky Survey (SDSS).  In order to exploit over 100 million
stellar objects with $r < 22.5$~mag observed by SDSS, we need to
understand the characteristics of stars in the SDSS $ugriz$ filters.
While star clusters provide important calibration samples for stellar
colors, the regions close to globular clusters, where the fraction of
field stars is smallest, are too crowded for the standard SDSS
photometric pipeline to process.  To complement the SDSS imaging survey,
we reduce the SDSS imaging data for crowded cluster fields using the
DAOPHOT/ALLFRAME suite of programs and present photometry for 17
globular clusters and 3 open clusters in a SDSS value-added catalog.
Our photometry and cluster fiducial sequences are on the native SDSS
2.5-meter $ugriz$ photometric system, and the fiducial sequences can be
directly applied to the SDSS photometry without relying upon any transformations.
Model photometry for red giant branch and main-sequence stars obtained by
Girardi et~al.\ cannot be matched simultaneously to fiducial sequences;
their colors differ by $\sim0.02$--$0.05$~mag.  Good agreement
($\la0.02$~mag in colors) is found with Clem et~al.\ empirical fiducial
sequences in $u'g'r'i'z'$ when using the transformation equations in Tucker et~al.
\end{abstract}

\section{Introduction}

As single-age and (in most cases) single-metallicity populations,
Galactic star clusters provide important calibration samples for
exploring the relationships between stellar colors and absolute
magnitudes as functions of stellar age and heavy-element content.
These two observable properties of a star are related to fundamental
physical parameters, such as the effective temperature ($T_{\rm eff}$)
and surface gravity ($\log{g}$), as well as the metallicity.  The color
and magnitude relations can be used to test stellar evolutionary
theories, to interpret the observed distribution of stars in color-color
and color-magnitude diagrams (CMDs), and to derive distances to stars and
star clusters via photometric parallax or main-sequence (MS) fitting
techniques \citep[e.g.,][]{johnson:57}.

Because the relationships between magnitude, color, and fundamental
stellar properties depend on the filters used, it is necessary to
characterize these relations for each filter system.  Galactic globular
and open clusters provide an ideal opportunity to achieve this goal
because the same distance can be assumed for cluster members with
a wide range of stellar masses.  Furthermore, observations of a large
number of Galactic clusters can cover a wide range of the heavy-element
content, providing an opportunity to explore the effects of metallicity
on magnitudes and colors for each set of filter bandpasses.

Among previous and ongoing optical surveys, the Sloan Digital Sky Survey
\citep[SDSS;][]{york:00,edr,dr1,dr2,dr3,dr4,dr5,dr6} is the largest and
most homogeneous database of stellar brightnesses currently available.
The original goal of the SDSS was to survey large numbers of galaxies
and quasars.  However, in the first five years of operation, SDSS-I has
made remarkable contributions to our understanding of the Milky Way and
its stellar populations
\citep[e.g.,][]{newberg:02,allendeprieto:06,belokurov:06,dejong:08,juric:08}.
These successes have initiated the Galactic structure program SEGUE
(Sloan Extension for Galactic Understanding and Exploration;
B.\ Yanny et al.\ 2008, in preparation), one of the surveys being
conducted in the ongoing three year extension of the survey (SDSS-II).
When SDSS-II finishes, it will provide imaging data for approximately
10,000 square degrees of the northern sky.

SDSS measures the brightnesses of stars using a dedicated 2.5-m telescope
\citep{gunn:06} in five broadband filters $u$, $g$, $r$, $i$, and $z$,
with average wavelengths of 3551\AA, 4686\AA, 6165\AA, 7481\AA, and 8931\AA,
respectively \citep{fukugita:96,edr}.  The 95\% detection repeatability
limits are 22.0~mag, 22.2~mag, 22.2~mag, 21.3~mag, and 20.5~mag for point
sources in $u$, $g$, $r$, $i$, and $z$, respectively.  The rms photometric
precision is $0.02$~mag for sources not limited by photon statistics \citep{ivezic:03},
and the photometric calibration is accurate to $\sim2\%$ in the $g$, $r$,
$i$ bands, and $\sim3\%$ in $u$ and $z$ \citep{ivezic:04}.
The SDSS filters represent a new filter set for stellar observations, and
therefore it is important to understand the properties of stars in this
system.  Furthermore, future imaging surveys such as the Panoramic Survey
Telescope \& Rapid Response System \citep[Pan-STARRS;][]{kaiser:02} and
the Large Synoptic Survey Telescope \citep[LSST;][]{stubbs:04} will use
similar photometric bandpasses, providing even deeper data in $ugriz$
than SDSS over a larger fraction of the sky.

During the course of SDSS-I, about 15 globular clusters and several open
clusters were observed.  Several more clusters were imaged in SDSS-II
including M71.  These clusters together provide accurate calibration samples for
stellar colors and magnitudes in the SDSS filters.  The SDSS images are processed
using the standard SDSS photometric pipelines \citep[{\it Photo};][]{lupton:02}.
{\it Photo} pre-processes the raw images, determines the point spread
function (PSF), detects objects, and measures their properties.  Photometric
calibration is then carried out using observations of stars in the secondary
patch transfer fields \citep{tucker:06,davenport:07}.  In this paper, we
simply refer to these calibrated magnitudes as {\it Photo} magnitudes.

\begin{figure*}
\epsscale{0.9}
\plotone{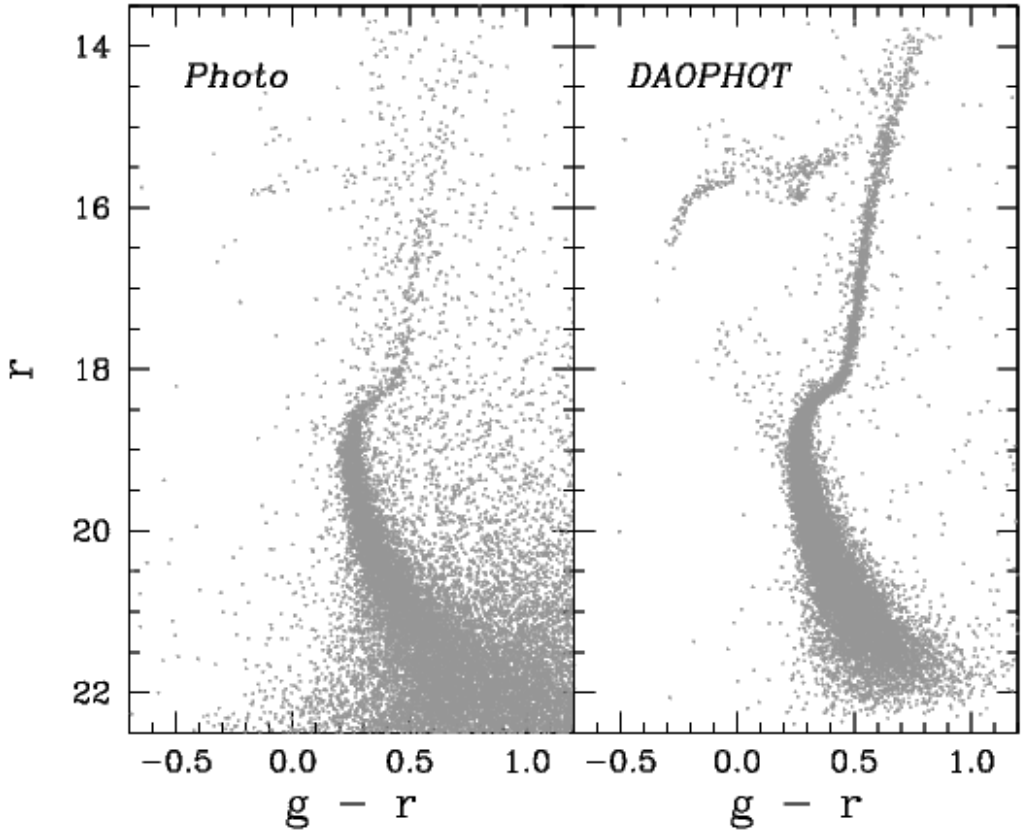}
\caption{CMDs of M3 from the SDSS photometric pipeline ({\it Photo};
{\it left}) and DAOPHOT reduction in this paper ({\it right}).
Stars within a $30\arcmin$ radius from the cluster center are shown in
the left panel, but the {\it Photo} photometry is only available on the
outskirts of the cluster.  In the right panel, RR Lyraes are scattered
off the cluster horizontal branch.
\label{fig:photo}}
\end{figure*}

{\it Photo} was originally designed to handle high Galactic latitude
fields with relatively low densities of Galactic field stars (owing to
the primarily extragalactic mission of SDSS-I); however, there are some
concerns about its photometry derived in crowded fields \citep{dr6}.
In particular, stellar clusters present a challenge to {\it Photo}.
Firstly, {\it Photo} slows down dramatically in the high density cluster
cores, which are too crowded for {\it Photo} to process, so it does not
provide photometry for the most crowded regions of these scans.
Figure~\ref{fig:photo} compares a CMD for the globular cluster M3
obtained from {\it Photo} photometry to that obtained from a DAOPHOT
\citep{stetson:87} reduction in this paper, which is specifically designed
for crowded field photometry.  The {\it Photo} photometry is only available
on the outskirts of the cluster, and it provides a considerably less
well-defined subgiant branch (SGB), red giant branch (RGB), and horizontal
branch (HB).  Secondly, there is a concern that the photometry in the area
surrounding clusters, and in low Galactic latitude fields, may also be
affected by inaccurate modeling of the PSF if stars in crowded regions
were selected as PSF stars by {\it Photo}.

Photometric information in crowded fields can be extracted from the
original SDSS imaging data.  For example, \citet{smolcic:07} used the
DoPHOT \citep{schechter:93} photometry package to explore the structure
of the Leo~I dwarf spheroidal galaxy (dSph).  Similarly, \citet{coleman:07}
used the DAOPHOT package to study the stellar distribution of the dSph Leo~II.
In this paper, we employ the DAOPHOT/ALLFRAME \citep{stetson:87,stetson:94}
suite of programs to derive photometry for 17 globular clusters and 3
open clusters that have been observed with SDSS.  We derive photometry by
running DAOPHOT for SDSS imaging frames where {\it Photo} did not run.
In addition, we reduce imaging data for fields farther away from the
clusters, where the {\it Photo} results are expected to be reliable, in
order to set up photometric zero points for the DAOPHOT photometry.  We
also compare DAOPHOT and {\it Photo} results for the open cluster fields
to verify the accuracy of the {\it Photo} magnitudes in these semi-crowded
fields.

An overview of the SDSS imaging survey and our sample clusters are
presented in \S~2.  In \S~3 we describe the preparation of imaging
data from the SDSS database.  In \S~4 we describe the method of crowded
field photometry using DAOPHOT/ALLFRAME, and evaluate the photometric
accuracy.  In \S~5 we present cluster photometry and fiducial sequences,
and compare them with theoretical stellar isochrones and fiducial
sequences in $u'g'r'i'z'$.

\section{SDSS Observations of Galactic Clusters}

\begin{figure*}
\epsscale{1.0}
\plotone{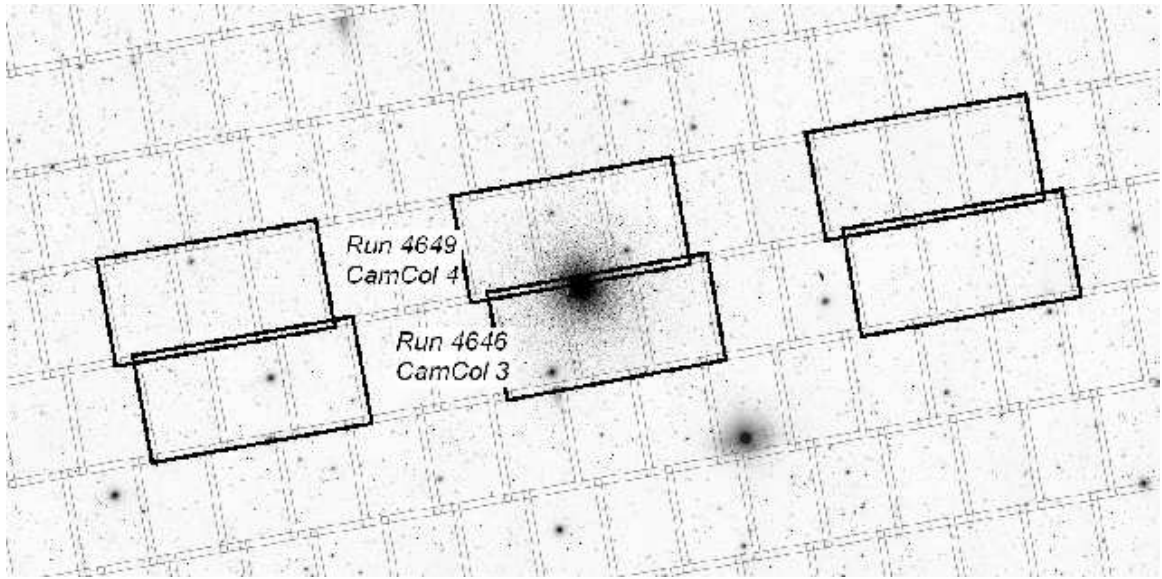}
\caption{SDSS scans over the $2\arcdeg \times 1\arcdeg$ region surrounding M3,
generated using the SDSS Finding Chart Tool.  Each of the horizontal strips
represents a scanning footprint for each CCD.  This strip is divided into
rectangular frames with small overlapping regions.  Adjacent strips in
different runs also overlap with each other.  Frames reduced in this
paper are indicated as thick boxes.  For each run and camcol, flanking
areas are shown on each side of the cluster.  North is to the top, and
the east to the left.\label{fig:m3scan}}
\end{figure*}

The SDSS images are taken in drift-scan or time-delay-and-integrate (TDI)
mode, with an effective exposure time of $54.1$ seconds per band.  The
imaging is carried out on moonless nights of good seeing (better than
$1.6\arcsec$) under photometric conditions \citep{hogg:01}.  A portion
of the sky (along great circles) is imaged in each run by 6 columns of
CCDs \citep{gunn:98}.  Each CCD observes $13.52\arcmin$ of sky,
forming a scanline or camcol, with a gap of $11.68\arcmin$ between the
columns.  A second scan or strip in a different run fills in the gap,
overlapping the first scan by 8\% on each side \citep{york:00}.  An
example of the scanning pattern is shown in Figure~\ref{fig:m3scan} for
the globular cluster M3.  Each of the rectangular regions represent a
SDSS field.  Frames reduced in this paper are indicated as thick boxes.

Table~\ref{tab:propgc} lists our sample of globular clusters observed
by SDSS, and summarizes estimates of the reddening, distance moduli,
and metallicity measurements for these clusters reported in the recent
literature.  A number of the properties are taken from the catalog of
\citet[][February 2003 revision]{harris:96}.  We also include the [Fe/H]
values reported by \citet{kraft:03,kraft:04}.  These are based on
\ion{Fe}{2} lines from high-resolution spectra, which are expected to be
less affected by non-local thermodynamic equilibrium (non-LTE) effects.
In their study a consistent technique was employed to derive
metallicities for giants in 16 key well-known globular clusters.  Seven
clusters in our sample are included in their sample of 16 key clusters.
For the remaining sample clusters we list their [Fe/H] determinations
based on the correlation between ${\rm [Fe/H]_{II}}$ and the reduced
strength of the near-infrared \ion{Ca}{2} triplet.  Although these
accurate metallicities make Galactic clusters useful calibrators, some
of the lighter elements (always C and N, but sometimes also O, Na, Mg,
and Al as well) vary from star to star \citep[e.g.,][]{kraft:94}.
However, extensive studies have
shown that the abundances of most elements (in particular Fe) are the
same for all cluster stars (in the sample we consider), and the overall
effect on the colors of stars in broadband filters, such as $ugriz$,
from variations of these lighter elements should be small.

\begin{deluxetable*}{llrrccccccc}
\tablewidth{0pt}
\tablecaption{Globular Cluster Properties\label{tab:propgc}}
\tablehead{
  \colhead{} &
  \colhead{Alternate} &
  \colhead{$l$} &
  \colhead{$b$} &
  \colhead{$V_{\rm HB}$} &
  \multicolumn{2}{c}{$E(B - V)$} &
  \multicolumn{2}{c}{$(m - M)_0$} &
  \multicolumn{2}{c}{[Fe/H]} \nl
  \cline{6-7} \cline{8-9} \cline{10-11}
  \colhead{NGC} &
  \colhead{Name} &
  \colhead{(deg)} &
  \colhead{(deg)} &
  \colhead{Harris} &
  \colhead{Harris}  &
  \colhead{Schlegel et al.}  &
  \colhead{Harris\tablenotemark{a}}  &
  \colhead{Kraft \& Ivans}  &
  \colhead{Harris} &
  \colhead{Kraft \& Ivans\tablenotemark{b}}
}
\startdata
2419 &        & $180.370$ & $+25.242$ & $20.45$ & $0.11$ & $0.06$ & $19.63$ & \nodata & $-2.12$ & \nodata   \nl
     & Pal~3  & $240.139$ & $+41.861$ & $20.51$ & $0.04$ & $0.04$ & $19.84$ & \nodata & $-1.66$ & $(-1.66)$ \nl
     & Pal~4  & $202.311$ & $+71.803$ & $20.80$ & $0.01$ & $0.02$ & $20.19$ & \nodata & $-1.48$ & $(-1.43)$ \nl
4147 &        & $252.849$ & $+77.189$ & $17.01$ & $0.02$ & $0.03$ & $16.42$ & \nodata & $-1.83$ & $(-1.79)$ \nl
5024 & M53    & $332.967$ & $+79.765$ & $16.81$ & $0.02$ & $0.02$ & $16.25$ & \nodata & $-1.99$ & $(-2.02)$ \nl
5053 &        & $335.690$ & $+78.944$ & $16.65$ & $0.04$ & $0.02$ & $16.07$ & \nodata & $-2.29$ & $(-2.41)$ \nl
5272 & M3     & $ 42.208$ & $+78.708$ & $15.68$ & $0.01$ & $0.01$ & $15.09$ & $15.02$ & $-1.57$ & $-1.50$   \nl
5466 &        & $ 42.150$ & $+73.592$ & $16.47$ & $0.00$ & $0.02$ & $16.00$ & \nodata & $-2.22$ & \nodata   \nl
     & Pal~5  & $  0.852$ & $+45.860$ & $17.51$ & $0.03$ & $0.06$ & $16.83$ & \nodata & $-1.41$ & \nodata   \nl
5904 & M5     & $  3.863$ & $+46.796$ & $15.07$ & $0.03$ & $0.04$ & $14.37$ & $14.42$ & $-1.27$ & $-1.26$   \nl
     & Pal~14 & $ 28.747$ & $+42.199$ & $20.04$ & $0.04$ & $0.03$ & $19.35$ & \nodata & $-1.52$ & $(-1.61)$ \nl
6205 & M13    & $ 59.008$ & $+40.912$ & $15.05$ & $0.02$ & $0.02$ & $14.42$ & $14.42$ & $-1.54$ & $-1.60$   \nl
6341 & M92    & $ 68.339$ & $+34.859$ & $15.10$ & $0.02$ & $0.02$ & $14.58$ & $14.75$ & $-2.28$ & $-2.38$   \nl
6838 & M71    & $ 56.744$ & $ -4.564$ & $14.48$ & $0.25$ & $0.32$ & $13.02$ & \nodata & $-0.73$ & $-0.81$   \nl
7006 &        & $ 63.770$ & $-19.407$ & $18.80$ & $0.05$ & $0.08$ & $18.09$ & \nodata & $-1.63$ & $-1.48$   \nl
7078 & M15    & $ 65.013$ & $-27.313$ & $15.83$ & $0.10$ & $0.11$ & $15.06$ & $15.25$ & $-2.26$ & $-2.42$   \nl
7089 & M2     & $ 53.371$ & $-35.770$ & $16.05$ & $0.06$ & $0.04$ & $15.30$ & \nodata & $-1.62$ & $(-1.56)$ \nl
\enddata
\tablenotetext{a}{True distance modulus assuming $A_V / E(B - V) = 3.1$ and reddening values in Harris.}
\tablenotetext{b}{[Fe/H] estimates \citep{kraft:04} in parentheses are derived from the
${\rm [Fe/H]}_{\rm II}$ correlation with the reduced strength of the \ion{Ca}{2} triplet
\citep{kraft:03}.}
\end{deluxetable*}

\begin{deluxetable*}{llrrccccc}
\tablewidth{0pt}
\tablecaption{Open Cluster Properties\label{tab:propoc}}
\tablehead{
  \colhead{} &
  \colhead{Alternate} &
  \colhead{$l$} &
  \colhead{$b$} &
  \multicolumn{2}{c}{$E(B - V)$} &
  \colhead{} &
  \colhead{} &
  \colhead{} \nl
  \cline{5-6}
  \colhead{NGC} &
  \colhead{Name} &
  \colhead{(deg)} &
  \colhead{(deg)} &
  \colhead{Stellar} &
  \colhead{Schlegel et al.}  &
  \colhead{$(m - M)_0$} &
  \colhead{[Fe/H]} &
  \colhead{ref}
}
\startdata
2420 &     & $198.107$ & $+19.634$ & $0.05$ & $0.04$ & $12.00$\tablenotemark{a} & $-0.37$  & 1\nl
2682 & M67 & $215.696$ & $+31.896$ & $0.04$ & $0.03$ & $ 9.61$ & $+0.00$  & 2\nl
6791 &     & $ 69.958$ & $+10.904$ & $0.10$ & $0.16$ & $13.02$ & $+0.40$  & 3\nl
\enddata
\tablerefs{Estimates for stellar $E(B - V)$, $(m - M)_0$, and [Fe/H]: 
(1) Anthony-Twarog et~al. 2006; (2) An et~al. 2007b, and
references therein; (3) Pinsonneault et~al. 2008 (in preparation) and
references therein.}
\tablenotetext{a}{True distance modulus assuming $A_V / E(B - V) = 3.1$.}.
\end{deluxetable*}

The distances and reddenings to clusters have also been the subject of
much research.  Kraft \& Ivans used the {\it Hipparcos} subdwarfs (see
references therein for their sample selection) to derive distances to
five key globular clusters from their application of the MS fitting
technique.  These {\it Hipparcos}-based distances are expected to be
accurate to $\sim10-15\%$ \citep[e.g.,][]{gratton:97,reid:97}, which
will hopefully be improved greatly from upcoming astrometric missions
such as {\it Gaia} \citep{perryman:01}.  They also provided estimates of
reddening for the clusters by comparing colors derived from
high-resolution spectroscopic determinations of $T_{\rm eff}$ with
the observed colors of the same stars.

Table~\ref{tab:propoc} lists reddening, distance, and metallicity
estimates for our sample open clusters.  For NGC~2420 we list those
given by \citet{anthony-twarog:06}, which are based on intermediate-band
$vbyCaH\beta$ photometry.  For M67 we take the reddening,
distance, and metallicity reported by \citet{an:07b}, which is 
an average between literature values and those estimated using empirically
calibrated sets of isochrones \citep[see also][]{pinsono:03,pinsono:04}.
The latter set of authors also used an extended set of calibrated
isochrones to estimate these parameters for NGC~6791 (M.\ H.\ Pinsonneault
et~al.\ 2008, in preparation); these values are listed in
Table~\ref{tab:propoc}.  Their metallicity estimate for NGC~6791 is
consistent with recent results from high-resolution spectroscopic studies
\citep{carraro:06,gratton:06,origlia:06}.  Their reddening estimate based
on the stellar sequence [$E(B - V) = 0.10\pm0.01$] is lower than the
\citet{schlegel:98} value.

Although typical SDSS imaging scans involve small overlaps between adjacent
stripes in most of the survey area, occasionally two runs from adjacent
stripes overlap by a larger fraction of each field.  This results in a
large number of stars with repeated flux measurements, providing an
opportunity to estimate realistic photometric errors (\S~\ref{sec:error}).
Five clusters in our sample (M67, NGC~2420, NGC~5466, NGC~6791, and Pal~14)
have been scanned in such a manner, covering most of the cluster fields twice.

\section{Data Acquisition and Preparation}

\begin{deluxetable}{lcll}
\tablewidth{0pt}
\tablecaption{Frames Reduced\label{tab:fields}}
\tablehead{
  \colhead{Cluster} & \colhead{} &
  \multicolumn{2}{c}{Fields} \nl
  \cline{3-4}
  \colhead{Name} &
  \colhead{Run-ReRun-CamCol} &
  \colhead{Cluster} & \colhead{Flanking}
}
\startdata
NGC~2420 & 2888-40-3 & 024-026 & 020-022/028-030 \nl
         & 3462-40-6 & 024-026 & 019-021/028-030 \nl
         & 3513-40-6 & 021-023 & 016-018/025-027 \nl
NGC~2419 & 1350-40-4 & 054     & 049-051/057-059 \nl
         & 1402-40-6 & 044-046 & 040-042/048-050 \nl
M67      & 5935-40-2 & 108-110 & 103-105/112-114 \nl
         & 5972-40-6 & 117-119 & 112-114/121-123 \nl
         & 6004-40-5 & 114-116 & 109-111/118-120 \nl
Pal~3    & 2141-40-4 & 059     & 054-056/062-064 \nl
Pal~4    & 5061-40-3 & 375     & 370-372/378-380 \nl
         & 5071-40-2 & 373     & 368-370/376-378 \nl
NGC~4147 & 5360-40-6 & 239     & 234-236/243-245 \nl
         & 5381-40-6 & 186     & 181-183/189-191 \nl
M53      & 5360-40-6 & 338-340 & 333-335/343-345 \nl
         & 5390-40-6 & 161-163 & 156-158/166-168 \nl
NGC~5053 & 5360-40-5 & 344-345 & 339-341/348-350 \nl
         & 5390-40-5 & 167-168 & 162-164/171-173 \nl
M3       & 4646-40-3 & 080-082 & 075-077/085-087 \nl
         & 4649-40-4 & 145-147 & 140-142/150-152 \nl
NGC~5466 & 4623-40-1 & 324-325 & 319-321/327-329 \nl
         & 4646-40-6 & 114-115 & 109-111/118-120 \nl
Pal~5    & 0756-44-3 & 755-757 & 751-753/760-762 \nl
M5       & 1458-40-4 & 699-701 & 694-696/704-706 \nl
         & 2327-40-4 & 048-049 & 043-045/052-054 \nl
Pal~14   & 4670-40-2 & 302     & 297-299/305-307 \nl
         & 5323-40-6 & 095     & 090-092/098-100 \nl
M13      & 3225-40-4 & 264-266 & 259-261/268-270 \nl
         & 3226-40-5 & 126-128 & 121-123/131-133 \nl
M92      & 4682-40-6 & 207-209 & 202-204/211-213 \nl
         & 5327-40-6 & 013-014 & 017-019         \nl
NGC~6791 & 5403-40-4 & 190-192 & 185-187/195-197 \nl
         & 5416-40-3 & 190-192 & 186-188/194-196 \nl
         & 6177-40-3 & 043-045 & 039-041/047-049 \nl
M71      & 6895-40-3 & 052-053 & 056-058         \nl
NGC~7006 & 4879-40-2 & 091     & 086-088/094-096 \nl
M2       & 2583-40-2 & 135-137 & 131-133/140-142 \nl
         & 2662-40-1 & 037-038 & 032-034/041-043 \nl
M15      & 1739-40-6 & 058-059 & 053-055/062-064 \nl
         & 2566-40-6 & 064-065 & 059-061/068-070 \nl
\enddata
\end{deluxetable}

We retrieved the {\tt fpC} corrected imaging frames and the {\tt fpM}
mask frames for the cluster fields from the Data Archive Server (DAS)
for all five bandpasses.  We also downloaded the best version {\tt tsField}
and {\tt asTrans} files for each field.  Table~\ref{tab:fields} lists
the SDSS run, rerun, camcol, and field numbers for each cluster field
analyzed. In addition to the cluster fields, we reduced flanking fields
(\S~\ref{sec:zeropoint}) belonging to the same run, rerun, and camcol,
which had considerably lower stellar densities. These fields had been
successfully run through {\it Photo}, and are used to set the zero points
for the DAOPHOT photometry by comparing the magnitudes of the stars in
the two different reductions.  This insures that our reductions
are securely tied to the 2.5-meter $ugriz$ photometric system.

Table~\ref{tab:fields} also lists the SDSS fields reduced using DAOPHOT. For
cluster fields we typically combined two or three contiguous fields to form a
single field, using the IRAF\footnote{IRAF is distributed by the National
Optical Astronomy Observatory, which is operated by the Association of
Universities for Research in Astronomy, Inc., under cooperative agreement with
the National Science Foundation.} package {\tt imtile} for DAOPHOT
reductions.\footnote{We use the term ``field'' to represent the combined SDSS
fields, each of which is defined as the data processing unit in {\it Photo}.}
Some of the globular clusters subtend a small enough angle that an entire cluster
fits within a single SDSS field. For these cases we did not attempt to include
adjacent fields. For the flanking fields we always combined three SDSS fields to
form a single field, and reduced it as a single data processing unit in DAOPHOT.

\begin{deluxetable*}{crrrrrrrrrr}
\tablewidth{0pt}
\tablecaption{Detector Properties\label{tab:ccd}}
\tablehead{
  \colhead{} &
  \multicolumn{5}{c}{GAIN (e$^-$/DN)} &
  \multicolumn{5}{c}{READ NOISE (DN$^2$)} \nl
  \cline{2-6} \cline{7-11}
  \colhead{CamCol} &
  \colhead{$u$} & \colhead{$g$} & \colhead{$r$}  & \colhead{$i$} & \colhead{$z$} &
  \colhead{$u$} & \colhead{$g$} & \colhead{$r$}  & \colhead{$i$} & \colhead{$z$} 
}
\startdata
$1$&$1.62$&$3.32$&$4.71$&$5.17$&$4.75$&$ 9.61$&$15.60$&$1.82$&$7.84$&$0.81$\nl
$2$&$1.60$&$3.86$&$4.60$&$6.57$&$5.16$&$12.60$&$ 1.44$&$1.00$&$5.76$&$1.00$\nl
$3$&$1.59$&$3.85$&$4.72$&$4.86$&$4.89$&$ 8.70$&$ 1.32$&$1.32$&$4.62$&$1.00$\nl
$4$&$1.60$&$4.00$&$4.76$&$4.89$&$4.78$&$12.60$&$ 1.96$&$1.32$&$6.25$&$9.61$\nl
$5$&$1.47$&$4.05$&$4.73$&$4.64$&$3.48$&$ 9.30$&$ 1.10$&$0.81$&$7.84$&$1.82$\nl
$6$&$2.17$&$4.04$&$4.90$&$4.76$&$4.69$&$ 7.02$&$ 1.82$&$0.90$&$5.06$&$1.21$\nl
\enddata
\end{deluxetable*}

Before running DAOPHOT, we removed the softbias of 1000~DN and masked
pixels affected by saturation.  DAOPHOT identifies pixels above a
``high good datum'' value as saturated.  However, for SDSS a saturated
pixel will overflow and pour charge into its neighbors.  This results
in distorted shapes for the PSF of the brightest stars, but without
necessarily setting the counts in an affected pixel above a certain value.
It is also noted that the full well depth varies from chip to chip.
To set the pixel value to a large number, we used the SDSS {\it readAtlasImages}
code\footnote{www.sdss.org/dr6/products/images/read\_mask.html} 
to set the pixels flagged as saturated, as well as the pixels within a
radius of 3 pixels, equal to 70000~DN. The bad pixel value in DAOPHOT was
then set to 65000~DN.  The gain and readnoise values for each chip and 
filter are listed in Table~\ref{tab:ccd}.

\section{Crowded Field Photometry}

\subsection{DAOPHOT/ALLFRAME Reduction}

The goal of this study is to obtain accurate photometry for stars in the
crowded cluster fields using the same final aperture radius as the SDSS
data, which is $18.584$ pixels ($7.43\arcsec$).  We used DAOPHOT and its
accompanying program ALLSTAR to find stars, derive a spatially varying
PSF, and perform the first measurements for all stars in a single field
and a single filter.  We then matched the stars from all the filters
in a single field, using DAOMATCH and DAOMASTER, to form a master list
that served as the input into ALLFRAME, which simultaneously reduces all the
data for a particular field.  The remainder of this section describes the
reduction process in detail.

Stars were identified in each frame using DAOPHOT/FIND, with a threshold
of 4$\sigma$, a low sharpness\footnote{The index sharpness in DAOPHOT/FIND
is defined as the ratio of the height of the bivariate delta-function,
which best fits the brightness peak, to the height of the bivariate
Gaussian function, which best fits the peak.}
cutoff of 0.30, a high sharpness cutoff of
1.40, a low roundness cutoff of $-1.00$, and a high roundness cutoff of
1.00.  The FWHM parameter, which is used to define the Gaussian that
detects stars, was set to the average FWHM of stars
in each frame.  Since our goal is to derive cluster fiducial sequences
rather than cluster luminosity functions, no effort was made to correct for
incompleteness, which obviously increases in the more crowded regions.

For each frame, a large number ($\sim 100-300$) of relatively isolated
stars spread across the frame were chosen as PSF stars. The fitting
radius of the PSF in the $\chi^2$ minimization was set to 5 pixels.
The PSF was calculated out to a radius of 15 pixels, which defines how
far out the light from the star was subtracted on the image.  DAOPHOT
first determined a constant, analytic PSF across the entire
field. Neighboring stars of the PSF stars were identified. Next, the
neighboring stars were subtracted using ALLSTAR with the first pass
PSF. The PSF was then re-determined with both the analytic and look-up
table components. This new PSF was then used to subtract the
neighbors. The next iteration allowed the PSF to vary linearly across
the frame, and subsequent iterations increased the variability to
quadratic and finally cubic. ALLSTAR was then run on the entire field.
The subtracted image was searched for additional stars, this time with
a threshold of 10$\sigma$, before the final ALLSTAR run was performed
on all stars.

Before we can run ALLFRAME, we need to provide a master list of  stars
so that the same stars can be used to reduce each frame.  In contrast to
the usual method of crowded field photometry, where multiple long and
short exposures are taken in all filters, SDSS scans most regions
once, with only small parts of the frames overlapping with a separate
scan.  Therefore, the standard technique of using exposures in the same
filter to eliminate cosmic rays and spurious detections (e.g., in the
bright wings of badly subtracted stars) is not possible.

Instead, we relied on multiple detections in different filter bandpasses
to reduce the chances of spurious detections.  We first used DAOMATCH
and DAOMASTER to match stars among different filter frames.  We then
made a master star list, requiring that a star should appear in at
least two of the frames.  This leads to the possibility that a star
could be a real detection in one of the bandpasses, but be eliminated
because of a lack of detection in the other four filters.
However, our selection criterion, with a minimum two detections in
different filter frames, insures that each star has at least one color,
which is crucial in the derivation of the fiducial sequences.  The master
list served as the input for ALLFRAME, which simultaneously determines the
separate brightnesses for stars in all frames while enforcing one set of
centroids and one transformation between images.

Finally, we applied aperture corrections to obtain the instrumental
magnitude of a star within an $18.584$ pixel aperture radius.  After
subtracting all other stars, we measured the aperture magnitudes for
the PSF stars through 12 different radii.  We used HSTDAOGROW, which
is a modified version of DAOGROW \citep{stetson:90}, to calculate the
total magnitude by the curve of growth method.  The difference between
these programs is that HSTDAOGROW does not extrapolate to twice
the largest aperture, as is done by DAOGROW.  The difference between the PSF
magnitude and the aperture magnitude for each star defines the aperture
correction.  We calculated the average difference after iterating twice
and discarding stars that were more than $1.8$ times the rms away from
the mean.  This final average aperture correction was applied to all of
the PSF magnitudes.

We converted (aperture-corrected PSF) DAOPHOT (Pogson) magnitudes into
the SDSS asinh magnitude system \citep[luptitude;][]{lupton:99} using
the photometric zero point ({\tt aa}), extinction coefficient ({\tt kk}),
and air-mass ({\tt airmass}) values from the {\tt tsField} files:
\begin{equation}
{\rm mag} = -(2.5/ln(10)) [asinh((f/f_0)/2b)+ln(b)]
\label{eq:luptitude1}
\end{equation}
where $b$ is the softening parameter for the photometric band
in question\footnote{The following softening parameter $b$ values are used in SDSS:
$1.4 \times 10^{-10}$,
$0.9 \times 10^{-10}$,
$1.2 \times 10^{-10}$,
$1.8 \times 10^{-10}$, and
$7.4 \times 10^{-10}$ in $u$, $g$, $r$, $i$, and $z$, respectively.
See also\\ {\tt http://www.sdss.org/dr6/algorithms/fluxcal.html}.} and
\begin{equation}
f/f_0 = ({\rm counts}/53.907456\ {\rm sec})
      \times 10^{0.4({\tt aa} + {\tt kk} \times {\tt airmass})}.
\label{eq:luptitude2}
\end{equation}
The air-mass value used was either the one for the central frame, if
three frames were used, or the eastern frame of the cluster field set.
Any changes in air mass during the time of the 2-3 frame scan were
negligible.

We used {\it cvtcoords} in the SDSS astrotools suite of programs
as well as the information in the {\tt asTrans} files from the
DAS to determine celestial coordinates of right ascension and
declination of the stars in the $r$-band images.  Astrometric positions
in SDSS are accurate to $< 0.1\arcsec$ for sources with $r < 20.5$~mag
\citep{pier:03}.

\subsection{Photometric Zero Points}
\label{sec:zeropoint}

Our initial DAOPHOT reductions in relatively low stellar-density fields
showed that there exist $\sim0.02$ mag differences between DAOPHOT and
{\it Photo} magnitudes.  Since the DAOPHOT reduction in this study does
not include photometric standard fields to independently calibrate the
data, we put the DAOPHOT magnitudes onto the {\it Photo} scale as described
below.

\subsubsection{Data}

In order to place our DAOPHOT photometry on the same scale as that
determined by {\it Photo}, it was necessary to compare results for stars
that are far enough from the clusters' dense stellar fields to avoid
crowding effects, but close enough to represent the local photometric
properties near the clusters.  These comparisons are accomplished by
using stars contained in a set of flanking fields that lie at least
two frames away ($\approx20\arcmin$) from the crowded cluster fields.
An example of flanking fields is shown in Figure~\ref{fig:m3scan} for M3.

The location of these flanking fields is largely based on how
{\it Photo} computes the model PSF.  In {\it Photo}, the PSF is modeled
using a Karhunen-Lo\`eve (KL) transform, where stars lying in $\pm2$
adjacent frames (in the scan direction) are used to determine the KL
basis functions \citep{lupton:02}.  At the same time, {\it Photo} also
relies upon stars in $\pm0.5$ adjacent frames to follow the spatial
(temporal) variation of the PSF.  Therefore, the two closest fields to a
cluster, where {\it Photo} has modeled the PSF without using stars in
the crowded region, are those that are two frames away from the
crowded fields.

For each run and camcol we selected flanking areas on each side of
a cluster, which we refer to as western and eastern flanking fields.  We
combined three contiguous fields to form a single flanking field and
derived stellar magnitudes following the same procedure as cluster
photometry using DAOPHOT.  By analyzing three combined fields instead of
just one, we had a larger number of bright PSF stars, especially in
relative sparse halo fields.  We typically selected $50-100$ PSF stars in
each flanking field with good signal-to-noise ratios ($r\la18$~mag), and
used a model PSF that varies cubically with position.  Although we had
only $\sim30$ PSF stars in the $u$-band frames for about one third of
our flanking fields, we found that the cubically varying PSF is
necessary to adequately reduce data in these fields.  We used HSTDAOGROW
to determine aperture corrections, and converted DAOPHOT magnitudes
into the SDSS asinh magnitudes.

To derive an accurate photometric zero point, we used photometry with
errors smaller than 0.05~mag in each band (errors reported from DAOPHOT)
for stars brighter than $15.5$~mag in $u$, $16.0$~mag in $gri$, and $15.0$~mag
in $z$.  We additionally filtered data based on the sharp\footnote{The index
sharp used here is defined differently from the sharpness index in DAOPHOT/FIND:
sharp$^2 \approx |s_{\rm obs}^2 - s_{\rm PSF}^2 |$ where $s_{\rm obs}$
is a characteristic radius of the measured image profile, and $s_{\rm PSF}$
is a characteristic radius of the PSF.  The sign of the sharp index is positive if
$s_{\rm obs} > s_{\rm PSF}$.} and $\chi$ values from DAOPHOT.  We adopted
$| {\rm sharp} | < 1$ and $\chi < 1.5+4.5\times 10^{-0.4 (m - m_0)}$
\citep{stetson:03}, where $m_0 = 15.5$~mag in $u$, $m_0 = 16.0$~mag in $gri$,
and $m_0 = 15.0$~mag in $z$, in order to remove objects that have relatively poor
goodness-of-fit values.

We retrieved {\it Photo} PSF magnitudes either from the Catalog Archive
Server (CAS) in the Sixth Data Release, or directly from the {\tt tsObject}
files when the data were not yet available through the data release.
We used {\it Photo} magnitudes for stars that passed a set of photometric
criteria to obtain clean photometry.  We selected objects that are
classified as {\tt STAR} (unresolved point sources) and used SDSS primary
or secondary detections with photometric errors smaller than $0.05$~mag.
For the $r$-band image of run {\tt 5071}, camcol {\tt 2}, field {\tt 376}
(hereafter we use the format {\tt 5071-r2-376} to represent a specific
frame) and {\tt 6895-i3-56}, we relaxed the threshold to $0.06$~mag
because all of the {\it Photo} magnitudes in these fields have errors
larger than $0.05$~mag.  We ignored photometry for objects that have
the following flags set:  {\tt EDGE, NOPROFILE, PEAKCENTER, NOTCHECKED,
PSF\_FLUX\_INTERP, SATURATED, BAD\_COUNTS\_ERROR, DEBLEND\_NOPEAK,
INTERP\_CENTER, or COSMIC\_RAY} \citep[e.g.,][]{ivezic:07}.\footnote{See
also\\ {\tt http://www.sdss.org/dr6/products/catalogs/flags.html.}}  We
employed these selection criteria in each filter bandpass, in order not
to eliminate a star from all of the filter frames even if it was flagged
or had a large error in one of the bandpasses.  This helps to keep many
of the point sources that were poorly detected in the $u$-band frames.
We then cross-matched photometry in {\it Photo} and DAOPHOT, using
a search radius of $3$~pixels.

\subsubsection{Comparison with {\it Photo}}

For the flanking fields we first compared the number of stellar objects
found by DAOPHOT and {\it Photo}.  In the $r$ band, DAOPHOT detected
$\sim50$ to $\sim5000$ stellar sources on each (SDSS) field with
$r < 20$~mag.  Among these, {\it Photo} recovered on average $75\%\pm9\%$
of the sources classified as {\tt STAR}, except in the case of M71
flanking field {\tt 6895-r3-56/57/58} (see below).  All detections in
{\it Photo} were recovered by DAOPHOT; there is no apparent trend of the
detection rate in {\it Photo} as a function of the total number of
detected sources in DAOPHOT.  The average recovery fraction in
{\it Photo} becomes $87\%\pm4\%$ when we matched sources in DAOPHOT with
the above $\chi$ and sharp index selections.

\begin{figure}
\epsscale{1.1}
\plotone{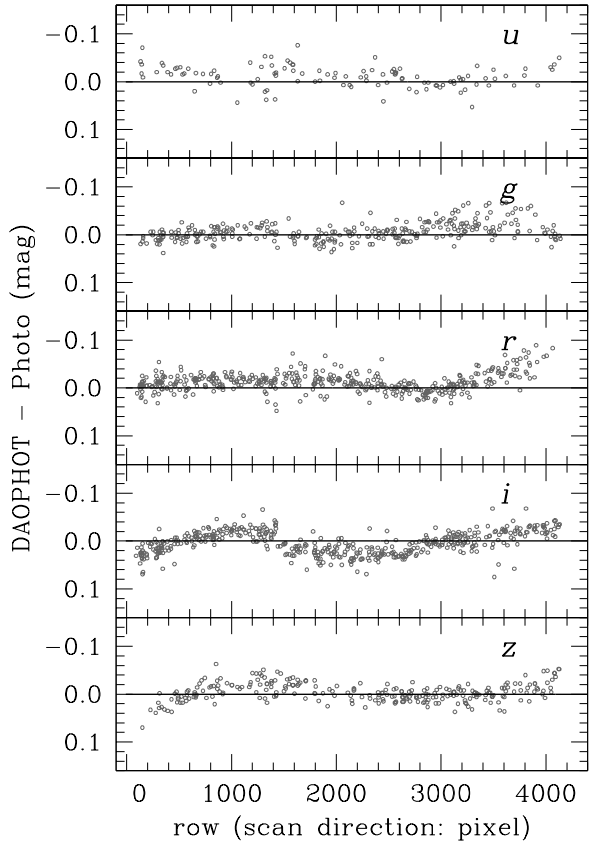}
\caption{Comparison between the DAOPHOT and the {\it Photo} magnitudes in one
of the M13 flanking fields.  Comparisons are shown for $u$, $g$, $r$, $i$,
and $z$, from the top to the bottom panels.\label{fig:daophotvsphoto}}
\end{figure}

Figure~\ref{fig:daophotvsphoto} shows the comparison between DAOPHOT
and {\it Photo} magnitudes in one of the flanking fields for M13,
{\tt 3226-5-121/122/123}.  The constant offsets between the two are the
zero-point corrections that will be applied to the cluster photometry.
However, we noted that there are systematic variations in the difference,
at a level of $\sim2\%$, in the scan direction or over time, in
addition to photon noise.  These high spatial frequency structures are
likely due to fast PSF variations, which were not followed by the PSF
of {\it Photo} on a rapid enough spatial or temporal scale.

The accuracy of PSF photometry can be best tested from the comparison
with aperture photometry for isolated bright stars.  Specifically,
we can use the individual aperture corrections to test how accurately
our model PSF accounted for the variability of the PSF by plotting the
difference between the PSF magnitude and the aperture magnitude as
a function of position.  If we have modeled the variability of the PSF
sufficiently well, there will be no dependence of that difference on the
position of the stars on the chip.

\begin{figure}
\epsscale{1.1}
\plotone{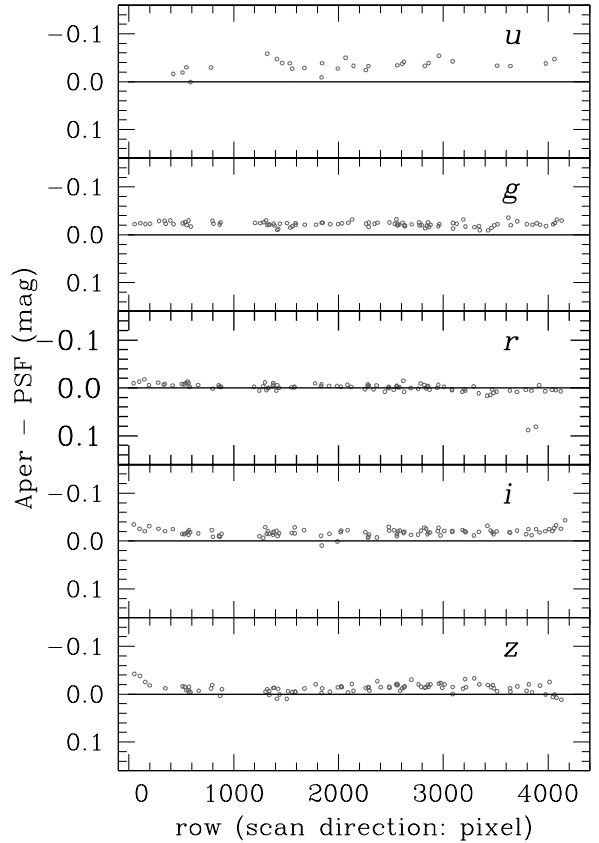}
\caption{Comparison between the DAOPHOT and the aperture photometry in the same
flanking field as in Fig.~\ref{fig:daophotvsphoto}.  The aperture
photometry was derived using HSTDAOGROW for PSF stars, shown as circles.
Comparisons are shown for $u$, $g$, $r$, $i$, and $z$, from the top to the bottom
panels.\label{fig:daophotaper}}
\end{figure}

Figure~\ref{fig:daophotaper} shows the differences between the DAOPHOT PSF and
aperture magnitudes, with aperture radii of $18.584$~pixels from the
HSTDAOGROW analysis.  Individual points are those used in our PSF modeling,
and an average offset in each panel represents the aperture correction
for DAOPHOT PSF photometry in each filter bandpass.  The same fields are
shown as in the above comparison with {\it Photo}.  However, in contrast
to the systematic variations seen in Figure~\ref{fig:daophotvsphoto},
the differences between the DAOPHOT and aperture magnitudes are quite
uniform, to better than $1\%$ accuracy, over a time scale covered by
at least one flanking field ($\sim30\arcmin$ or $\sim3$~min in time).
This test shows that the systematic residuals in Figure~\ref{fig:daophotvsphoto}
are due to errors in the {\it Photo} magnitudes, presumably due to imperfect
modeling of the PSFs.  In support of this conclusion, we found  similar
high spatial frequency structures (in size and amplitude) as those in
Figure~\ref{fig:daophotvsphoto}, obtained from the comparison between the {\it Photo}
PSF and aperture 7 magnitudes (flux measurements with an 18.584~pixel
aperture from {\it Photo}).

\begin{figure}
\epsscale{1.1}
\plotone{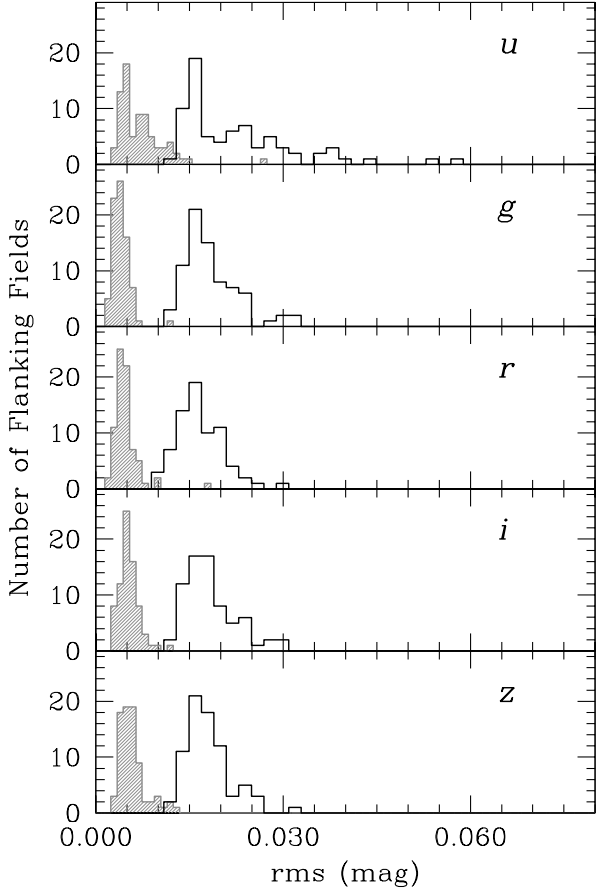}
\caption{Distribution of the rms differences between the DAOPHOT and
the aperture photometry ({\it grey shaded histogram}) and between the DAOPHOT
and the {\it Photo} photometry ({\it black histogram}) from all of the flanking
fields.\label{fig:rms}}
\end{figure}

Other flanking fields also exhibit stable differences between DAOPHOT
magnitudes and aperture photometry, but exhibit systematic variations
of {\it Photo} magnitudes, seen most clearly in the scanning
direction.  After iterative $3\sigma$ clipping, we computed a rms
dispersion for each field and estimated a median of the rms from all
of our flanking fields (except for the few cases discussed below).  From a
comparison between the HSTDAOGROW aperture magnitudes and the DAOPHOT
magnitudes, we found a median rms of $0.0061$~mag, $0.0039$~mag, $0.0045$~mag,
$0.0052$~mag, and $0.0054$~mag in $ugriz$, respectively, yielding
a precise aperture correction and its spatial uniformity.  From the
comparison between {\it Photo} and DAOPHOT, however, we found a factor of
three larger rms values:
$0.0198$~mag, $0.0172$~mag, $0.0162$~mag, $0.0176$~mag, and $0.0173$~mag
in $ugriz$, respectively (Fig.~\ref{fig:rms}).
We note that \citet{smolcic:07} compared the DoPHOT and {\it Photo}
photometry in an uncrowded field and estimated the rms differences of
$0.029$~mag, $0.013$~mag, $0.027$~mag in $gri$, respectively.

The $2\%$ variation of {\it Photo} is consistent with the specified
size of the photometric errors in the SDSS project \citep{ivezic:03,ivezic:04}.
While this level of accuracy already makes SDSS one of the most
successful optical surveys \citep[see also][]{sesar:06}, the spatial
variations of the {\it Photo} PSF magnitudes clearly indicates that
there is room for future improvement in the photometric accuracy
\citep[e.g.,][]{ivezic:07,padmanabhan:08}.

There is a small fraction of cases where DAOPHOT magnitudes vary
significantly with respect to aperture photometry over a given frame.
In some of these cases the difference between the DAOPHOT and the aperture
photometry jumps by $\sim0.02$~mag systematically in some parts of the
flanking fields.  These abrupt variations are strongly correlated
with the change in the PSF shapes, which may be caused by a sudden
change in the telescope focus or tracking.

We initially attempted to model these spatially varying PSFs by reducing
individual frames of each flanking field.  However, we found that, in most cases,
the sudden PSF variations could not be adequately modeled by cubically
varying the PSFs in DAOPHOT, so we decided not to include
such fields in the following analyses.  The problematic flanking fields
have significantly large rms values in the comparison with aperture
photometry, so we used rms cuts of $0.040$~mag in the $u$ band and
$0.020$~mag in $griz$, after initial $5\sigma$ clipping, to identify
them.  A total of nine flanking fields ($2\%$ of all of the 380
flanking fields in this study) were rejected using these cuts:
{\tt 3462-r6-19/20/21, 4649-i4-150/151/152, 5071-u2-368/369/370,
5071-i2-368/369/370, 5360-u5-339/340/341, 5360-i5-339/340/341,
5360-r6-343/344/345, 6004-r5-109/110/111, 6004-i5-109/110/111}.
For these runs we used photometry in a flanking field on the other side
of a cluster to set the photometric zero points.

Among the fields with small rms differences between the aperture and the DAOPHOT
photometry, the DAOPHOT magnitudes show particularly large rms differences
with {\it Photo} ($>0.050$~mag after initial $5\sigma$ clipping)
in two flanking fields: {\tt 5403-r4-185/186/187} (NGC~6791
run) and {\tt 6895-i3-56/57/58} (M71 run).  In the former case, most of
the dispersion comes from a strong discontinuity in the magnitude difference
of the field {\tt 187}, when compared with the preceding two fields. In the last
field, the {\tt PSP\_FIELD\_PSF11} flag in {\tt PspStatus} was set, indicating
that {\it Photo} magnitudes were derived using a spatially (temporally) constant
PSF. We did not use this field in the following analyses.

The {\tt 6895-3-56/57/58} (M71 run) has the highest stellar density
among cluster flanking fields in this study.  Approximately $15000$ point
sources were found in one flanking field with $r < 20$~mag from DAOPHOT.  The
large rms observed in this field was caused by a magnitude-dependent trend
in the difference between DAOPHOT and {\it Photo} (which is not seen in
other flanking fields): {\it Photo} detections become fainter with increasing
apparent magnitude.  One likely
explanation is that {\it Photo} over-estimates the sky brightness in crowded
fields, where it fails in the detection and subtraction of faint objects.
In fact, {\it Photo} detected only $\sim700$ sources classified as {\tt STAR}
with $r < 20$~mag in this flanking field and does not report flux measurements
in the other flanking field.  We derived zero-point differences between
{\it Photo} and DAOPHOT using bright stars with $u<20$~mag, $g<19$~mag,
$r<18$~mag, $i<18$~mag, and $z<17$~mag.

\begin{figure}
\epsscale{1.1}
\plotone{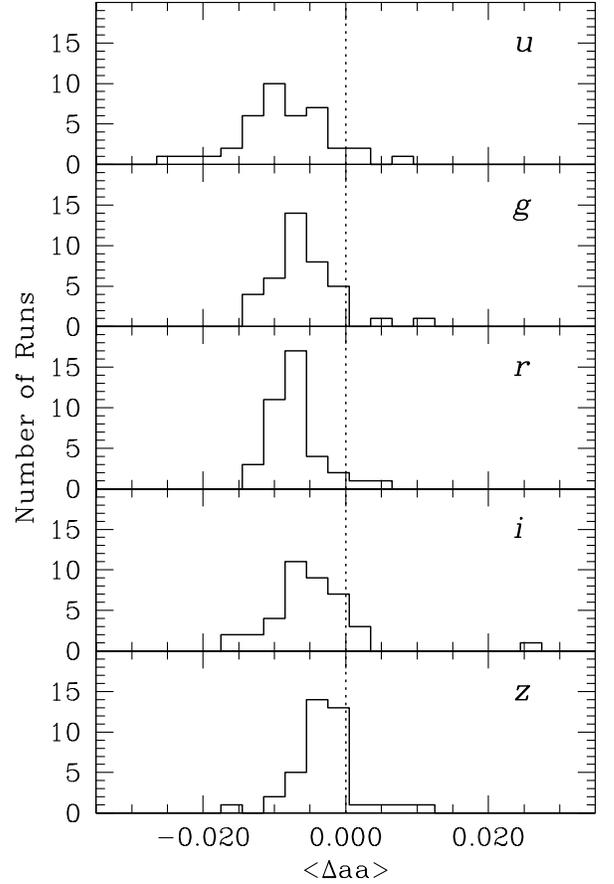}
\caption{Distribution of the zero-point corrections for the DAOPHOT photometry.
\label{fig:zp.comp}}
\end{figure}

\begin{deluxetable*}{rccccccl}
\tablewidth{0pt}
\tablecaption{The {\tt aa} Coefficients for DAOPHOT\label{tab:aa}}
\tablehead{
  \colhead{Run} &
  \colhead{CamCol} &
  \colhead{\tt aa\_u} &
  \colhead{\tt aa\_g} &
  \colhead{\tt aa\_r} &
  \colhead{\tt aa\_i} &
  \colhead{\tt aa\_z} &
  \colhead{Cluster}
}
\startdata
 756 & 3 & $-24.1047$ & $-24.5849$ & $-24.1775$ & $-23.7702$ & $-21.8598$ & Pal5    \nl
1350 & 4 & $-24.0218$ & $-24.5221$ & $-24.0788$ & $-23.7370$ & $-21.8173$ & NGC2419 \nl
1402 & 3 & $-24.0123$ & $-24.4928$ & $-24.0966$ & $-23.7079$ & $-21.8016$ & NGC2419 \nl
1458 & 4 & $-23.9525$ & $-24.4824$ & $-24.0408$ & $-23.6922$ & $-21.7321$ & M5      \nl
1739 & 6 & $-23.5430$ & $-24.3924$ & $-24.0260$ & $-23.6522$ & $-21.8577$ & M15     \nl
2141 & 4 & $-23.9682$ & $-24.4864$ & $-24.0662$ & $-23.7309$ & $-21.8065$ & Pal3    \nl
2327 & 4 & $-23.8256$ & $-24.4737$ & $-24.0553$ & $-23.7266$ & $-21.8064$ & M5      \nl
2566 & 6 & $-23.4285$ & $-24.3798$ & $-23.9759$ & $-23.6428$ & $-21.8381$ & M15     \nl
2583 & 2 & $-23.7293$ & $-24.3516$ & $-23.9822$ & $-23.5582$ & $-21.8425$ & M2      \nl
2662 & 1 & $-23.9651$ & $-24.5883$ & $-24.0796$ & $-23.6851$ & $-22.0001$ & M2      \nl
2888 & 3 & $-23.8460$ & $-24.4556$ & $-24.0773$ & $-23.7461$ & $-21.8830$ & NGC2420 \nl
3225 & 4 & $-23.7044$ & $-24.4065$ & $-23.9744$ & $-23.6809$ & $-21.7117$ & M13     \nl
3226 & 5 & $-23.9337$ & $-24.3186$ & $-24.0345$ & $-23.6193$ & $-22.2724$ & M13     \nl
3462 & 6 & $-23.5143$ & $-24.4859$ & $-24.0565$ & $-23.6935$ & $-21.9792$ & NGC2420 \nl
3513 & 6 & $-23.5261$ & $-24.4528$ & $-24.0475$ & $-23.6957$ & $-21.9535$ & NGC2420 \nl
4623 & 1 & $-23.8607$ & $-24.5685$ & $-24.0320$ & $-23.6199$ & $-21.9555$ & NGC5466 \nl
4646 & 3 & $-23.7777$ & $-24.4222$ & $-23.9935$ & $-23.6680$ & $-21.7882$ & M3      \nl
4646 & 6 & $-23.4123$ & $-24.3961$ & $-23.9683$ & $-23.6237$ & $-21.9051$ & NGC5466 \nl
4649 & 4 & $-23.7366$ & $-24.4195$ & $-23.9497$ & $-23.6788$ & $-21.7534$ & M3      \nl
4670 & 2 & $-23.7051$ & $-24.4017$ & $-23.9825$ & $-23.5853$ & $-21.9417$ & Pal14   \nl
4682 & 6 & $-23.4981$ & $-24.4152$ & $-23.9959$ & $-23.6507$ & $-21.9315$ & M92     \nl
4879 & 2 & $-23.7270$ & $-24.4404$ & $-24.0645$ & $-23.6569$ & $-21.9531$ & NGC7006 \nl
5061 & 3 & $-23.8128$ & $-24.4389$ & $-24.0376$ & $-23.7196$ & $-21.8324$ & Pal4    \nl
5071 & 2 & $-23.7610$ & $-24.4030$ & $-24.0285$ & $-23.6435$ & $-22.0033$ & Pal4    \nl
5323 & 6 & $-23.3916$ & $-24.4289$ & $-24.0283$ & $-23.6781$ & $-21.9806$ & Pal14   \nl
5327 & 6 & $-23.4375$ & $-24.4386$ & $-24.0427$ & $-23.6906$ & $-21.9649$ & M92     \nl
5360 & 5 & $-23.9117$ & $-24.3206$ & $-24.0069$ & $-23.6301$ & $-22.3503$ & NGC5053 \nl
5360 & 6 & $-23.3901$ & $-24.3960$ & $-23.9773$ & $-23.6301$ & $-21.9294$ & NGC4147 \nl
5360 & 6 & $-23.3998$ & $-24.4025$ & $-23.9855$ & $-23.6411$ & $-21.9304$ & M53     \nl
5381 & 6 & $-23.3332$ & $-24.4052$ & $-23.9971$ & $-23.6417$ & $-21.9385$ & NGC4147 \nl
5390 & 5 & $-23.8879$ & $-24.3443$ & $-24.0286$ & $-23.6446$ & $-22.3340$ & NGC5053 \nl
5390 & 6 & $-23.3636$ & $-24.4049$ & $-23.9863$ & $-23.6236$ & $-21.8995$ & M53     \nl
5403 & 4 & $-23.6659$ & $-24.4256$ & $-23.9299$ & $-23.7138$ & $-21.7897$ & NGC6791 \nl
5416 & 3 & $-23.7120$ & $-24.3963$ & $-24.0259$ & $-23.6750$ & $-21.8010$ & NGC6791 \nl
5935 & 2 & $-23.6510$ & $-24.3622$ & $-23.9817$ & $-23.5872$ & $-21.9635$ & M67     \nl
5972 & 6 & $-23.3384$ & $-24.3808$ & $-23.9791$ & $-23.6422$ & $-21.9654$ & M67     \nl
6004 & 5 & $-23.8660$ & $-24.3285$ & $-24.0502$ & $-23.6612$ & $-22.3708$ & M67     \nl
6177 & 3 & $-23.9108$ & $-24.4583$ & $-24.0660$ & $-23.7181$ & $-21.8399$ & NGC6791 \nl
6895 & 3 & $-23.7923$ & $-24.4366$ & $-24.0421$ & $-23.6497$ & $-21.7717$ & M71     \nl
\enddata
\end{deluxetable*}

\begin{deluxetable}{ccccccc}
\tablewidth{0pt}
\tablecaption{Statistical Properties of Zero-Point Corrections in Flanking Fields\label{tab:rms}}
\tablehead{
  \colhead{} &
  \multicolumn{3}{c}{$\langle \Delta {\tt aa} \rangle$} &
  \multicolumn{3}{c}{$\Delta {\rm {\tt aa}^{west}} - \Delta {\rm {\tt aa}^{east}}$} \nl
  \cline{2-4} \cline{5-7}
  \colhead{Filter} &
  \colhead{average} &
  \colhead{rms} &
  \colhead{$N_{\rm run}$} &
  \colhead{average} &
  \colhead{rms} &
  \colhead{$N_{\rm run}$\tablenotemark{a}}
}
\startdata
$u$ & $-0.0092$ & $0.0063$ &39& $+0.0023$ & $0.0082$ &35 \nl
$g$ & $-0.0062$ & $0.0046$ &39& $+0.0019$ & $0.0060$ &37 \nl
$r$ & $-0.0068$ & $0.0040$ &39& $-0.0004$ & $0.0074$ &33 \nl
$i$ & $-0.0053$ & $0.0069$ &39& $+0.0025$ & $0.0052$ &33 \nl
$z$ & $-0.0034$ & $0.0044$ &39& $-0.0003$ & $0.0057$ &37
\enddata
\tablenotetext{a}{The total number of comparisons in this column is
smaller than the total number of runs in this study ($N_{\rm run} = 39$)
because some of the runs have only one flanking field available in the
analysis.}
\end{deluxetable}

For each flanking field we took an average over three fields to
determine a zero-point correction, $\Delta {\tt aa}$, for DAOPHOT:
\begin{equation}
\Delta {\tt aa} \equiv {\tt aa}^{\rm DAOPHOT} - {\tt aa}^{\it Photo}
= \langle m^{\rm DAOPHOT}_i - m^{\it Photo}_i \rangle.
\label{eq:aa}
\end{equation}
We then averaged results from two flanking fields on each run, filter,
and camcol, with weights given by the errors in $\Delta {\tt aa}$.  These
are either the propagated error from the {\it Photo} comparison
on each flanking field, or the difference in $\Delta {\tt aa}$ between
western and eastern flanking fields divided by two, whichever is larger.
Table~\ref{tab:aa} lists new {\tt aa} coefficients for our fields.
The second and third columns in Table~\ref{tab:rms} list the average
zero-point correction and rms dispersion in each bandpass, respectively,
from all of our flanking fields.  As also shown in Figure~\ref{fig:zp.comp},
the $u$ band has the largest average correction among bandpasses, with
$\langle \Delta {\tt aa} \rangle \approx -0.009$, while longer wavelength
filters have smaller $\langle \Delta {\tt aa} \rangle$ values.  These
zero-point corrections are systematic in nature and statistically
significant.

The cause of the zero-point difference remains unclear.  It could
be an error in the aperture correction, or it could be due to different
ways of determining sky values in the two data reduction procedures.  To
scrutinize this issues, one may wish to reduce secondary patches using
DAOPHOT and independently calibrate DAOPHOT magnitudes.  However, this
requires a significant amount of data reduction with human intervention
(e.g., PSF selection).  Therefore, we chose to put DAOPHOT photometry onto the
{\it Photo} system, which is internally defined in a self-consistent manner,
and avoid any discussion of the absolute calibration in this paper.  The new zero
points (${\tt aa^{\rm DAOPHOT}}$) derived from the {\it Photo} comparisons
were used, along with the extinction coefficient ({\tt kk}) and air-mass
({\tt airmass}) values, to derive magnitudes for stars in the crowded
cluster fields (Eqs.~\ref{eq:luptitude1} and \ref{eq:luptitude2}).

\subsection{Photometric Errors}
\label{sec:error}

To assess the accuracy of our photometry, we need to know both how well
we can determine the zero point of the calibration and how well we can
measure the brightness of a particular star, which can be affected by
the degree of crowding as well as by photon noise.  We first consider
zero-point errors, followed by a discussion of the random star-by-star errors.

\subsubsection{Zero-Point Accuracy}

Systematic zero-point errors for the DAOPHOT photometry are the results
of uncertainties in the aperture correction, the derivation of
zero-point differences between DAOPHOT and {\it Photo} magnitudes, and
the intrinsic zero-point errors in the reference {\it Photo} system.
The zero-point errors in {\it Photo} can be further divided into an absolute
calibration error, which can be reduced to a problem of tying the SDSS
magnitude system to an AB system \citep{dr2}, and a relative zero-point
calibration error.  The latter is exhibited as spatial variations in the
calibration on the sky, or differences in flux measurements for stars
observed in overlapping runs.

To assess the zero-point errors we took the following two approaches.
Firstly, we compared the DAOPHOT photometry in flanking fields to the
{\it Photo} values, in order to check the spatial stability of the
{\it Photo} zero points over several frames.  We took into account the
fact that all of the above error components, except the absolute
calibration error, will manifest themselves collectively as differences
in the {\it Photo} comparisons.  Secondly, we used multiple measurements
for sources detected in more than two different runs to assess the
zero-point errors.  A discussion on absolute calibration errors is
beyond the scope of this paper.

In \S~\ref{sec:zeropoint} we showed that the DAOPHOT PSF magnitudes are
spatially and temporally uniform to $\sim0.005$~mag with respect to
the aperture photometry.  Given this high internal precision of the
photometry, $\sim2\%$ spatial (temporal) variations in the difference
between the DAOPHOT and the {\it Photo} magnitudes on a sub-field scale
(e.g., Fig.~\ref{fig:daophotvsphoto}) were attributed to the PSF modeling
errors in {\it Photo}.  Nevertheless, these high spatial frequency
structures are not a significant error component of the DAOPHOT zero
point because we took an average of the difference between the two
photometric measurements with a large number of comparison stars in each
flanking field ($<0.001$~mag).

We investigated an additional error component of the photometric zero
point by comparing results from the western and eastern flanking fields.
The two flanking fields on each run and camcol are separated by about
10 fields.  Therefore, a zero-point variation over a large angular
scale ($\sim1.5\arcdeg$ or $\sim10$~min in scanning time), or the
difference in zero-point corrections in these two fields ($\Delta
{\tt aa}^{\rm west}$ and $\Delta {\tt aa}^{\rm east}$ for the western and
eastern fields, respectively) can be used as a measure of the zero-point
shifts.

\begin{figure}
\epsscale{1.1}
\plotone{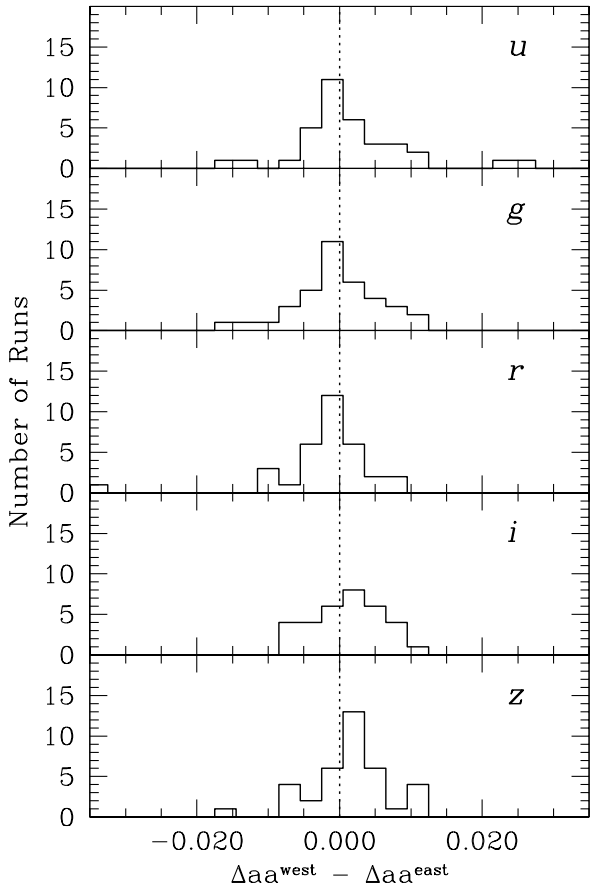}
\caption{Distribution of zero-point differences between the western and
the eastern flanking fields.
\label{fig:we.comp}}
\end{figure}

Figure~\ref{fig:we.comp} shows the distribution of zero-point differences
in the western and eastern fields for all pairs of the flanking fields.
The fifth column in Table~\ref{tab:rms} lists the global average of the
zero-point difference, which is generally consistent with zero in
all of the filter bandpasses.  However, zero points from the western and
eastern flanking fields typically differ by $\sim0.006$~mag on each
run, as shown in the sixth column.  This indicates that DAOPHOT magnitudes
will have a mild zero-point variation over an $\sim1.5\arcdeg$ scale,
which is smaller than the $\sim0.02$~mag fluctuations seen on a sub-field
scale.

However, it should be kept in mind that the above comparisons are based
on flux measurements in each run and camcol, which have the same
{\tt aa} and {\tt kk} coefficients from the {\tt tsField} files. As a
second approach, we assessed the zero-point errors by comparing flux
measurements from overlapping regions between different strips and runs.
A small fraction of stars are found in overlapping strips, and their
fluxes were individually measured and calibrated to the {\it Photo}
system for each run in the DAOPHOT reduction.  Therefore, the net
magnitude difference between the two runs directly measures the
reliability of our photometric zero points.

\begin{deluxetable*}{lccccccc}
\tablewidth{0pt}
\tablecaption{Photometric Comparisons in Overlapping Runs\label{tab:overlap}}
\tablehead{
  \colhead{} &
  \multicolumn{2}{c}{Runs} &
  \multicolumn{5}{c}{Ref. $-$ Comp. (mag)} \nl
  \cline{2-3}
  \cline{4-8}
  \colhead{Cluster} &
  \colhead{Ref.} &
  \colhead{Comp.} &
  \colhead{$\langle \Delta u \rangle$} &
  \colhead{$\langle \Delta g \rangle$} &
  \colhead{$\langle \Delta r \rangle$} &
  \colhead{$\langle \Delta i \rangle$} &
  \colhead{$\langle \Delta z \rangle$}
}
\startdata
M2       & 2583 & 2662 & $+0.051\pm0.005$ & $-0.016\pm0.001$ & $+0.005\pm0.002$ & $+0.002\pm0.002$ & $+0.027\pm0.002$ \nl
M3       & 4646 & 4649 & $-0.041\pm0.006$ & $-0.018\pm0.002$ & $-0.044\pm0.001$ & $-0.033\pm0.001$ & $-0.056\pm0.003$ \nl
M5       & 1458 & 2327 & $+0.006\pm0.007$ & $+0.001\pm0.001$ & $-0.004\pm0.001$ & $-0.014\pm0.001$ & $+0.013\pm0.004$ \nl
M13      & 3225 & 3226 & $-0.003\pm0.005$ & $-0.020\pm0.001$ & $-0.023\pm0.002$ & $-0.006\pm0.002$ & $-0.012\pm0.002$ \nl
M15      & 2566 & 1739 & $+0.048\pm0.008$ & $+0.020\pm0.002$ & $+0.005\pm0.002$ & $+0.000\pm0.002$ & $-0.017\pm0.006$ \nl
M53      & 5390 & 5360 & $-0.085\pm0.018$ & $+0.000\pm0.002$ & $+0.014\pm0.002$ & $+0.017\pm0.003$ & $+0.014\pm0.006$ \nl
M67      & 5935 & 5972 & $+0.011\pm0.003$ & $-0.001\pm0.002$ & $-0.009\pm0.002$ & $+0.004\pm0.002$ & $-0.017\pm0.002$ \nl
M67      & 5935 & 6004 & $-0.003\pm0.005$ & $-0.004\pm0.002$ & $-0.004\pm0.002$ & $-0.004\pm0.002$ & $-0.021\pm0.002$ \nl
M92      & 4682 & 5327 & $+0.026\pm0.013$ & $-0.009\pm0.002$ & $-0.028\pm0.002$ & $-0.040\pm0.002$ & $-0.002\pm0.005$ \nl
NGC~2419 & 1402 & 1350 & $+0.036\pm0.020$ & $+0.026\pm0.002$ & $+0.029\pm0.003$ & $+0.033\pm0.002$ & $+0.046\pm0.006$ \nl
NGC~2420 & 2888 & 3462 & $-0.019\pm0.003$ & $-0.001\pm0.001$ & $-0.008\pm0.001$ & $-0.002\pm0.001$ & $+0.005\pm0.002$ \nl
NGC~2420 & 2888 & 3513 & $+0.002\pm0.002$ & $+0.005\pm0.001$ & $-0.013\pm0.001$ & $-0.003\pm0.001$ & $+0.010\pm0.001$ \nl
NGC~4147 & 5381 & 5360 & $+0.104\pm0.018$ & $+0.006\pm0.004$ & $+0.072\pm0.005$ & $+0.045\pm0.005$ & $+0.033\pm0.007$ \nl
NGC~5053 & 5390 & 5360 & $+0.016\pm0.031$ & $+0.027\pm0.004$ & $+0.037\pm0.005$ & $+0.002\pm0.005$ & $+0.006\pm0.014$ \nl
NGC~5466 & 4623 & 4646 & $+0.008\pm0.003$ & $+0.019\pm0.001$ & $+0.023\pm0.001$ & $+0.009\pm0.001$ & $+0.002\pm0.002$ \nl
NGC~6791 & 5416 & 5403 & $-0.012\pm0.006$ & $-0.053\pm0.001$ & $-0.027\pm0.001$ & $-0.047\pm0.001$ & $-0.051\pm0.002$ \nl
NGC~6791 & 5416 & 6177 & $-0.041\pm0.001$ & $-0.035\pm0.000$ & $-0.027\pm0.000$ & $-0.032\pm0.000$ & $-0.032\pm0.001$ \nl
Pal~4    & 5061 & 5071 & \nodata          & $-0.021\pm0.006$ & $+0.004\pm0.010$ & $-0.017\pm0.006$ & $+0.000\pm0.013$ \nl
Pal~14   & 4670 & 5323 & $+0.017\pm0.031$ & $+0.004\pm0.002$ & $+0.007\pm0.002$ & $-0.005\pm0.002$ & $-0.002\pm0.005$ \nl
rms &\nodata &\nodata  & $0.042$          & $0.021$          & $0.027$          & $0.024$          & $0.026$          \nl
\enddata
\end{deluxetable*}

Table~\ref{tab:overlap} shows comparisons for the DAOPHOT magnitudes from
all of the overlapping runs.  The second and third column list two runs
in the comparison.  A ``reference'' run was selected if it covers a
larger fraction of a cluster than a ``comparison'' run, otherwise a
small-numbered run was chosen.  In the derivation of the fiducial
sequences (\S~\ref{sec:fiducial}) we use the local zero point set by the
reference runs to combine photometry from two different runs.  The fourth
through eighth columns list weighted average magnitude differences in
the five passbands.  We matched stars from separate runs using the
celestial coordinates with a match radius of $1\arcsec$.  We used stars
that satisfy the same selection criteria on magnitudes, errors, $\chi$,
and sharp index values as in \S~\ref{sec:zeropoint}.  However, in the
$u$-band and $z$-band frames, we relaxed the thresholds on magnitude
errors to $0.10$~mag to include more comparison stars in these small
overlapping regions.  We further increased the threshold for the
$u$-band matches to $0.30$~mag in some clusters (M53, M92, NGC~4147,
NGC~5053, and Pal~14), which have an even smaller number of comparison
stars.  For Pal~4 we did not compute the zero-point difference in the
$u$ band because no stars were found in that filter frame that satisfy
our $\chi$ and sharp index criteria.

The rms of these differences from all comparisons are $0.042$~mag, $0.021$~mag,
$0.027$~mag, $0.024$~mag, and $0.026$~mag in $ugriz$, respectively.  In
Table~\ref{tab:overlap} we also found that the rms differences in colors are
$0.040$~mag in $u - g$, $0.021$~mag in $g - r$, $0.017$~mag in $g - i$,
and $0.021$~mag in $g - z$.  Although there is a mild zero-point variation over
$\sim10$ fields ($\sim0.6\%$), the calibration accuracy of the DAOPHOT photometry
is predominantly limited by these $\sim2\%$ run-to-run zero-point variations.
Our results are consistent with a zero-point uncertainty of $\sim2\%$--$3\%$ in
{\it Photo} \citep{ivezic:04}.

\subsubsection{Random Errors}

Repeated flux measurements in overlapping strips/runs also provide
an opportunity to determine the star-to-star uncertainties in the
photometry.  We used the same matched list of stars as in the previous
section, but without the cuts on magnitudes and magnitude errors.
We adjusted the net zero-point differences between the runs
(Table~\ref{tab:overlap}) before making the photometric comparisons and
then estimated
the standard deviations of individual measurements.  We only considered
double measurements, although some of the stars in the open cluster fields
have been detected in three runs.

\begin{figure*}
\epsscale{1.0}
\plotone{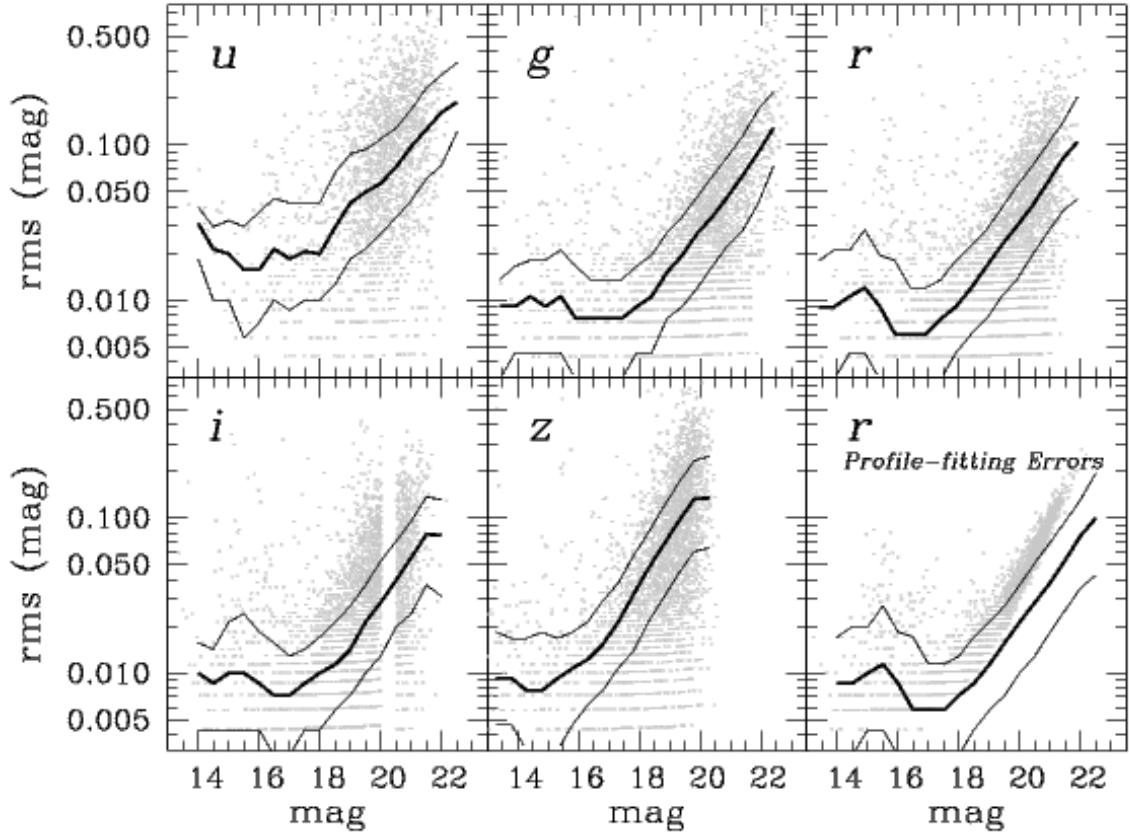}
\caption{The rms magnitude errors from repeated flux measurements in
overlapping strips in five bandpasses.  For clarity, only 10\% of points
are shown.  The thick solid line shows the median of these with intervals
of $0.5$~mag; thin lines on either side are the first and third quartiles.
The bottom right panel compares the profile-fitting errors from DAOPHOT
({\it points}) to the same curves in the top right panel.  Apparently
discrete values on the y axis are due to round-off approximations.
\label{fig:stripsdbl}}
\end{figure*}

Figure~\ref{fig:stripsdbl} shows the standard deviations of individual
measurements for stars that have been observed twice in overlapping strips.
The thick solid line shows the median of these with intervals of
$0.5$~mag; thin lines on either side are the first and third
quartiles. The  error distributions at the bright ends indicate errors of
$\sim1\%$ in $griz$ and $\sim2\%$ in the $u$ band, while the {\it Photo}
magnitudes have $2\%$ rms photometric precision for sources not
limited by photon statistics \citep{ivezic:03}.  The photometric
precision of DAOPHOT is about a factor of two better than that of {\it Photo}.

The bottom right panel in Figure~\ref{fig:stripsdbl} shows the reported
DAOPHOT errors in the $r$-band.  DAOPHOT estimates standard errors in the
individual (instrumental) magnitudes, which are obtained either from the PSF
profile-fitting residuals or from the star and sky flux measurements
\citep{stetson:03}.  For stars observed in different strips, we estimated
the standard error in the mean as $1/\sigma^2 = \Sigma_i (1/\sigma_i)^2$,
where $\sigma_i$ is the error reported from DAOPHOT, and multiplied it
by the square root of the number of measurements ($=2$).  As seen in
Figure~\ref{fig:stripsdbl}, most of the points are between the median
and third quartile of the error distribution.  Although DAOPHOT errors are
slightly larger than the errors estimated from repeat measurements, they
represent the approximate size of the errors well.  The comparisons in other
bands are similar to that in the $r$ band.

\subsection{Comparison with {\it Photo} in Semi-Crowded Fields}

The stellar densities in open cluster fields are significantly lower
than in the cores of globular clusters, and {\it Photo} reports
magnitudes for many objects.  On the other hand, open cluster fields
are more crowded than the typical high Galactic latitude fields in SDSS.
Therefore, DAOPHOT magnitudes can be used to test the accuracy of the
SDSS imaging pipelines near the Galactic plane, which is directly
related to the quality assurance for the SEGUE imaging outputs.  The
systematic errors in these semi-crowded fields cannot be fully accounted
for using the method based on the stellar locus \citep{ivezic:04} because
the extinction corrections from \citet{schlegel:98} become uncertain
near the Galactic plane \citep{dr6}.

In three open clusters in our sample, DAOPHOT detected $\sim100$,
$\sim300$, and $\sim1500$ sources in M67 ($b = +31.9\arcdeg$),
NGC~2420 ($b = +19.6\arcdeg$), and NGC~6791 ($b = +10.9\arcdeg$),
respectively, with $r < 20$~mag on each frame ($10\arcmin \times 13\arcmin$).
Stellar densities in NGC~2420 and NGC~6791 are about $2-10$ times higher
than the median density in the flanking fields for globular clusters,
which is $\sim150$ per frame.  {\it Photo} detected a comparable number of 
sources classified as {\tt STAR} ($\sim80\%$).  However, it is noted again that
{\it Photo} failed to detect many stellar sources in fields near the
globular cluster M71, where the stellar density is about four times
higher than that of NGC~6791.  We discuss the issue of M71 again in
\S~\ref{sec:prime}.

\begin{figure}
\epsscale{1.1}
\plotone{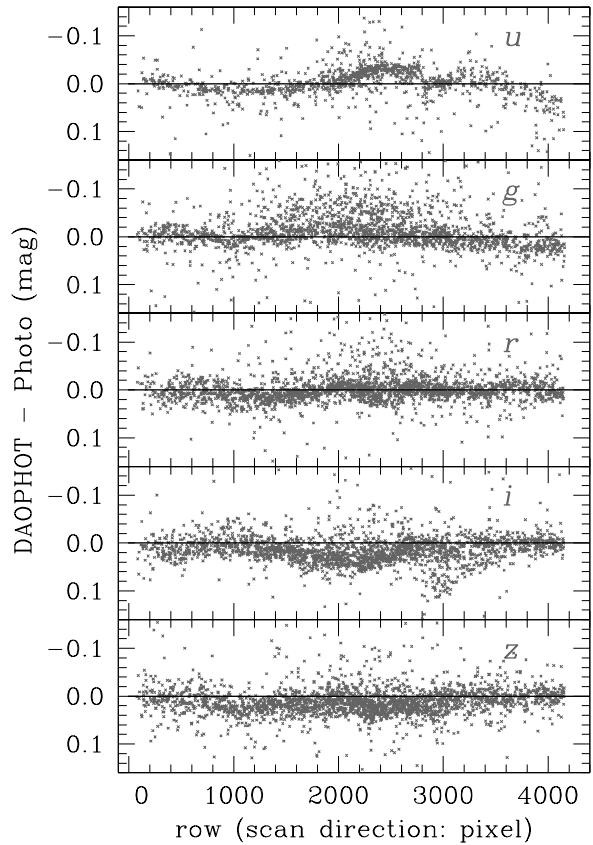}
\caption{Comparison between the DAOPHOT and the {\it Photo} magnitudes in one
of the NGC~2420 cluster fields.  Comparisons are shown for $u$, $g$, $r$,
$i$, and $z$, from the top to the bottom panels.\label{fig:compoc}}
\end{figure}

\begin{deluxetable}{ll}
\tablewidth{0pt}
\tablecaption{Quantities in the Value-Added DAOPHOT Photometric Catalog
\label{tab:phot}}
\tablehead{
  \colhead{Column} &
  \colhead{Description}
}
\startdata
{\tt Run      } & SDSS Run number                                  \nl
{\tt ReRun    } & SDSS ReRun number                                \nl
{\tt CamCol   } & SDSS CamCol 1-6                                  \nl
{\tt DAOPHOTID} & DAOPHOT ID number                                \nl
{\tt RA       } & RA (J2000) in deg                                \nl
{\tt DEC      } & DEC (J2000) in deg                               \nl
{\tt x\_u     } & $u$ band x pixel (9999.99 = no detection)        \nl
{\tt y\_u     } & $u$ band y pixel                                 \nl
{\tt u        } & $u$ DAOPHOT magnitude (99.999 = no detection) in mag   \nl
{\tt uErr     } & $u$ band magnitude error in mag                  \nl
{\tt chi      } & DAOPHOT $\chi$ for $u$ band                      \nl
{\tt sharp    } & DAOPHOT sharp for $u$ band                \nl
{\tt flag     } & 0=ok, 1=near saturated pixel, 9=no detection in $u$ band        \nl
{\tt x\_g     } & $g$ band x pixel (9999.99 = no detection)        \nl
{\tt y\_g     } & $g$ band y pixel                                 \nl
{\tt g        } & $g$ DAOPHOT magnitude (99.999 = no detection) in mag   \nl
{\tt gErr     } & $g$ band magnitude error in mag                  \nl
{\tt chi      } & DAOPHOT $\chi$ for $g$ band                      \nl
{\tt sharp    } & DAOPHOT sharp for $g$ band                \nl
{\tt flag     } & 0=ok, 1=near saturated pixel, 9=no detection in $g$ band        \nl
{\tt x\_r     } & $r$ band x pixel (9999.99 = no detection)        \nl
{\tt y\_r     } & $r$ band y pixel                                 \nl
{\tt r        } & $r$ DAOPHOT magnitude (99.999 = no detection) in mag   \nl
{\tt rErr     } & $r$ band magnitude error in mag                  \nl
{\tt chi      } & DAOPHOT $\chi$ for $r$ band                      \nl
{\tt sharp    } & DAOPHOT sharp for $r$ band                \nl
{\tt flag     } & 0=ok, 1=near saturated pixel, 9=no detection in $r$ band        \nl
{\tt x\_i     } & $i$ band x pixel (9999.99 = no detection)        \nl
{\tt y\_i     } & $i$ band y pixel                                 \nl
{\tt i        } & $i$ DAOPHOT magnitude (99.999 = no detection) in mag   \nl
{\tt iErr     } & $i$ band magnitude error in mag                  \nl
{\tt chi      } & DAOPHOT $\chi$ for $i$ band                      \nl
{\tt sharp    } & DAOPHOT sharp for $i$ band                \nl
{\tt flag     } & 0=ok, 1=near saturated pixel, 9=no detection in $i$ band        \nl
{\tt x\_z     } & $z$ band x pixel (9999.99 = no detection)        \nl
{\tt y\_z     } & $z$ band y pixel                                 \nl
{\tt z        } & $z$ DAOPHOT magnitude (99.999 = no detection) in mag   \nl
{\tt zErr     } & $z$ band magnitude error in mag                  \nl
{\tt chi      } & DAOPHOT $\chi$ for $z$ band                      \nl
{\tt sharp    } & DAOPHOT sharp for $z$ band                \nl
{\tt flag     } & 0=ok, 1=near saturated pixel, 9=no detection in $z$ band        \nl
\enddata
\end{deluxetable}

While {\it Photo} detected a comparable number of stellar sources in the
cluster fields, its photometry is less accurate than obtained in high
Galactic latitude fields.  Figure~\ref{fig:compoc} shows differences between
the DAOPHOT and the {\it Photo} magnitudes in one of the NGC~2420 runs,
{\tt 2888-3-24/25/26}.  The comparisons are shown with no corrections on
{\tt aa} to DAOPHOT magnitudes.  The spatial variations of the difference
are stronger than those in the typical flanking fields for
globular clusters (e.g., Fig.~\ref{fig:daophotvsphoto}).  The average
rms differences from all of the open cluster runs are $0.016$~mag, $0.028$~mag,
$0.021$~mag, $0.023$~mag, and $0.018$~mag in $ugriz$, respectively.  The rms
values for $gri$ are marginally larger than those from the globular cluster
flanking fields.  The most likely explanation for the large differences
is that {\it Photo} has trouble finding isolated bright objects for the
PSF modeling in these semi-crowded fields.  Thus, caution should be used for
{\it Photo} results in open clusters and for those in low Galactic
latitude fields.

\section{Results}

In this section we present the DAOPHOT/ALLFRAME photometry for 20
clusters and derive cluster fiducial sequences over a wide range of
metal abundances, $\Delta \log(Z/Z_\odot) \sim3$~dex.  We then use
these fiducial sequences to perform a preliminary test of theoretical
isochrones, and to compare with fiducial sequences in $u'g'r'i'z'$ from
\citet{clem:08}.

\subsection{Value-Added Catalog for DAOPHOT Cluster Photometry}

We present the cluster photometry for this paper as a SDSS
value-added catalog.\footnote{Available at
{\tt http://www.sdss.org/dr6/products/\\value\_added/anjohnson08\_clusterphotometry.htm}}
There is a file for each cluster for each run, labeled {\tt cluster\_run.phot}.
For example, the M92 data are in {\tt m92\_4682.phot} and {\tt m92\_5327.phot}.
Table~\ref{tab:phot} lists the columns of data.  We note that the DAOPHOT
identification number is unique for each cluster/run combination, and that the
x and y pixel positions are for the tiled images.  We flag saturated stars, i.e.,
those stars, which have a pixel set to $65000$~DN within a 10~pixel radius
from a stellar centroid.  Although DAOPHOT will determine magnitudes for
these stars based on the non-saturated pixels, these magnitudes should
be treated with extreme caution.  The flag is set to $1$ if the star is
near a saturated pixel, $9$ if the star is not detected in a given frame,
and $0$ otherwise.

\begin{figure*}
\epsscale{1.0}
\plotone{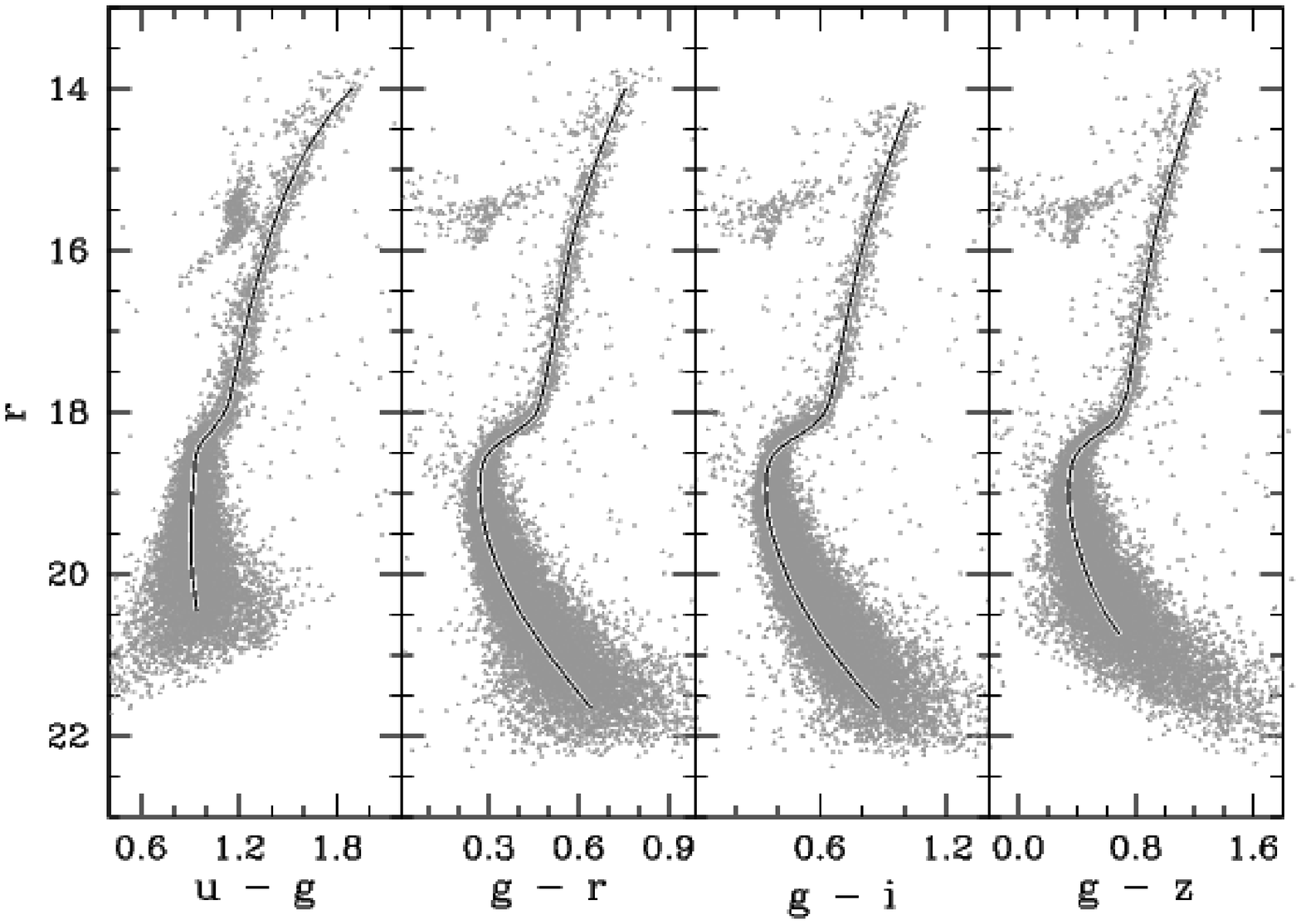}
\caption{CMDs for M3.  The points represent stars that passed the selection
criteria based on the $\chi$, sharp, and separation indices (see text).
The solid lines are the cluster fiducial sequences.  Cluster RR Lyraes
are scattered off the horizontal branch.\label{fig:cmdm3}}
\end{figure*}

\begin{figure*}
\epsscale{1.0}
\plotone{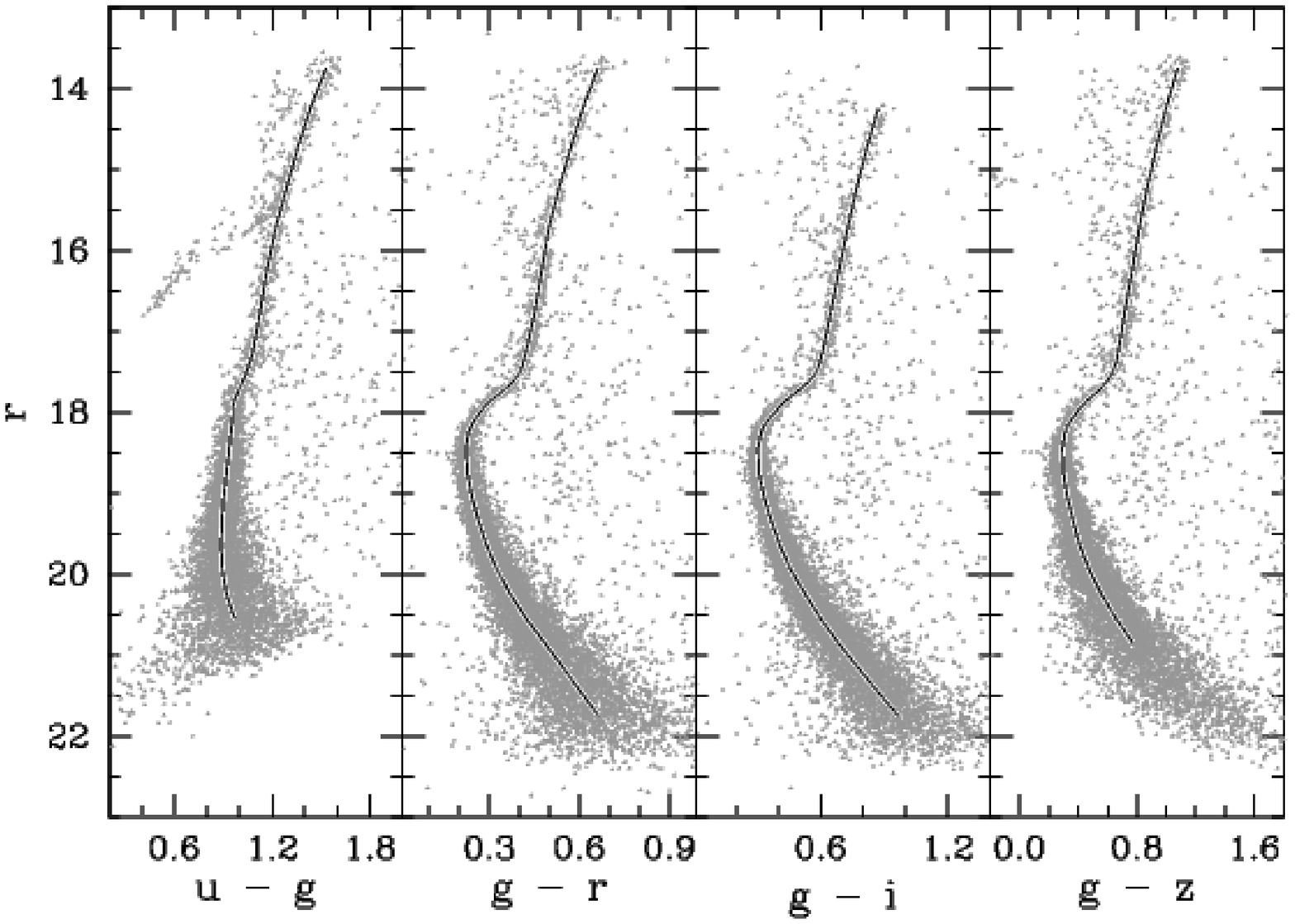}
\caption{Same as in Fig.~\ref{fig:cmdm3}, but for M92.\label{fig:cmdm92}}
\end{figure*}

\begin{figure*}
\epsscale{1.0}
\plotone{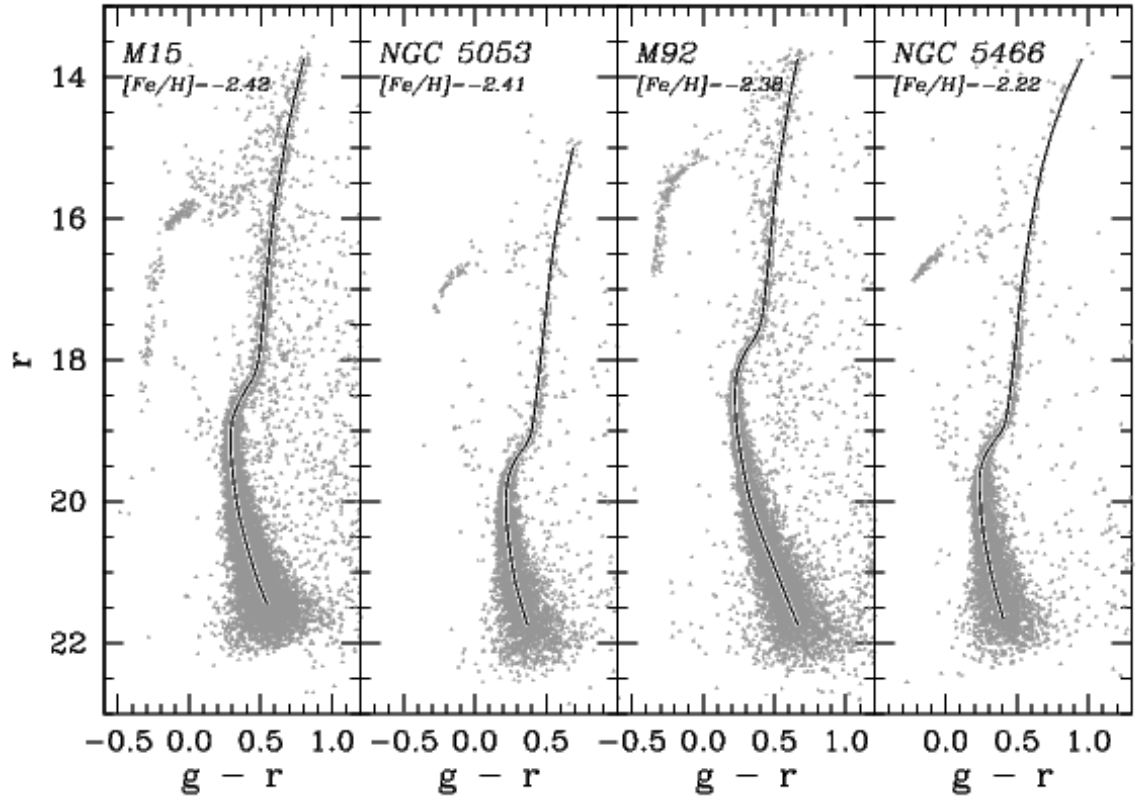}
\caption{CMDs for clusters with ${\rm [Fe/H]} < -2.2$ with $g - r$ as
color indices.  Points are stars that passed the selection criteria
based on the $\chi$, sharp, and separation indices (see text).
\label{fig:all.cmd.1}}
\end{figure*}

\begin{figure*}
\epsscale{1.0}
\plotone{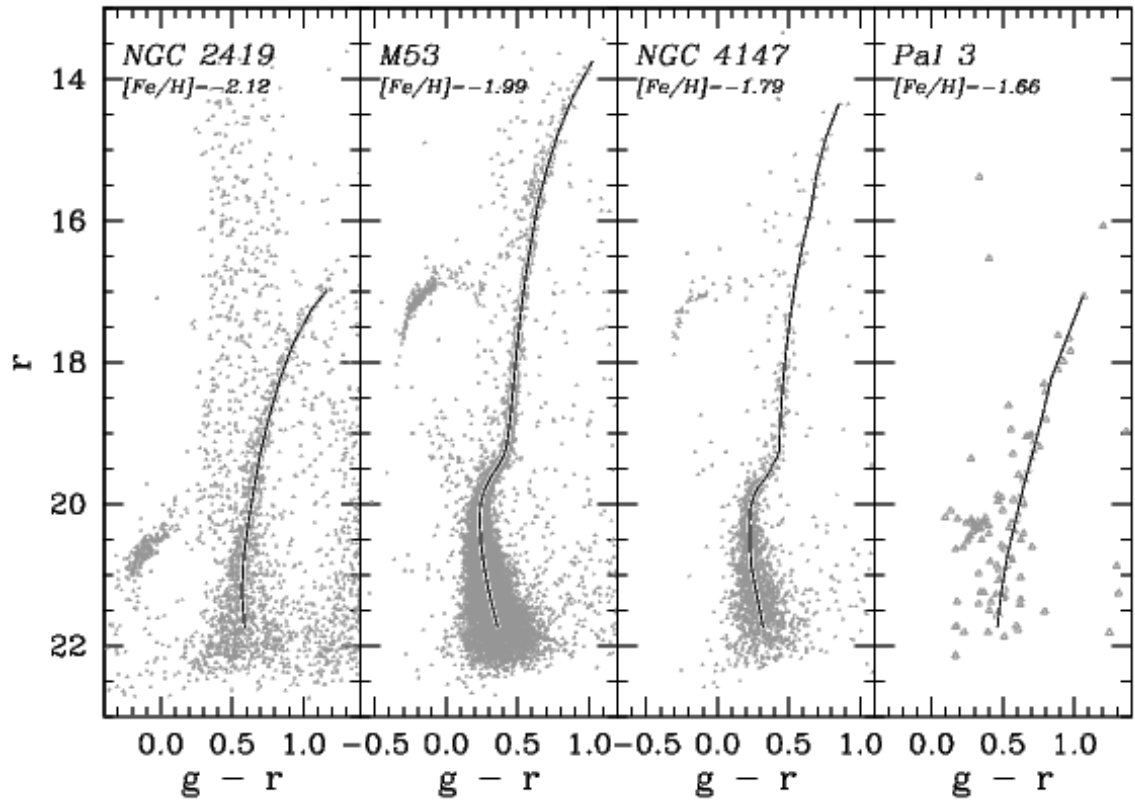}
\caption{Same as in Fig.~\ref{fig:all.cmd.1}, but for clusters with
$-2.2 < {\rm [Fe/H]} < -1.65$.  A separation index was not used to
filter data for Pal~3.\label{fig:all.cmd.2}}
\end{figure*}

\begin{figure*}
\epsscale{1.0}
\plotone{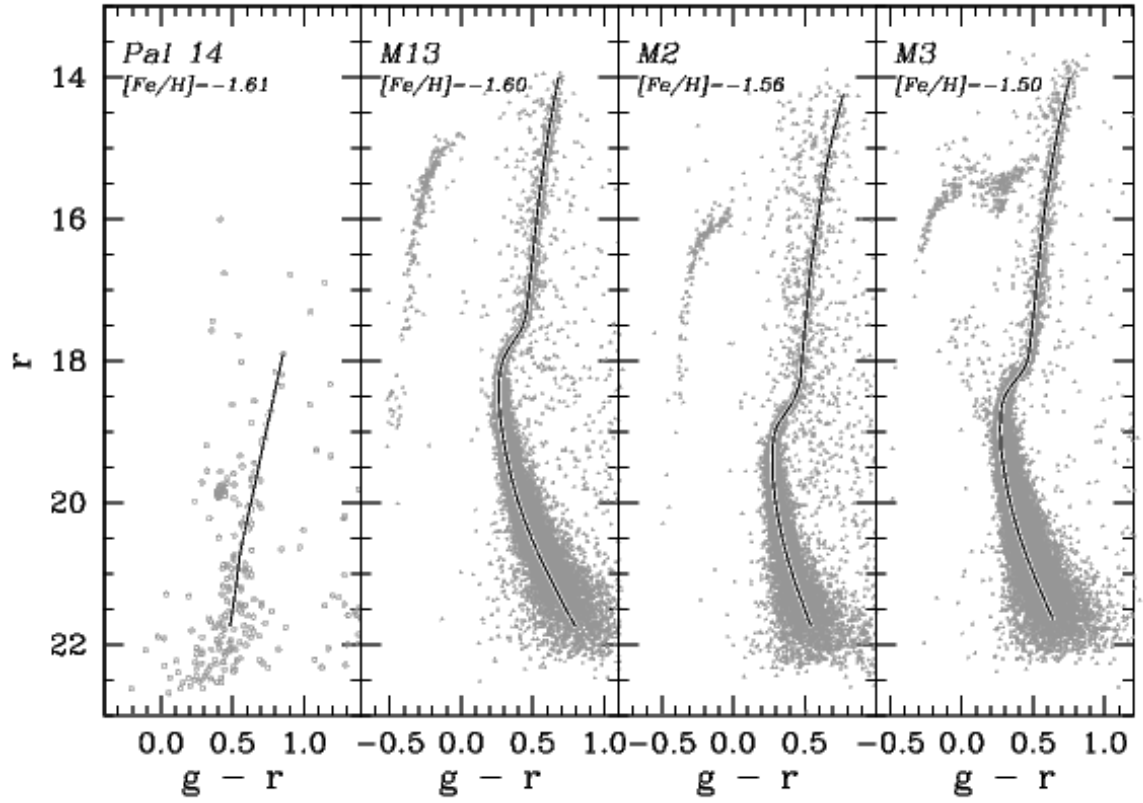}
\caption{Same as in Fig.~\ref{fig:all.cmd.1}, but for clusters with
$-1.65 < {\rm [Fe/H]} \leq -1.50$.  A separation index was not used to
filter data for Pal~14.\label{fig:all.cmd.3}}
\end{figure*}

\begin{figure*}
\epsscale{1.0}
\plotone{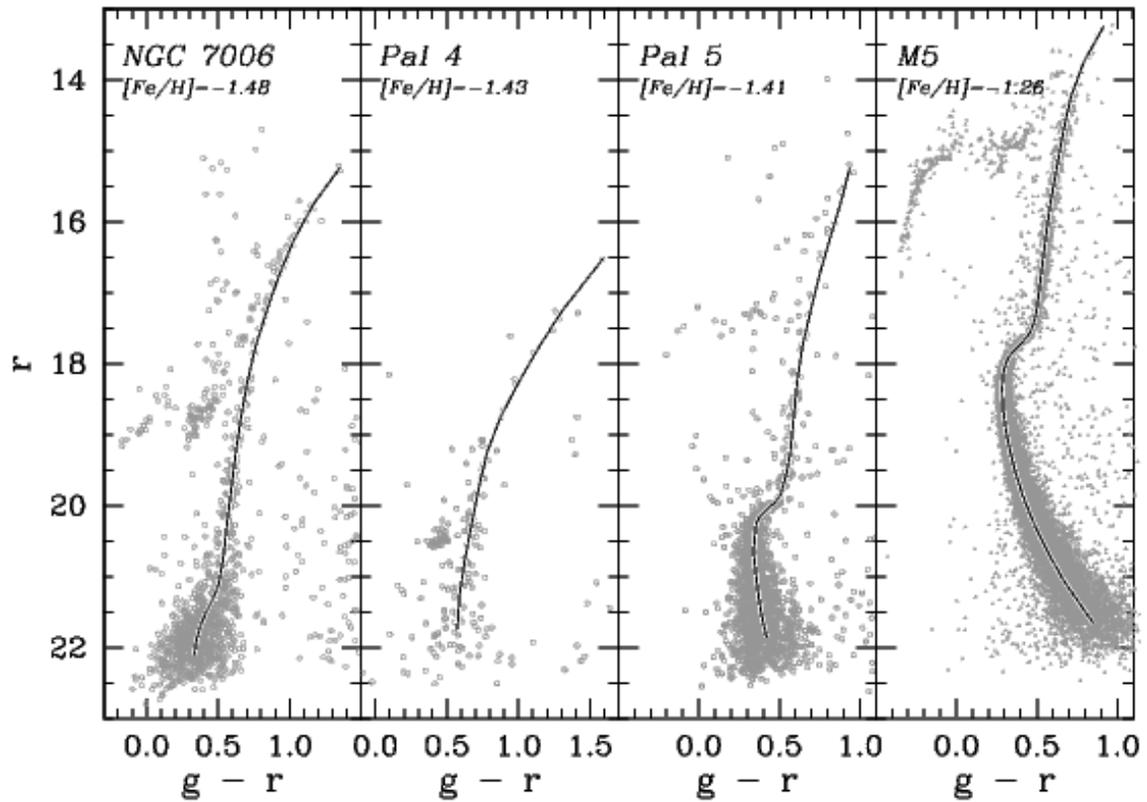}
\caption{Same as in Fig.~\ref{fig:all.cmd.1}, but for clusters with
$-1.5 < {\rm [Fe/H]} < -1.0$.  A separation index was not used to
filter data except for M5.\label{fig:all.cmd.4}}
\end{figure*}

\begin{figure*}
\epsscale{1.0}
\plotone{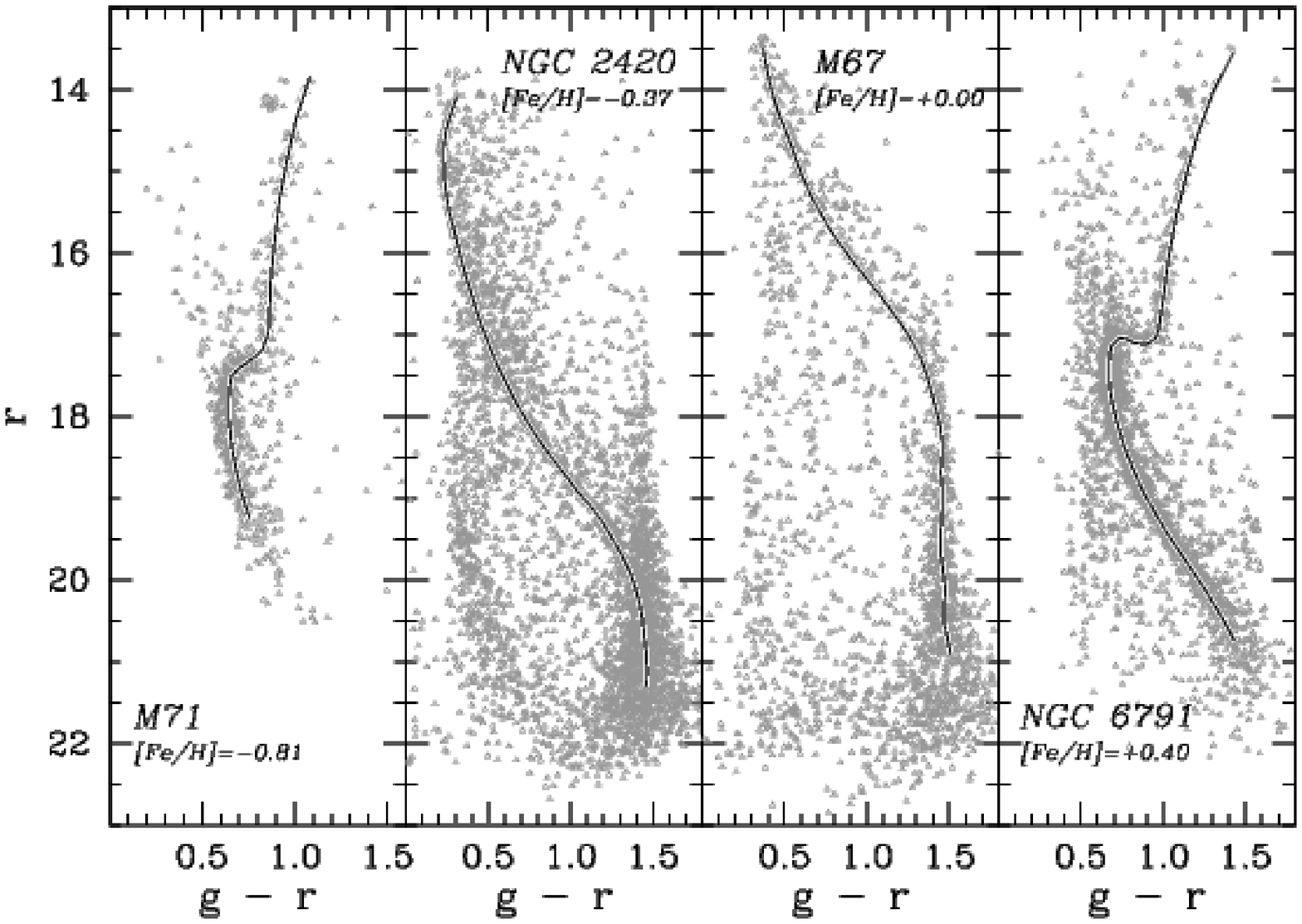}
\caption{Same as in Fig.~\ref{fig:all.cmd.1}, but for clusters with
${\rm [Fe/H]} > -1.0$.
\label{fig:all.cmd.5}}
\end{figure*}

Figures~\ref{fig:cmdm3} and \ref{fig:cmdm92} show CMDs of M3 and M92,
respectively, with $u - g$, $g - r$, $g - i$, and $g - z$ as color indices,
and $r$ as a luminosity index; hereafter $(u - g, r)$, $(g - r, r)$,
$(g - i, r)$, and $(g - z, r)$, respectively.  The $(g - r, r)$ CMDs for
all analyzed clusters are shown in Figures~\ref{fig:all.cmd.1}-\ref{fig:all.cmd.5}.
Stars brighter than $r\sim14$~mag are saturated in the SDSS CCDs, so the
brightest portions of the RGB are not seen in many cluster CMDs.  A detailed
description on the data and fiducial sequences is presented in the following section.

RR Lyraes show the most notable variations of fluxes in globular cluster studies.
This can be seen in Figures~\ref{fig:cmdm3}-\ref{fig:all.cmd.4}, where RR Lyraes
are scattered off the HBs, because most of them were observed only once in SDSS.
They also stand out with large rms magnitude errors in the repeat flux measurements
at $r\sim16$~mag (Fig.~\ref{fig:stripsdbl}).

\subsection{Cluster Fiducial Sequences}
\label{sec:fiducial}

The individual points in Figures~\ref{fig:cmdm3}-\ref{fig:all.cmd.5} are
those stars that satisfy $\chi$ and sharp index selection criteria in
\S~\ref{sec:zeropoint}.  To identify stars in less-crowded regions, we
further used the separation index \citep{stetson:03}, which is defined
as the logarithmic ratio of the surface brightness of a star to the
summed brightness from all neighboring stars.  We followed the detailed
procedure of computing a separation index in \citet{clem:08}.  That is,
we assumed the Moffat stellar profile of the surface brightness and
considered the light contribution from those stars lying within 10 times
the assumed FWHM in the $r$-band images.  We adopted $1.4\arcsec$ for the
typical FWHM seeing for all fields using {\tt RunQA} \citep{ivezic:04}.
In most of the cases we accepted those stars with a separation index
larger than $3.5$~dex.  For M71 we used stars with a separation index larger
than $2$~dex, which produces the best looking sequences on the CMDs.  However, we
did not apply the above criterion based on the separation index for the relatively
sparse clusters NGC~4147, NGC~7006, Pal~3, Pal~4, Pal~5, and Pal~14.

Given the $\sim2\%$ zero-point differences between different runs, we
adjusted the zero point for the photometry in one of the runs to match
the others before combining them together.  Our selected runs for the
local photometric standards are listed in the second column of
Table~\ref{tab:overlap}.  To reduce the contamination from
background stars, we selected stars within a $2.5\arcmin$ radius from
a cluster center for Pal~3, Pal~4, and NGC~7006, those within a $5.0\arcmin$
radius for NGC~6791, and Pal~5, and those within a $2.0\arcmin$ for M71.

The curves in Figures~\ref{fig:cmdm3}-\ref{fig:all.cmd.5} represent
cluster fiducial sequences derived from the above data sample.
A cluster fiducial sequence is defined as the locus of the number density
peaks on a CMD.  Representing a color-magnitude relation for single
stars in a cluster, fiducial sequences can be used to derive relative
distances to stars and star clusters and to test theoretical stellar
isochrones.  However, most of Galactic clusters lack a complete census
of cluster membership and binarity, and the observed CMDs typically contain
a non-uniform distribution of foreground/background stars and cluster
binaries.  In particular, low mass-ratio binaries are difficult to
identify because their colors and magnitudes are similar to those of
single stars.  Therefore, careful selection of data points is required
in order to derive accurate cluster fiducial sequences.

We adopted a photometric filtering scheme, as in \citet[][see also An
et al.\ 2007a]{an:07b}, in order to reduce the number of cluster binaries
and non-cluster members in the CMDs.  The photometric filtering is an
automated process with a least amount of human intervention.  The filtering
process iteratively identifies the MS, SGB, and RGB ridgelines independently
of the isochrones, determines the spread of points, and rejects stars if
they are too far away from the ridgeline for a given magnitude.  We combined
the results from $(g - r, r)$, $(g - i, r)$, and $(g - z, r)$, and rejected
stars if they were tagged as an outlier at least in one of the CMDs.  We
started with a $3\sigma$ rejection and reduced the threshold until it was
limited to $2.5\sigma$.  About $20\%$ of stars were rejected from each set
of CMDs, including cluster HB stars.  For extremely sparse clusters
(Pal~3, Pal~4, and Pal~14), we handpicked probable single star members from
$(g - r, r)$, $(g - i, r)$, and $(g - z, r)$.

\begin{deluxetable}{ccccc}
\tablewidth{0pt}
\tablecaption{Fiducial Sequences for NGC~2419\label{tab:ngc2419}}
\tablehead{
  \colhead{$r$} &
  \colhead{$u - g$} &
  \colhead{$g - r$} &
  \colhead{$g - i$} &
  \colhead{$g - z$}
}
\startdata
 $ 17.000 $&$ 2.657 $&$ 1.160 $&$ 1.657 $&$ 1.938 $\nl
 $ 17.250 $&$ 2.433 $&$ 1.058 $&$ 1.505 $&$ 1.765 $\nl
 $ 17.750 $&$ 2.154 $&$ 0.925 $&$ 1.315 $&$ 1.536 $\nl
 $ 18.250 $&$ 1.929 $&$ 0.832 $&$ 1.188 $&$ 1.383 $\nl
 $ 18.750 $&$ 1.750 $&$ 0.760 $&$ 1.086 $&$ 1.265 $\nl
 $ 19.250 $&$ 1.622 $&$ 0.701 $&$ 1.006 $&$ 1.168 $\nl
 $ 19.750 $&$ 1.480 $&$ 0.655 $&$ 0.941 $&$ 1.091 $\nl
 $ 20.250 $&$ 1.328 $&$ 0.615 $&$ 0.881 $&$ 1.025 $\nl
 $ 20.750 $&$ 1.179 $&$ 0.580 $&$ 0.826 $&$ 0.976 $\nl
 $ 21.250 $&\nodata&$ 0.567 $&$ 0.784 $&\nodata\nl
 $ 21.750 $&\nodata&$ 0.588 $&$ 0.776 $&\nodata\nl
\enddata
\end{deluxetable}

\begin{deluxetable}{ccccc}
\tablewidth{0pt}
\tablecaption{Fiducial Sequences for NGC~2420\label{tab:ngc2420}}
\tablehead{
  \colhead{$r$} &
  \colhead{$u - g$} &
  \colhead{$g - r$} &
  \colhead{$g - i$} &
  \colhead{$g - z$}
}
\startdata
$14.100$&$ 1.162 $&$0.310 $&$0.382 $&$0.384 $\nl
$14.300$&$ 1.150 $&$0.273 $&$0.343 $&$0.333 $\nl
$14.500$&$ 1.133 $&$0.245 $&$0.308 $&$0.292 $\nl
$14.700$&$ 1.122 $&$0.229 $&$0.285 $&$0.271 $\nl
$14.900$&$ 1.113 $&$0.229 $&$0.282 $&$0.268 $\nl
$15.100$&$ 1.108 $&$0.235 $&$0.291 $&$0.275 $\nl
$15.300$&$ 1.104 $&$0.246 $&$0.311 $&$0.297 $\nl
$15.500$&$ 1.100 $&$0.265 $&$0.339 $&$0.333 $\nl
$15.700$&$ 1.103 $&$0.290 $&$0.373 $&$0.375 $\nl
$15.900$&$ 1.115 $&$0.315 $&$0.408 $&$0.417 $\nl
$16.100$&$ 1.133 $&$0.340 $&$0.445 $&$0.462 $\nl
$16.300$&$ 1.159 $&$0.367 $&$0.482 $&$0.512 $\nl
$16.500$&$ 1.191 $&$0.398 $&$0.523 $&$0.562 $\nl
$16.700$&$ 1.230 $&$0.429 $&$0.566 $&$0.614 $\nl
$16.900$&$ 1.287 $&$0.463 $&$0.614 $&$0.670 $\nl
$17.100$&$ 1.362 $&$0.500 $&$0.667 $&$0.734 $\nl
$17.300$&$ 1.449 $&$0.539 $&$0.723 $&$0.804 $\nl
$17.500$&$ 1.551 $&$0.586 $&$0.788 $&$0.882 $\nl
$17.700$&$ 1.663 $&$0.636 $&$0.857 $&$0.966 $\nl
$17.900$&$ 1.777 $&$0.689 $&$0.926 $&$1.051 $\nl
$18.100$&$ 1.903 $&$0.748 $&$1.003 $&$1.144 $\nl
$18.300$&$ 2.016 $&$0.812 $&$1.094 $&$1.255 $\nl
$18.500$&$ 2.129 $&$0.884 $&$1.194 $&$1.375 $\nl
$18.700$&$ 2.294 $&$0.958 $&$1.294 $&$1.488 $\nl
$18.900$&\nodata&$1.035 $&$1.404 $&$1.609 $\nl
$19.100$&\nodata&$1.120 $&$1.531 $&$1.758 $\nl
$19.300$&\nodata&$1.195 $&$1.651 $&$1.904 $\nl
$19.500$&\nodata&$1.255 $&$1.748 $&$2.027 $\nl
$19.700$&\nodata&$1.312 $&$1.842 $&$2.153 $\nl
$19.900$&\nodata&$1.360 $&$1.942 $&$2.284 $\nl
$20.100$&\nodata&$1.394 $&$2.032 $&$2.398 $\nl
$20.300$&\nodata&$1.422 $&$2.107 $&$2.489 $\nl
$20.500$&\nodata&$1.438 $&$2.179 $&$2.575 $\nl
$20.700$&\nodata&$1.446 $&$2.249 $&$2.678 $\nl
$20.900$&\nodata&$1.457 $&$2.309 $&$2.777 $\nl
$21.100$&\nodata&$1.464 $&$2.353 $&$2.851 $\nl
$21.300$&\nodata&$1.462 $&$2.389 $&$2.928 $\nl
\enddata
\end{deluxetable}

\begin{deluxetable}{ccccc}
\tablewidth{0pt}
\tablecaption{Fiducial Sequences for M67 (NGC~2682)\label{tab:m67}}
\tablehead{
  \colhead{$r$} &
  \colhead{$u - g$} &
  \colhead{$g - r$} &
  \colhead{$g - i$} &
  \colhead{$g - z$}
}
\startdata
$13.500 $&$ 1.246 $&$ 0.369 $&\nodata&$ 0.499 $\nl
$13.700 $&$ 1.256 $&$ 0.388 $&$ 0.522 $&$ 0.535 $\nl
$13.900 $&$ 1.287 $&$ 0.408 $&$ 0.539 $&$ 0.569 $\nl
$14.100 $&$ 1.327 $&$ 0.431 $&$ 0.566 $&$ 0.612 $\nl
$14.300 $&$ 1.378 $&$ 0.465 $&$ 0.610 $&$ 0.663 $\nl
$14.500 $&$ 1.444 $&$ 0.507 $&$ 0.665 $&$ 0.719 $\nl
$14.700 $&$ 1.519 $&$ 0.545 $&$ 0.719 $&$ 0.785 $\nl
$14.900 $&$ 1.601 $&$ 0.579 $&$ 0.766 $&$ 0.849 $\nl
$15.100 $&$ 1.692 $&$ 0.616 $&$ 0.816 $&$ 0.911 $\nl
$15.300 $&$ 1.794 $&$ 0.661 $&$ 0.878 $&$ 0.992 $\nl
$15.500 $&$ 1.896 $&$ 0.713 $&$ 0.949 $&$ 1.082 $\nl
$15.700 $&$ 2.011 $&$ 0.772 $&$ 1.036 $&$ 1.186 $\nl
$15.900 $&$ 2.143 $&$ 0.841 $&$ 1.136 $&$ 1.303 $\nl
$16.100 $&$ 2.259 $&$ 0.912 $&$ 1.230 $&$ 1.406 $\nl
$16.300 $&$ 2.368 $&$ 0.988 $&$ 1.338 $&$ 1.535 $\nl
$16.500 $&$ 2.467 $&$ 1.070 $&$ 1.465 $&$ 1.696 $\nl
$16.700 $&$ 2.533 $&$ 1.147 $&$ 1.578 $&$ 1.828 $\nl
$16.900 $&$ 2.591 $&$ 1.216 $&$ 1.674 $&$ 1.943 $\nl
$17.100 $&$ 2.638 $&$ 1.279 $&$ 1.773 $&$ 2.066 $\nl
$17.300 $&$ 2.660 $&$ 1.330 $&$ 1.866 $&$ 2.182 $\nl
$17.500 $&$ 2.655 $&$ 1.363 $&$ 1.953 $&$ 2.297 $\nl
$17.700 $&$ 2.628 $&$ 1.392 $&$ 2.048 $&$ 2.422 $\nl
$17.900 $&$ 2.625 $&$ 1.419 $&$ 2.130 $&$ 2.539 $\nl
$18.100 $&$ 2.655 $&$ 1.440 $&$ 2.188 $&$ 2.628 $\nl
$18.300 $&\nodata&$ 1.455 $&$ 2.241 $&$ 2.704 $\nl
$18.500 $&\nodata&$ 1.454 $&$ 2.292 $&$ 2.776 $\nl
$18.700 $&\nodata&$ 1.448 $&$ 2.339 $&$ 2.841 $\nl
$18.900 $&\nodata&$ 1.458 $&$ 2.387 $&$ 2.908 $\nl
$19.100 $&\nodata&$ 1.460 $&$ 2.430 $&$ 2.974 $\nl
$19.300 $&\nodata&$ 1.451 $&$ 2.474 $&$ 3.043 $\nl
$19.500 $&\nodata&$ 1.445 $&$ 2.520 $&$ 3.116 $\nl
$19.700 $&\nodata&$ 1.447 $&$ 2.569 $&$ 3.189 $\nl
$19.900 $&\nodata&$ 1.463 $&$ 2.612 $&$ 3.253 $\nl
$20.100 $&\nodata&$ 1.476 $&$ 2.635 $&$ 3.285 $\nl
$20.300 $&\nodata&$ 1.467 $&$ 2.671 $&$ 3.338 $\nl
$20.500 $&\nodata&$ 1.464 $&$ 2.711 $&$ 3.415 $\nl
$20.700 $&\nodata&$ 1.487 $&$ 2.714 $&$ 3.417 $\nl
$20.900 $&\nodata&$ 1.505 $&$ 2.680 $&$ 3.333 $\nl
\enddata
\end{deluxetable}

\begin{deluxetable}{ccccc}
\tablewidth{0pt}
\tablecaption{Fiducial Sequences for Pal~3\label{tab:pal3}}
\tablehead{
  \colhead{$r$} &
  \colhead{$u - g$} &
  \colhead{$g - r$} &
  \colhead{$g - i$} &
  \colhead{$g - z$}
}
\startdata
$ 17.071$&$ 2.431$&$1.059$&$ 1.461$&$ 1.697 $\nl
$ 17.750$&$ 2.231$&$0.929$&$ 1.309$&$ 1.528 $\nl
$ 18.250$&$ 2.079$&$0.836$&$ 1.189$&$ 1.388 $\nl
$ 18.750$&$ 1.915$&$0.777$&$ 1.097$&$ 1.271 $\nl
$ 19.250$&$ 1.682$&$0.712$&$ 1.018$&$ 1.154 $\nl
$ 19.750$&$ 1.485$&$0.643$&$ 0.945$&$ 1.055 $\nl
$ 20.250$&$ 1.376$&$0.585$&$ 0.882$&$ 0.999 $\nl
$ 20.750$&\nodata&$0.528$&$ 0.805$&$ 0.944 $\nl
$ 21.250$&\nodata&$0.487$&$ 0.709$&$ 0.929 $\nl
$ 21.750$&\nodata&$0.461$&$ 0.598$&\nodata\nl
\enddata
\end{deluxetable}

\begin{deluxetable}{ccccc}
\tablewidth{0pt}
\tablecaption{Fiducial Sequences for Pal~4\label{tab:pal4}}
\tablehead{
  \colhead{$r$} &
  \colhead{$u - g$} &
  \colhead{$g - r$} &
  \colhead{$g - i$} &
  \colhead{$g - z$}
}
\startdata
$16.526$&$3.305 $&$1.589 $&$2.416 $&$2.800$\nl
$17.250$&$3.224 $&$1.307 $&$1.915 $&$2.225$\nl
$17.750$&$3.014 $&$1.149 $&$1.641 $&$1.905$\nl
$18.250$&$2.662 $&$1.005 $&$1.418 $&$1.645$\nl
$18.750$&$2.346 $&$0.876 $&$1.230 $&$1.419$\nl
$19.250$&$2.099 $&$0.781 $&$1.094 $&$1.252$\nl
$19.750$&$1.858 $&$0.722 $&$1.013 $&$1.166$\nl
$20.250$&$1.564 $&$0.673 $&$0.951 $&$1.092$\nl
$20.750$&\nodata&$0.627 $&$0.884 $&$1.050$\nl
$21.250$&\nodata&$0.589 $&$0.820 $&$1.105$\nl
$21.750$&\nodata&$0.573 $&$0.793 $&$1.201$\nl
\enddata
\end{deluxetable}

Although a cluster ridgeline was obtained as a by-product from the
photometric filtering, it has many small scale structures, which are
mostly not physical.  To obtain a smooth cluster sequence we used
wider magnitude bins and estimated median colors in the last stage of
the filtering process.  We adjusted the magnitude bin size to adequately
follow the shapes of the MS, SGB, and RGB, and smoothed the curve by
averaging each point with a linear interpolation between adjacent points.

Although the above method worked well in most of the cases, it often
showed a deviation at the top of the RGB for sparsely populated clusters
(e.g., NGC~4147, Pal~5).  We adjusted these sequences by hand to match
the observed RGB.  In addition, we drew by hand the SGB of NGC~6791,
which exhibits double color peaks at a given $r$ magnitude.
Fiducial sequences for the 20 clusters in our study are provided in
Tables~\ref{tab:ngc2419}-\ref{tab:m2}.

\begin{deluxetable}{ccccc}
\tablewidth{0pt}
\tablecaption{Fiducial Sequences for NGC~4147\label{tab:ngc4147}}
\tablehead{
  \colhead{$r$} &
  \colhead{$u - g$} &
  \colhead{$g - r$} &
  \colhead{$g - i$} &
  \colhead{$g - z$}
}
\startdata
$14.350$&$2.250$&$0.850$&$1.160$&$1.450$\nl
$14.850$&$2.010$&$0.757$&$1.079$&$1.289$\nl
$15.350$&$1.819$&$0.696$&$1.016$&$1.187$\nl
$15.850$&$1.676$&$0.652$&$0.959$&$1.121$\nl
$16.350$&$1.546$&$0.597$&$0.892$&$1.046$\nl
$16.850$&$1.451$&$0.548$&$0.822$&$0.964$\nl
$17.350$&$1.387$&$0.511$&$0.757$&$0.891$\nl
$17.850$&$1.321$&$0.480$&$0.705$&$0.828$\nl
$18.350$&$1.255$&$0.457$&$0.671$&$0.776$\nl
$18.850$&$1.189$&$0.440$&$0.643$&$0.735$\nl
$19.150$&$1.132$&$0.435$&$0.624$&$0.702$\nl
$19.250$&$1.122$&$0.432$&$0.612$&$0.685$\nl
$19.350$&$1.134$&$0.414$&$0.582$&$0.659$\nl
$19.450$&$1.140$&$0.386$&$0.546$&$0.620$\nl
$19.550$&$1.107$&$0.363$&$0.526$&$0.578$\nl
$19.650$&$1.055$&$0.331$&$0.491$&$0.518$\nl
$19.750$&$1.024$&$0.291$&$0.433$&$0.447$\nl
$19.850$&$0.995$&$0.263$&$0.389$&$0.407$\nl
$19.950$&$0.980$&$0.244$&$0.359$&$0.378$\nl
$20.050$&$0.983$&$0.232$&$0.337$&$0.342$\nl
$20.250$&$0.959$&$0.224$&$0.318$&$0.316$\nl
$20.550$&$0.921$&$0.227$&$0.313$&$0.322$\nl
$20.850$&$0.888$&$0.236$&$0.331$&$0.358$\nl
$21.150$&\nodata&$0.261$&$0.370$&$0.445$\nl
$21.450$&\nodata&$0.293$&$0.413$&\nodata\nl
$21.750$&\nodata&$0.319$&$0.455$&\nodata\nl
\enddata
\end{deluxetable}

\begin{deluxetable}{ccccc}
\tablewidth{0pt}
\tablecaption{Fiducial Sequences for M53 (NGC~5024)\label{tab:m53}}
\tablehead{
  \colhead{$r$} &
  \colhead{$u - g$} &
  \colhead{$g - r$} &
  \colhead{$g - i$} &
  \colhead{$g - z$}
}
\startdata
 $  13.750 $&$ 2.461 $&$ 1.026 $&\nodata&$ 1.707 $\nl
 $  14.250 $&$ 2.117 $&$ 0.890 $&$ 1.254 $&$ 1.479 $\nl
 $  14.750 $&$ 1.869 $&$ 0.788 $&$ 1.132 $&$ 1.313 $\nl
 $  15.250 $&$ 1.686 $&$ 0.708 $&$ 1.022 $&$ 1.193 $\nl
 $  15.750 $&$ 1.539 $&$ 0.645 $&$ 0.936 $&$ 1.093 $\nl
 $  16.250 $&$ 1.422 $&$ 0.600 $&$ 0.871 $&$ 1.015 $\nl
 $  16.750 $&$ 1.328 $&$ 0.560 $&$ 0.816 $&$ 0.947 $\nl
 $  17.250 $&$ 1.254 $&$ 0.525 $&$ 0.765 $&$ 0.887 $\nl
 $  17.750 $&$ 1.190 $&$ 0.497 $&$ 0.724 $&$ 0.837 $\nl
 $  18.250 $&$ 1.137 $&$ 0.475 $&$ 0.692 $&$ 0.794 $\nl
 $  18.750 $&$ 1.087 $&$ 0.451 $&$ 0.658 $&$ 0.751 $\nl
 $  19.150 $&$ 1.034 $&$ 0.425 $&$ 0.621 $&$ 0.708 $\nl
 $  19.250 $&$ 1.014 $&$ 0.414 $&$ 0.605 $&$ 0.689 $\nl
 $  19.350 $&$ 0.989 $&$ 0.397 $&$ 0.576 $&$ 0.657 $\nl
 $  19.450 $&$ 0.963 $&$ 0.369 $&$ 0.532 $&$ 0.606 $\nl
 $  19.550 $&$ 0.938 $&$ 0.334 $&$ 0.479 $&$ 0.537 $\nl
 $  19.650 $&$ 0.915 $&$ 0.305 $&$ 0.431 $&$ 0.468 $\nl
 $  19.750 $&$ 0.897 $&$ 0.280 $&$ 0.393 $&$ 0.414 $\nl
 $  19.850 $&$ 0.890 $&$ 0.260 $&$ 0.362 $&$ 0.375 $\nl
 $  19.950 $&$ 0.894 $&$ 0.247 $&$ 0.339 $&$ 0.351 $\nl
 $  20.050 $&$ 0.895 $&$ 0.239 $&$ 0.327 $&$ 0.341 $\nl
 $  20.250 $&$ 0.877 $&$ 0.235 $&$ 0.324 $&$ 0.343 $\nl
 $  20.550 $&$ 0.845 $&$ 0.243 $&$ 0.335 $&$ 0.373 $\nl
 $  20.850 $&$ 0.809 $&$ 0.263 $&$ 0.360 $&$ 0.433 $\nl
 $  21.150 $&\nodata&$ 0.291 $&$ 0.400 $&\nodata\nl
 $  21.450 $&\nodata&$ 0.324 $&$ 0.453 $&\nodata\nl
 $  21.750 $&\nodata&$ 0.362 $&$ 0.510 $&\nodata\nl
\enddata
\end{deluxetable}

\begin{deluxetable}{ccccc}
\tablewidth{0pt}
\tablecaption{Fiducial Sequences for NGC~5053\label{tab:ngc5053}}
\tablehead{
  \colhead{$r$} &
  \colhead{$u - g$} &
  \colhead{$g - r$} &
  \colhead{$g - i$} &
  \colhead{$g - z$}
}
\startdata
 $  15.000 $&$ 1.555 $&$ 0.688 $&$ 0.995 $&$ 1.133 $\nl
 $  15.750 $&$ 1.404 $&$ 0.611 $&$ 0.881 $&$ 1.011 $\nl
 $  16.250 $&$ 1.313 $&$ 0.564 $&$ 0.818 $&$ 0.942 $\nl
 $  16.750 $&$ 1.235 $&$ 0.530 $&$ 0.770 $&$ 0.883 $\nl
 $  17.250 $&$ 1.167 $&$ 0.501 $&$ 0.727 $&$ 0.828 $\nl
 $  17.750 $&$ 1.113 $&$ 0.474 $&$ 0.689 $&$ 0.781 $\nl
 $  18.250 $&$ 1.067 $&$ 0.447 $&$ 0.651 $&$ 0.738 $\nl
 $  18.750 $&$ 1.026 $&$ 0.417 $&$ 0.605 $&$ 0.682 $\nl
 $  18.850 $&$ 1.015 $&$ 0.410 $&$ 0.595 $&$ 0.664 $\nl
 $  18.950 $&$ 0.998 $&$ 0.403 $&$ 0.584 $&$ 0.635 $\nl
 $  19.050 $&$ 0.968 $&$ 0.393 $&$ 0.566 $&$ 0.607 $\nl
 $  19.150 $&$ 0.934 $&$ 0.371 $&$ 0.532 $&$ 0.586 $\nl
 $  19.250 $&$ 0.929 $&$ 0.340 $&$ 0.483 $&$ 0.541 $\nl
 $  19.350 $&$ 0.928 $&$ 0.303 $&$ 0.427 $&$ 0.471 $\nl
 $  19.450 $&$ 0.913 $&$ 0.274 $&$ 0.381 $&$ 0.407 $\nl
 $  19.550 $&$ 0.905 $&$ 0.254 $&$ 0.347 $&$ 0.360 $\nl
 $  19.650 $&$ 0.896 $&$ 0.237 $&$ 0.322 $&$ 0.317 $\nl
 $  19.750 $&$ 0.886 $&$ 0.225 $&$ 0.308 $&$ 0.293 $\nl
 $  19.950 $&$ 0.878 $&$ 0.215 $&$ 0.296 $&$ 0.293 $\nl
 $  20.250 $&$ 0.864 $&$ 0.218 $&$ 0.297 $&$ 0.313 $\nl
 $  20.550 $&$ 0.849 $&$ 0.234 $&$ 0.319 $&$ 0.339 $\nl
 $  20.850 $&$ 0.822 $&$ 0.256 $&$ 0.354 $&$ 0.382 $\nl
 $  21.150 $&$ 0.796 $&$ 0.284 $&$ 0.397 $&$ 0.452 $\nl
 $  21.450 $&\nodata&$ 0.323 $&$ 0.448 $&\nodata\nl
 $  21.750 $&\nodata&$ 0.369 $&$ 0.508 $&\nodata\nl
\enddata
\end{deluxetable}

\begin{deluxetable}{ccccc}
\tablewidth{0pt}
\tablecaption{Fiducial Sequences for M3 (NGC~5272)\label{tab:m3}}
\tablehead{
  \colhead{$r$} &
  \colhead{$u - g$} &
  \colhead{$g - r$} &
  \colhead{$g - i$} &
  \colhead{$g - z$}
}
\startdata
 $ 14.000 $&$ 1.895 $&$ 0.756 $&\nodata&$ 1.219 $\nl
 $ 14.250 $&$ 1.784 $&$ 0.726 $&$ 1.015 $&$ 1.178 $\nl
 $ 14.750 $&$ 1.616 $&$ 0.673 $&$ 0.943 $&$ 1.089 $\nl
 $ 15.250 $&$ 1.494 $&$ 0.627 $&$ 0.878 $&$ 1.007 $\nl
 $ 15.750 $&$ 1.402 $&$ 0.587 $&$ 0.821 $&$ 0.938 $\nl
 $ 16.250 $&$ 1.327 $&$ 0.556 $&$ 0.773 $&$ 0.882 $\nl
 $ 16.750 $&$ 1.262 $&$ 0.531 $&$ 0.733 $&$ 0.833 $\nl
 $ 17.250 $&$ 1.208 $&$ 0.507 $&$ 0.696 $&$ 0.787 $\nl
 $ 17.750 $&$ 1.154 $&$ 0.480 $&$ 0.656 $&$ 0.742 $\nl
 $ 17.850 $&$ 1.143 $&$ 0.473 $&$ 0.646 $&$ 0.724 $\nl
 $ 17.950 $&$ 1.126 $&$ 0.465 $&$ 0.633 $&$ 0.701 $\nl
 $ 18.050 $&$ 1.096 $&$ 0.451 $&$ 0.612 $&$ 0.678 $\nl
 $ 18.150 $&$ 1.061 $&$ 0.424 $&$ 0.575 $&$ 0.635 $\nl
 $ 18.250 $&$ 1.022 $&$ 0.387 $&$ 0.521 $&$ 0.561 $\nl
 $ 18.350 $&$ 0.979 $&$ 0.346 $&$ 0.458 $&$ 0.483 $\nl
 $ 18.450 $&$ 0.948 $&$ 0.312 $&$ 0.406 $&$ 0.424 $\nl
 $ 18.550 $&$ 0.935 $&$ 0.291 $&$ 0.374 $&$ 0.387 $\nl
 $ 18.650 $&$ 0.929 $&$ 0.279 $&$ 0.355 $&$ 0.364 $\nl
 $ 18.750 $&$ 0.926 $&$ 0.273 $&$ 0.345 $&$ 0.350 $\nl
 $ 18.950 $&$ 0.919 $&$ 0.269 $&$ 0.339 $&$ 0.340 $\nl
 $ 19.250 $&$ 0.908 $&$ 0.275 $&$ 0.348 $&$ 0.352 $\nl
 $ 19.550 $&$ 0.904 $&$ 0.295 $&$ 0.376 $&$ 0.386 $\nl
 $ 19.850 $&$ 0.908 $&$ 0.322 $&$ 0.416 $&$ 0.436 $\nl
 $ 20.150 $&$ 0.921 $&$ 0.357 $&$ 0.464 $&$ 0.496 $\nl
 $ 20.450 $&$ 0.942 $&$ 0.400 $&$ 0.526 $&$ 0.580 $\nl
 $ 20.750 $&\nodata&$ 0.451 $&$ 0.599 $&$ 0.690 $\nl
 $ 21.050 $&\nodata&$ 0.509 $&$ 0.682 $&\nodata\nl
 $ 21.350 $&\nodata&$ 0.574 $&$ 0.775 $&\nodata\nl
 $ 21.650 $&\nodata&$ 0.640 $&$ 0.872 $&\nodata\nl
\enddata
\end{deluxetable}

\begin{deluxetable}{ccccc}
\tablewidth{0pt}
\tablecaption{Fiducial Sequences for NGC~5466\label{tab:ngc5466}}
\tablehead{
  \colhead{$r$} &
  \colhead{$u - g$} &
  \colhead{$g - r$} &
  \colhead{$g - i$} &
  \colhead{$g - z$}
}
\startdata
$13.750 $&$ 2.227 $&$ 0.954 $&$ 1.366 $&$ 1.576 $\nl
$14.250 $&$ 1.918 $&$ 0.839 $&$ 1.197 $&$ 1.386 $\nl
$14.750 $&$ 1.707 $&$ 0.748 $&$ 1.072 $&$ 1.241 $\nl
$15.250 $&$ 1.556 $&$ 0.679 $&$ 0.976 $&$ 1.128 $\nl
$15.750 $&$ 1.430 $&$ 0.625 $&$ 0.898 $&$ 1.034 $\nl
$16.250 $&$ 1.340 $&$ 0.582 $&$ 0.840 $&$ 0.961 $\nl
$16.750 $&$ 1.262 $&$ 0.545 $&$ 0.786 $&$ 0.897 $\nl
$17.250 $&$ 1.195 $&$ 0.514 $&$ 0.738 $&$ 0.840 $\nl
$17.750 $&$ 1.145 $&$ 0.491 $&$ 0.701 $&$ 0.795 $\nl
$18.250 $&$ 1.099 $&$ 0.465 $&$ 0.662 $&$ 0.747 $\nl
$18.750 $&$ 1.046 $&$ 0.429 $&$ 0.609 $&$ 0.681 $\nl
$18.750 $&$ 1.046 $&$ 0.429 $&$ 0.609 $&$ 0.681 $\nl
$18.850 $&$ 1.034 $&$ 0.419 $&$ 0.595 $&$ 0.658 $\nl
$18.950 $&$ 1.016 $&$ 0.406 $&$ 0.570 $&$ 0.621 $\nl
$19.050 $&$ 0.981 $&$ 0.378 $&$ 0.526 $&$ 0.568 $\nl
$19.150 $&$ 0.947 $&$ 0.338 $&$ 0.469 $&$ 0.503 $\nl
$19.250 $&$ 0.925 $&$ 0.302 $&$ 0.418 $&$ 0.435 $\nl
$19.350 $&$ 0.912 $&$ 0.275 $&$ 0.378 $&$ 0.381 $\nl
$19.450 $&$ 0.903 $&$ 0.255 $&$ 0.348 $&$ 0.349 $\nl
$19.550 $&$ 0.897 $&$ 0.243 $&$ 0.328 $&$ 0.331 $\nl
$19.650 $&$ 0.895 $&$ 0.237 $&$ 0.316 $&$ 0.320 $\nl
$19.850 $&$ 0.891 $&$ 0.237 $&$ 0.312 $&$ 0.318 $\nl
$20.150 $&$ 0.870 $&$ 0.245 $&$ 0.322 $&$ 0.338 $\nl
$20.450 $&$ 0.848 $&$ 0.263 $&$ 0.347 $&$ 0.384 $\nl
$20.750 $&$ 0.831 $&$ 0.289 $&$ 0.387 $&$ 0.469 $\nl
$21.050 $&\nodata&$ 0.320 $&$ 0.437 $&$ 0.589 $\nl
$21.350 $&\nodata&$ 0.358 $&$ 0.493 $&\nodata\nl
$21.650 $&\nodata&$ 0.402 $&$ 0.551 $&\nodata\nl
\enddata
\end{deluxetable}

\begin{deluxetable}{ccccc}
\tablewidth{0pt}
\tablecaption{Fiducial Sequences for Pal~5\label{tab:pal5}}
\tablehead{
  \colhead{$r$} &
  \colhead{$u - g$} &
  \colhead{$g - r$} &
  \colhead{$g - i$} &
  \colhead{$g - z$}
}
\startdata
 $  15.250 $&$  2.236 $&$ 0.935 $&$ 1.332 $&$ 1.560 $\nl
 $  15.750 $&$  2.158 $&$ 0.879 $&$ 1.258 $&$ 1.481 $\nl
 $  16.250 $&$  2.020 $&$ 0.811 $&$ 1.169 $&$ 1.376 $\nl
 $  16.750 $&$  1.833 $&$ 0.749 $&$ 1.081 $&$ 1.271 $\nl
 $  17.250 $&$  1.636 $&$ 0.694 $&$ 1.000 $&$ 1.174 $\nl
 $  17.750 $&$  1.513 $&$ 0.644 $&$ 0.929 $&$ 1.086 $\nl
 $  18.250 $&$  1.440 $&$ 0.610 $&$ 0.873 $&$ 1.012 $\nl
 $  18.750 $&$  1.349 $&$ 0.587 $&$ 0.834 $&$ 0.956 $\nl
 $  19.250 $&$  1.267 $&$ 0.566 $&$ 0.802 $&$ 0.928 $\nl
 $  19.550 $&$  1.268 $&$ 0.543 $&$ 0.776 $&$ 0.905 $\nl
 $  19.650 $&$  1.297 $&$ 0.530 $&$ 0.754 $&$ 0.886 $\nl
 $  19.750 $&$  1.278 $&$ 0.519 $&$ 0.727 $&$ 0.860 $\nl
 $  19.850 $&$  1.188 $&$ 0.507 $&$ 0.705 $&$ 0.803 $\nl
 $  19.950 $&$  1.117 $&$ 0.474 $&$ 0.661 $&$ 0.723 $\nl
 $  20.050 $&$  1.088 $&$ 0.424 $&$ 0.588 $&$ 0.657 $\nl
 $  20.150 $&$  1.072 $&$ 0.380 $&$ 0.517 $&$ 0.599 $\nl
 $  20.250 $&$  1.047 $&$ 0.357 $&$ 0.484 $&$ 0.551 $\nl
 $  20.350 $&$  1.028 $&$ 0.351 $&$ 0.475 $&$ 0.528 $\nl
 $  20.450 $&$  1.015 $&$ 0.346 $&$ 0.466 $&$ 0.527 $\nl
 $  20.650 $&$  0.975 $&$ 0.339 $&$ 0.460 $&$ 0.531 $\nl
 $  20.950 $&$  0.921 $&$ 0.352 $&$ 0.468 $&$ 0.526 $\nl
 $  21.250 $&\nodata&$ 0.368 $&$ 0.496 $&\nodata\nl 
 $  21.550 $&\nodata&$ 0.388 $&$ 0.538 $&\nodata\nl
 $  21.850 $&\nodata&$ 0.418 $&$ 0.587 $&\nodata\nl
\enddata
\end{deluxetable}

\begin{deluxetable}{ccccc}
\tablewidth{0pt}
\tablecaption{Fiducial Sequences for M5 (NGC~5904)\label{tab:m5}}
\tablehead{
  \colhead{$r$} &
  \colhead{$u - g$} &
  \colhead{$g - r$} &
  \colhead{$g - i$} &
  \colhead{$g - z$}
}
\startdata
 $  13.250 $&$ 2.047 $&$ 0.915 $&\nodata&$ 1.395 $\nl
 $  13.750 $&$ 1.885 $&$ 0.792 $&\nodata&$ 1.295 $\nl
 $  14.250 $&$ 1.745 $&$ 0.710 $&$ 1.017 $&$ 1.190 $\nl
 $  14.750 $&$ 1.631 $&$ 0.661 $&$ 0.951 $&$ 1.094 $\nl
 $  15.250 $&$ 1.528 $&$ 0.621 $&$ 0.891 $&$ 1.016 $\nl
 $  15.750 $&$ 1.444 $&$ 0.584 $&$ 0.838 $&$ 0.953 $\nl
 $  16.250 $&$ 1.378 $&$ 0.552 $&$ 0.794 $&$ 0.901 $\nl
 $  16.750 $&$ 1.321 $&$ 0.525 $&$ 0.756 $&$ 0.855 $\nl
 $  17.250 $&$ 1.271 $&$ 0.498 $&$ 0.717 $&$ 0.808 $\nl
 $  17.250 $&$ 1.271 $&$ 0.498 $&$ 0.717 $&$ 0.808 $\nl
 $  17.350 $&$ 1.253 $&$ 0.490 $&$ 0.709 $&$ 0.794 $\nl
 $  17.450 $&$ 1.229 $&$ 0.477 $&$ 0.693 $&$ 0.774 $\nl
 $  17.550 $&$ 1.203 $&$ 0.459 $&$ 0.662 $&$ 0.737 $\nl
 $  17.650 $&$ 1.155 $&$ 0.429 $&$ 0.616 $&$ 0.680 $\nl
 $  17.750 $&$ 1.089 $&$ 0.386 $&$ 0.554 $&$ 0.604 $\nl
 $  17.850 $&$ 1.040 $&$ 0.343 $&$ 0.492 $&$ 0.530 $\nl
 $  17.950 $&$ 1.017 $&$ 0.315 $&$ 0.450 $&$ 0.475 $\nl
 $  18.050 $&$ 1.006 $&$ 0.299 $&$ 0.427 $&$ 0.444 $\nl
 $  18.150 $&$ 0.999 $&$ 0.290 $&$ 0.415 $&$ 0.429 $\nl
 $  18.350 $&$ 0.986 $&$ 0.284 $&$ 0.406 $&$ 0.420 $\nl
 $  18.650 $&$ 0.976 $&$ 0.290 $&$ 0.414 $&$ 0.428 $\nl
 $  18.950 $&$ 0.975 $&$ 0.308 $&$ 0.439 $&$ 0.456 $\nl
 $  19.250 $&$ 0.983 $&$ 0.335 $&$ 0.478 $&$ 0.502 $\nl
 $  19.550 $&$ 1.012 $&$ 0.368 $&$ 0.527 $&$ 0.560 $\nl
 $  19.850 $&$ 1.068 $&$ 0.411 $&$ 0.589 $&$ 0.635 $\nl
 $  20.150 $&$ 1.140 $&$ 0.463 $&$ 0.662 $&$ 0.725 $\nl
 $  20.450 $&$ 1.218 $&$ 0.522 $&$ 0.746 $&$ 0.831 $\nl
 $  20.750 $&\nodata&$ 0.588 $&$ 0.841 $&$ 0.950 $\nl
 $  21.050 $&\nodata&$ 0.665 $&$ 0.949 $&\nodata\nl
 $  21.350 $&\nodata&$ 0.754 $&$ 1.074 $&\nodata\nl
 $  21.650 $&\nodata&$ 0.850 $&$ 1.213 $&\nodata\nl
\enddata
\end{deluxetable}

\clearpage

\begin{deluxetable}{ccccc}
\tablewidth{0pt}
\tablecaption{Fiducial Sequences for Pal~14\label{tab:pal14}}
\tablehead{
  \colhead{$r$} &
  \colhead{$u - g$} &
  \colhead{$g - r$} &
  \colhead{$g - i$} &
  \colhead{$g - z$}
}
\startdata
$17.905 $&$2.009 $&$0.854 $&$1.221 $&$1.441$\nl
$18.250 $&$1.915 $&$0.817 $&$1.180 $&$1.368$\nl
$18.750 $&$1.865 $&$0.760 $&$1.110 $&$1.280$\nl
$19.250 $&$1.766 $&$0.706 $&$1.030 $&$1.201$\nl
$19.750 $&$1.626 $&$0.655 $&$0.949 $&$1.120$\nl
$20.250 $&\nodata&$0.600 $&$0.879 $&$1.031$\nl
$20.750 $&\nodata&$0.550 $&$0.816 $&$0.956$\nl
$21.250 $&\nodata&$0.521 $&$0.777 $&$0.944$\nl
$21.750 $&\nodata&$0.484 $&$0.775 $&\nodata\nl
\enddata
\end{deluxetable}

\begin{deluxetable}{ccccc}
\tablewidth{0pt}
\tablecaption{Fiducial Sequences for M13 (NGC~6205)\label{tab:m13}}
\tablehead{
  \colhead{$r$} &
  \colhead{$u - g$} &
  \colhead{$g - r$} &
  \colhead{$g - i$} &
  \colhead{$g - z$}
}
\startdata
 $   14.000 $&$ 1.669 $&$ 0.679 $&$ 0.971 $&$ 1.119 $\nl
 $   14.750 $&$ 1.500 $&$ 0.613 $&$ 0.862 $&$ 0.998 $\nl
 $   15.250 $&$ 1.402 $&$ 0.574 $&$ 0.805 $&$ 0.928 $\nl
 $   15.750 $&$ 1.329 $&$ 0.540 $&$ 0.761 $&$ 0.873 $\nl
 $   16.250 $&$ 1.275 $&$ 0.512 $&$ 0.722 $&$ 0.827 $\nl
 $   16.750 $&$ 1.222 $&$ 0.486 $&$ 0.686 $&$ 0.784 $\nl
 $   17.250 $&$ 1.164 $&$ 0.459 $&$ 0.652 $&$ 0.735 $\nl
 $   17.350 $&$ 1.148 $&$ 0.451 $&$ 0.643 $&$ 0.721 $\nl
 $   17.450 $&$ 1.123 $&$ 0.439 $&$ 0.623 $&$ 0.699 $\nl
 $   17.550 $&$ 1.092 $&$ 0.421 $&$ 0.591 $&$ 0.665 $\nl
 $   17.650 $&$ 1.053 $&$ 0.392 $&$ 0.545 $&$ 0.610 $\nl
 $   17.750 $&$ 1.014 $&$ 0.357 $&$ 0.493 $&$ 0.544 $\nl
 $   17.850 $&$ 0.986 $&$ 0.325 $&$ 0.447 $&$ 0.486 $\nl
 $   17.950 $&$ 0.966 $&$ 0.299 $&$ 0.410 $&$ 0.441 $\nl
 $   18.050 $&$ 0.951 $&$ 0.282 $&$ 0.383 $&$ 0.409 $\nl
 $   18.150 $&$ 0.942 $&$ 0.271 $&$ 0.364 $&$ 0.387 $\nl
 $   18.250 $&$ 0.938 $&$ 0.264 $&$ 0.355 $&$ 0.375 $\nl
 $   18.450 $&$ 0.929 $&$ 0.262 $&$ 0.352 $&$ 0.370 $\nl
 $   18.750 $&$ 0.917 $&$ 0.270 $&$ 0.363 $&$ 0.383 $\nl
 $   19.050 $&$ 0.913 $&$ 0.288 $&$ 0.388 $&$ 0.415 $\nl
 $   19.350 $&$ 0.923 $&$ 0.315 $&$ 0.427 $&$ 0.462 $\nl
 $   19.650 $&$ 0.952 $&$ 0.349 $&$ 0.476 $&$ 0.523 $\nl
 $   19.950 $&$ 0.997 $&$ 0.391 $&$ 0.535 $&$ 0.599 $\nl
 $   20.250 $&$ 1.053 $&$ 0.441 $&$ 0.608 $&$ 0.687 $\nl
 $   20.550 $&\nodata&$ 0.500 $&$ 0.692 $&$ 0.792 $\nl
 $   20.850 $&\nodata&$ 0.569 $&$ 0.791 $&$ 0.913 $\nl
 $   21.150 $&\nodata&$ 0.643 $&$ 0.895 $&\nodata\nl
 $   21.450 $&\nodata&$ 0.719 $&$ 1.004 $&\nodata\nl
 $   21.750 $&\nodata&$ 0.798 $&$ 1.127 $&\nodata\nl
\enddata
\end{deluxetable}

\begin{deluxetable}{ccccc}
\tablewidth{0pt}
\tablecaption{Fiducial Sequences for M92 (NGC~6342)\label{tab:m92}}
\tablehead{
  \colhead{$r$} &
  \colhead{$u - g$} &
  \colhead{$g - r$} &
  \colhead{$g - i$} &
  \colhead{$g - z$}
}
\startdata
 $ 13.750 $&$ 1.531 $&$ 0.661 $&\nodata&$ 1.079 $\nl
 $ 14.250 $&$ 1.432 $&$ 0.612 $&$ 0.870 $&$ 1.000 $\nl
 $ 14.750 $&$ 1.355 $&$ 0.570 $&$ 0.816 $&$ 0.929 $\nl
 $ 15.250 $&$ 1.285 $&$ 0.530 $&$ 0.766 $&$ 0.867 $\nl
 $ 15.750 $&$ 1.220 $&$ 0.497 $&$ 0.724 $&$ 0.814 $\nl
 $ 16.250 $&$ 1.169 $&$ 0.474 $&$ 0.685 $&$ 0.766 $\nl
 $ 16.750 $&$ 1.125 $&$ 0.451 $&$ 0.649 $&$ 0.720 $\nl
 $ 17.250 $&$ 1.081 $&$ 0.422 $&$ 0.609 $&$ 0.675 $\nl
 $ 17.350 $&$ 1.070 $&$ 0.415 $&$ 0.598 $&$ 0.664 $\nl
 $ 17.450 $&$ 1.053 $&$ 0.408 $&$ 0.586 $&$ 0.647 $\nl
 $ 17.550 $&$ 1.034 $&$ 0.394 $&$ 0.567 $&$ 0.622 $\nl
 $ 17.650 $&$ 1.015 $&$ 0.371 $&$ 0.532 $&$ 0.584 $\nl
 $ 17.750 $&$ 0.992 $&$ 0.341 $&$ 0.485 $&$ 0.529 $\nl
 $ 17.850 $&$ 0.970 $&$ 0.308 $&$ 0.437 $&$ 0.466 $\nl
 $ 17.950 $&$ 0.958 $&$ 0.280 $&$ 0.397 $&$ 0.412 $\nl
 $ 18.050 $&$ 0.955 $&$ 0.258 $&$ 0.364 $&$ 0.369 $\nl
 $ 18.150 $&$ 0.952 $&$ 0.242 $&$ 0.336 $&$ 0.333 $\nl
 $ 18.250 $&$ 0.948 $&$ 0.230 $&$ 0.317 $&$ 0.310 $\nl
 $ 18.450 $&$ 0.936 $&$ 0.221 $&$ 0.302 $&$ 0.292 $\nl
 $ 18.750 $&$ 0.918 $&$ 0.225 $&$ 0.307 $&$ 0.299 $\nl
 $ 19.050 $&$ 0.903 $&$ 0.240 $&$ 0.329 $&$ 0.326 $\nl
 $ 19.350 $&$ 0.891 $&$ 0.263 $&$ 0.363 $&$ 0.365 $\nl
 $ 19.650 $&$ 0.886 $&$ 0.291 $&$ 0.405 $&$ 0.414 $\nl
 $ 19.950 $&$ 0.894 $&$ 0.326 $&$ 0.458 $&$ 0.477 $\nl
 $ 20.250 $&$ 0.919 $&$ 0.369 $&$ 0.523 $&$ 0.557 $\nl
 $ 20.550 $&$ 0.969 $&$ 0.422 $&$ 0.597 $&$ 0.657 $\nl
 $ 20.850 $&\nodata&$ 0.483 $&$ 0.682 $&$ 0.771 $\nl
 $ 21.150 $&\nodata&$ 0.544 $&$ 0.774 $&\nodata\nl
 $ 21.450 $&\nodata&$ 0.604 $&$ 0.869 $&\nodata\nl
 $ 21.750 $&\nodata&$ 0.664 $&$ 0.964 $&\nodata\nl
\enddata
\end{deluxetable}

\begin{deluxetable}{ccccc}
\tablewidth{0pt}
\tablecaption{Fiducial Sequences for NGC~6791\label{tab:ngc6791}}
\tablehead{
  \colhead{$r$} &
  \colhead{$u - g$} &
  \colhead{$g - r$} &
  \colhead{$g - i$} &
  \colhead{$g - z$}
}
\startdata
$13.550$&$ 3.424 $&$1.424 $&\nodata&$2.465$\nl
$14.050$&$ 3.227 $&$1.299 $&$1.790 $&$2.147$\nl
$14.550$&$ 3.083 $&$1.206 $&$1.649 $&$1.936$\nl
$15.050$&$ 2.954 $&$1.138 $&$1.543 $&$1.808$\nl
$15.550$&$ 2.815 $&$1.078 $&$1.452 $&$1.697$\nl
$16.050$&$ 2.707 $&$1.034 $&$1.388 $&$1.614$\nl
$16.550$&$ 2.626 $&$1.001 $&$1.348 $&$1.561$\nl
$16.850$&$ 2.572 $&$0.978 $&$1.316 $&$1.524$\nl
$16.950$&$ 2.538 $&$0.968 $&$1.299 $&$1.506$\nl
$17.050$&$ 2.470 $&$0.950 $&$1.273 $&$1.488$\nl
$17.117$&$ 2.371 $&$0.896 $&$1.202 $&$1.380$\nl
$17.100$&$ 2.168 $&$0.820 $&$1.105 $&$1.260$\nl
$17.038$&$ 2.031 $&$0.766 $&$1.023 $&$1.162$\nl
$17.045$&$ 1.950 $&$0.732 $&$0.965 $&$1.110$\nl
$17.097$&$ 1.900 $&$0.702 $&$0.931 $&$1.055$\nl
$17.150$&$ 1.863 $&$0.685 $&$0.913 $&$1.029$\nl
$17.250$&$ 1.843 $&$0.675 $&$0.898 $&$1.012$\nl
$17.350$&$ 1.829 $&$0.670 $&$0.891 $&$1.003$\nl
$17.450$&$ 1.827 $&$0.669 $&$0.890 $&$1.002$\nl
$17.550$&$ 1.830 $&$0.670 $&$0.892 $&$1.002$\nl
$17.750$&$ 1.851 $&$0.682 $&$0.908 $&$1.022$\nl
$18.050$&$ 1.918 $&$0.715 $&$0.952 $&$1.075$\nl
$18.350$&$ 2.016 $&$0.760 $&$1.012 $&$1.145$\nl
$18.650$&$ 2.112 $&$0.815 $&$1.087 $&$1.236$\nl
$18.950$&\nodata&$0.886 $&$1.186 $&$1.356$\nl
$19.250$&\nodata&$0.969 $&$1.306 $&$1.502$\nl
$19.550$&\nodata&$1.061 $&$1.439 $&$1.663$\nl
$19.850$&\nodata&$1.162 $&$1.590 $&$1.843$\nl
$20.150$&\nodata&$1.263 $&$1.753 $&$2.039$\nl
$20.450$&\nodata&$1.354 $&$1.913 $&$2.229$\nl
$20.750$&\nodata&$1.431 $&$2.067 $&$2.425$\nl
\enddata
\end{deluxetable}

\begin{deluxetable}{ccccc}
\tablewidth{0pt}
\tablecaption{Fiducial Sequences for M71 (NGC~6838)\label{tab:m71}}
\tablehead{
  \colhead{$r$} &
  \colhead{$u - g$} &
  \colhead{$g - r$} &
  \colhead{$g - i$} &
  \colhead{$g - z$}
}
\startdata
$13.850$ & $2.464$ & $1.080$ & $9.000$ & $1.879$ \nl
$14.350$ & $2.294$ & $1.010$ & $1.500$ & $1.766$ \nl
$14.850$ & $2.170$ & $0.959$ & $1.433$ & $1.681$ \nl
$15.350$ & $2.074$ & $0.919$ & $1.374$ & $1.616$ \nl
$15.850$ & $1.991$ & $0.890$ & $1.337$ & $1.566$ \nl
$16.350$ & $1.927$ & $0.871$ & $1.309$ & $1.523$ \nl
$16.850$ & $1.871$ & $0.859$ & $1.276$ & $1.487$ \nl
$16.987$ & $1.854$ & $0.857$ & $1.268$ & $1.478$ \nl
$17.200$ & $1.821$ & $0.822$ & $1.240$ & $1.420$ \nl
$17.267$ & $1.776$ & $0.790$ & $1.204$ & $1.373$ \nl
$17.362$ & $1.634$ & $0.730$ & $1.111$ & $1.272$ \nl
$17.443$ & $1.508$ & $0.682$ & $1.045$ & $1.183$ \nl
$17.514$ & $1.456$ & $0.659$ & $1.019$ & $1.140$ \nl
$17.650$ & $1.444$ & $0.650$ & $1.000$ & $1.124$ \nl
$17.750$ & $1.441$ & $0.647$ & $0.991$ & $1.122$ \nl
$17.850$ & $1.437$ & $0.645$ & $0.989$ & $1.123$ \nl
$18.050$ & $1.437$ & $0.645$ & $0.992$ & $1.129$ \nl
$18.350$ & $1.477$ & $0.658$ & $1.006$ & $1.151$ \nl
$18.650$ & $1.540$ & $0.683$ & $1.040$ & $1.193$ \nl
$18.950$ & $1.601$ & $0.716$ & $1.098$ & $1.265$ \nl
$19.250$ & $1.675$ & $0.757$ & $1.171$ & $1.357$ \nl
\enddata
\end{deluxetable}

\begin{deluxetable}{ccccc}
\tablewidth{0pt}
\tablecaption{Fiducial Sequences for NGC~7006\label{tab:ngc7006}}
\tablehead{
  \colhead{$r$} &
  \colhead{$u - g$} &
  \colhead{$g - r$} &
  \colhead{$g - i$} &
  \colhead{$g - z$}
}
\startdata
 $15.250 $&\nodata&$ 1.348 $&$ 1.898 $&$ 2.220 $\nl
 $15.750 $&$ 2.858 $&$ 1.168 $&$ 1.659 $&$ 1.939 $\nl
 $16.250 $&$ 2.508 $&$ 1.030 $&$ 1.462 $&$ 1.709 $\nl
 $16.750 $&$ 2.238 $&$ 0.927 $&$ 1.316 $&$ 1.538 $\nl
 $17.250 $&$ 2.022 $&$ 0.839 $&$ 1.197 $&$ 1.401 $\nl
 $17.750 $&$ 1.825 $&$ 0.763 $&$ 1.095 $&$ 1.279 $\nl
 $18.250 $&$ 1.677 $&$ 0.703 $&$ 1.011 $&$ 1.178 $\nl
 $18.750 $&$ 1.568 $&$ 0.659 $&$ 0.945 $&$ 1.112 $\nl
 $19.250 $&$ 1.461 $&$ 0.622 $&$ 0.890 $&$ 1.059 $\nl
 $19.750 $&$ 1.380 $&$ 0.591 $&$ 0.840 $&$ 1.000 $\nl
 $20.250 $&$ 1.329 $&$ 0.562 $&$ 0.789 $&$ 0.940 $\nl
 $20.750 $&\nodata&$ 0.534 $&$ 0.745 $&$ 0.862 $\nl
 $21.100 $&\nodata&$ 0.504 $&$ 0.696 $&$ 0.801 $\nl
 $21.300 $&\nodata&$ 0.465 $&$ 0.629 $&$ 0.786 $\nl
 $21.500 $&\nodata&$ 0.416 $&$ 0.547 $&\nodata\nl
 $21.700 $&\nodata&$ 0.376 $&$ 0.493 $&\nodata\nl
 $21.900 $&\nodata&$ 0.347 $&$ 0.457 $&\nodata\nl
 $22.100 $&\nodata&$ 0.332 $&$ 0.419 $&\nodata\nl
\enddata
\end{deluxetable}

\begin{deluxetable}{ccccc}
\tablewidth{0pt}
\tablecaption{Fiducial Sequences for M15 (NGC~7078)\label{tab:m15}}
\tablehead{
  \colhead{$r$} &
  \colhead{$u - g$} &
  \colhead{$g - r$} &
  \colhead{$g - i$} &
  \colhead{$g - z$}
}
\startdata
 $  13.750 $&$ 1.776 $&$ 0.803 $&\nodata&$ 1.401 $\nl
 $  14.250 $&$ 1.645 $&$ 0.743 $&$ 1.100 $&$ 1.291 $\nl
 $  14.750 $&$ 1.531 $&$ 0.689 $&$ 1.016 $&$ 1.200 $\nl
 $  15.250 $&$ 1.442 $&$ 0.645 $&$ 0.949 $&$ 1.126 $\nl
 $  15.750 $&$ 1.363 $&$ 0.604 $&$ 0.890 $&$ 1.054 $\nl
 $  16.250 $&$ 1.294 $&$ 0.571 $&$ 0.843 $&$ 0.995 $\nl
 $  16.750 $&$ 1.236 $&$ 0.546 $&$ 0.806 $&$ 0.951 $\nl
 $  17.250 $&$ 1.188 $&$ 0.523 $&$ 0.771 $&$ 0.910 $\nl
 $  17.750 $&$ 1.148 $&$ 0.500 $&$ 0.734 $&$ 0.857 $\nl
 $  17.950 $&$ 1.137 $&$ 0.490 $&$ 0.718 $&$ 0.833 $\nl
 $  18.050 $&$ 1.123 $&$ 0.480 $&$ 0.702 $&$ 0.818 $\nl
 $  18.150 $&$ 1.092 $&$ 0.465 $&$ 0.678 $&$ 0.791 $\nl
 $  18.250 $&$ 1.073 $&$ 0.448 $&$ 0.651 $&$ 0.757 $\nl
 $  18.350 $&$ 1.063 $&$ 0.420 $&$ 0.612 $&$ 0.707 $\nl
 $  18.450 $&$ 1.044 $&$ 0.387 $&$ 0.562 $&$ 0.648 $\nl
 $  18.550 $&$ 1.025 $&$ 0.360 $&$ 0.523 $&$ 0.604 $\nl
 $  18.650 $&$ 1.015 $&$ 0.337 $&$ 0.486 $&$ 0.561 $\nl
 $  18.750 $&$ 1.016 $&$ 0.315 $&$ 0.449 $&$ 0.513 $\nl
 $  18.850 $&$ 1.016 $&$ 0.300 $&$ 0.427 $&$ 0.485 $\nl
 $  19.050 $&$ 1.003 $&$ 0.289 $&$ 0.413 $&$ 0.470 $\nl
 $  19.350 $&$ 0.987 $&$ 0.291 $&$ 0.415 $&$ 0.470 $\nl
 $  19.650 $&$ 0.973 $&$ 0.306 $&$ 0.435 $&$ 0.493 $\nl
 $  19.950 $&$ 0.960 $&$ 0.328 $&$ 0.468 $&$ 0.533 $\nl
 $  20.250 $&$ 0.941 $&$ 0.355 $&$ 0.506 $&$ 0.586 $\nl
 $  20.550 $&$ 0.916 $&$ 0.391 $&$ 0.556 $&$ 0.665 $\nl
 $  20.850 $&\nodata&$ 0.434 $&$ 0.622 $&$ 0.772 $\nl
 $  21.150 $&\nodata&$ 0.484 $&$ 0.697 $&\nodata\nl
 $  21.450 $&\nodata&$ 0.543 $&$ 0.772 $&\nodata\nl
\enddata
\end{deluxetable}

\begin{deluxetable}{ccccc}
\tablewidth{0pt}
\tablecaption{Fiducial Sequences for M2 (NGC~7089)\label{tab:m2}}
\tablehead{
  \colhead{$r$} &
  \colhead{$u - g$} &
  \colhead{$g - r$} &
  \colhead{$g - i$} &
  \colhead{$g - z$}
}
\startdata
 $ 14.250 $&$ 2.016 $&$ 0.771 $&$ 1.123 $&$ 1.285 $\nl
 $ 14.750 $&$ 1.839 $&$ 0.707 $&$ 1.023 $&$ 1.168 $\nl
 $ 15.250 $&$ 1.695 $&$ 0.650 $&$ 0.938 $&$ 1.069 $\nl
 $ 15.750 $&$ 1.587 $&$ 0.609 $&$ 0.878 $&$ 1.003 $\nl
 $ 16.250 $&$ 1.499 $&$ 0.573 $&$ 0.827 $&$ 0.944 $\nl
 $ 16.750 $&$ 1.418 $&$ 0.540 $&$ 0.779 $&$ 0.887 $\nl
 $ 17.250 $&$ 1.350 $&$ 0.515 $&$ 0.739 $&$ 0.840 $\nl
 $ 17.750 $&$ 1.294 $&$ 0.492 $&$ 0.703 $&$ 0.796 $\nl
 $ 18.250 $&$ 1.260 $&$ 0.472 $&$ 0.667 $&$ 0.753 $\nl
 $ 18.250 $&$ 1.260 $&$ 0.472 $&$ 0.667 $&$ 0.753 $\nl
 $ 18.350 $&$ 1.243 $&$ 0.460 $&$ 0.652 $&$ 0.737 $\nl
 $ 18.450 $&$ 1.208 $&$ 0.444 $&$ 0.631 $&$ 0.709 $\nl
 $ 18.550 $&$ 1.170 $&$ 0.426 $&$ 0.605 $&$ 0.672 $\nl
 $ 18.650 $&$ 1.125 $&$ 0.398 $&$ 0.563 $&$ 0.625 $\nl
 $ 18.750 $&$ 1.084 $&$ 0.361 $&$ 0.512 $&$ 0.566 $\nl
 $ 18.850 $&$ 1.050 $&$ 0.327 $&$ 0.466 $&$ 0.507 $\nl
 $ 18.950 $&$ 1.028 $&$ 0.302 $&$ 0.429 $&$ 0.454 $\nl
 $ 19.050 $&$ 1.021 $&$ 0.287 $&$ 0.401 $&$ 0.418 $\nl
 $ 19.150 $&$ 1.020 $&$ 0.280 $&$ 0.385 $&$ 0.405 $\nl
 $ 19.350 $&$ 1.011 $&$ 0.273 $&$ 0.374 $&$ 0.397 $\nl
 $ 19.650 $&$ 0.999 $&$ 0.276 $&$ 0.381 $&$ 0.400 $\nl
 $ 19.950 $&$ 0.997 $&$ 0.293 $&$ 0.405 $&$ 0.431 $\nl
 $ 20.250 $&$ 0.991 $&$ 0.318 $&$ 0.443 $&$ 0.481 $\nl
 $ 20.550 $&$ 0.975 $&$ 0.350 $&$ 0.492 $&$ 0.545 $\nl
 $ 20.850 $&$ 0.957 $&$ 0.389 $&$ 0.550 $&$ 0.631 $\nl
 $ 21.150 $&\nodata&$ 0.437 $&$ 0.617 $&$ 0.738 $\nl
 $ 21.450 $&\nodata&$ 0.493 $&$ 0.696 $&\nodata\nl
 $ 21.750 $&\nodata&$ 0.546 $&$ 0.790 $&\nodata\nl
\enddata
\end{deluxetable}

\subsection{Preliminary Test of Theoretical Models}

\citet{girardi:04} provided the first extensive sets of theoretical isochrones
in the $ugriz$ system.  They derived magnitudes in $ugriz$ using ATLAS9 synthetic
spectra \citep{castelli:97,bessell:98} for most regions of $T_{\rm eff}$ and
$\log{g}$ space.  Here we test the models constructed from the \citet{girardi:00}
evolutionary tracks with our fiducial sequences.  Detailed comparisons to theoretical
models will be presented in a companion paper (D.\ An et al.\ 2008, in preparation).

Isochrones in \citet{girardi:04} were constructed for a perfect AB
magnitude system, in which magnitudes can be translated directly into
physical flux units.  However, it is known that the SDSS photometry
slightly deviates from a true AB system \citep{dr2}.  To compare our
fiducial sequences to the isochrones, we adjusted model magnitudes
using AB corrections given by \citet{eisenstein:06}: $u_{\rm AB} = u - 0.040$,
$i_{\rm AB} = i + 0.015$, and $z_{\rm AB} = z + 0.030$, with no
corrections in $g$ and $r$ (see also Holberg \& Bergeron 2006).

We restricted our comparisons to five globular clusters (M3, M5, M13,
M15, and M92) and one solar-metallicity open cluster (M67).  These
clusters not only have well-defined sequences but also have relatively
well-studied distances and reddening estimates, which are necessary to
infer the absolute magnitudes of stars.  Furthermore, the
metallicities of these clusters are well-studied, so the model colors
can be tested more accurately for a given metallicity.

For globular clusters, we adopted the MS-fitting distances given by
\citet[][see Table~1]{kraft:03}, which are based on measurements of
{\it Hipparcos} subdwarfs.  We also adopted reddening values from
\citet{kraft:03}.  For M67 we adopted cluster distance
and reddening estimates from \citet{an:07b}.  The isochrone colors and
magnitudes were corrected for the assumed reddening using theoretical
computations of extinction coefficients ($A_{\lambda}/A_V$)
in \citet{girardi:04}.  Specifically, we used their model flux calculation
for dwarfs ($T_{\rm eff} = 5000$, $\log{g} = 4.50$, [m/H] = 0):
$A_u/A_V = 1.574$, $A_g/A_V = 1.189$, $A_r/A_V = 0.877$, $A_i/A_V = 0.673$,
and $A_z/A_V = 0.489$ where $V$ represents the Johnson $V$ filter.  The
differences in the extinction coefficients for different $T_{\rm eff}$,
$\log{g}$, and [m/H] are negligible for our cluster sample with
$E(B - V) \leq 0.10$.  We assumed $R_V \equiv A_V/E(B - V) = 3.1$ and
derived extinction and color-excess values in $ugriz$ for a given $E(B - V)$.

\begin{figure}
\epsscale{1.15}
\plotone{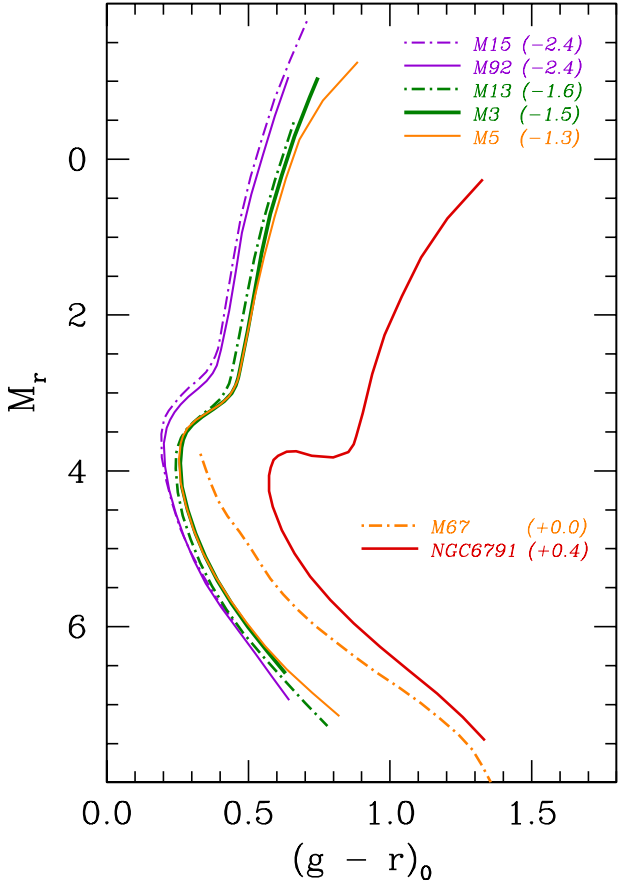}
\caption{Composite CMDs in $(g - r, M_r)_0$ for five globular clusters
(M3, M5, M13, M15, and M92) and two open clusters (M67 and NGC~6791).
Their metal abundances are shown in parenthesis (see text for details).
\label{fig:all.cmd}}
\end{figure}

Figure~\ref{fig:all.cmd} shows fiducial sequences for these clusters
on the absolute magnitude $M_r$ versus intrinsic color $(g - r)_0$
space, with the above adopted distances and reddening values.
We included a fiducial sequence for NGC~6791 in the plot, using its
parameters in Table~\ref{tab:propoc}.  These clusters cover a wide
range of metal abundances ($-2.4 < {\rm [Fe/H]} < +0.4$); their fiducial
sequences become redder at higher metallicities.  In the figure,
two groups of globular cluster sequences are distinguished by two
different colors according to their metal abundances in \citet{kraft:03}:
M15 and M92 (${\rm [Fe/H]} \approx -2.4$; {\it violet}), M3 and M13
(${\rm [Fe/H]} \approx -1.6$; {\it green}).  Cluster sequences in each
group of the clusters show $\la2\%$ agreement in color.  We note that
these differences are within the expected size of the errors from the
adopted distance, reddening, and photometric zero points
(\S~\ref{sec:zeropoint}).

\begin{figure*}
\epsscale{1.0}
\plotone{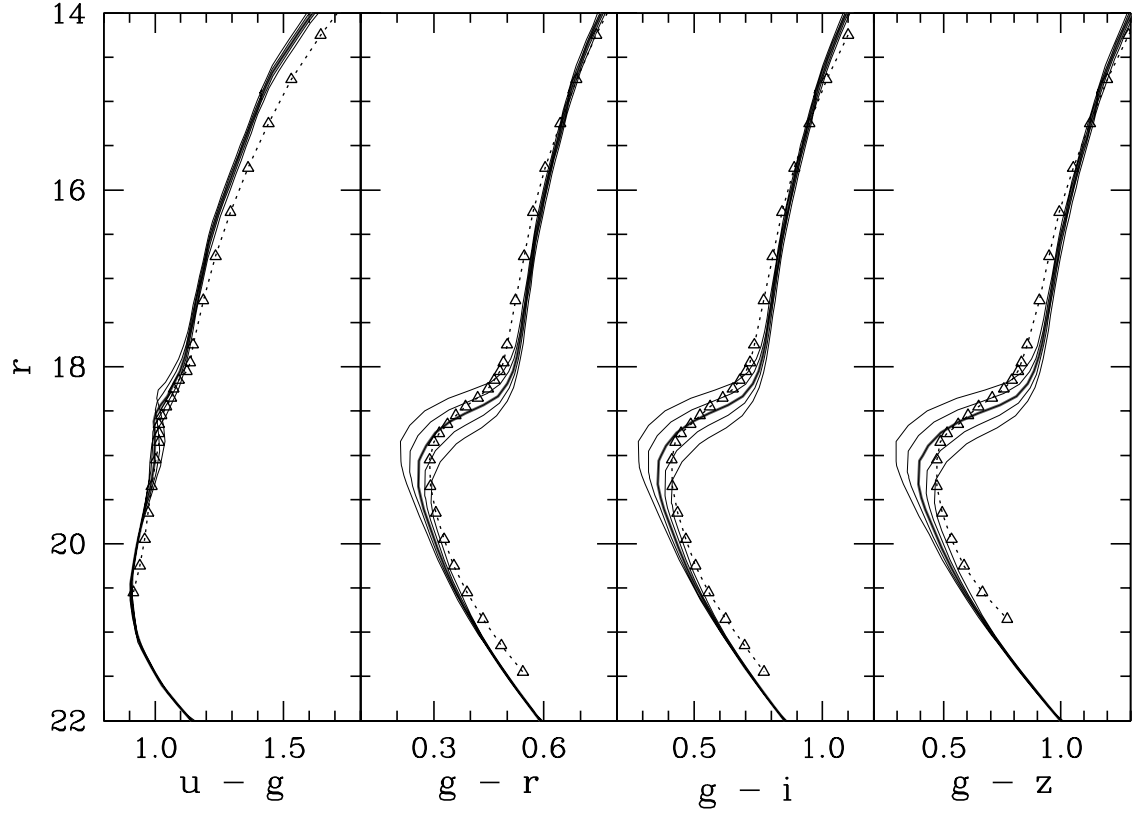}
\caption{Comparisons between fiducial sequences for M15 in this paper
({\it dotted line with triangles}) and the Girardi et~al.\ theoretical
isochrones ({\it solid lines}).  Models are shown with
$Z = 0.0001$ (${\rm [m/H]} \approx -2.3$)
at ages of 10.0, 11.2, 12.6, 14.1, and 15.9~Gyrs.\label{fig:cmdm15.pv}}
\end{figure*}

\begin{figure*}
\epsscale{1.0}
\plotone{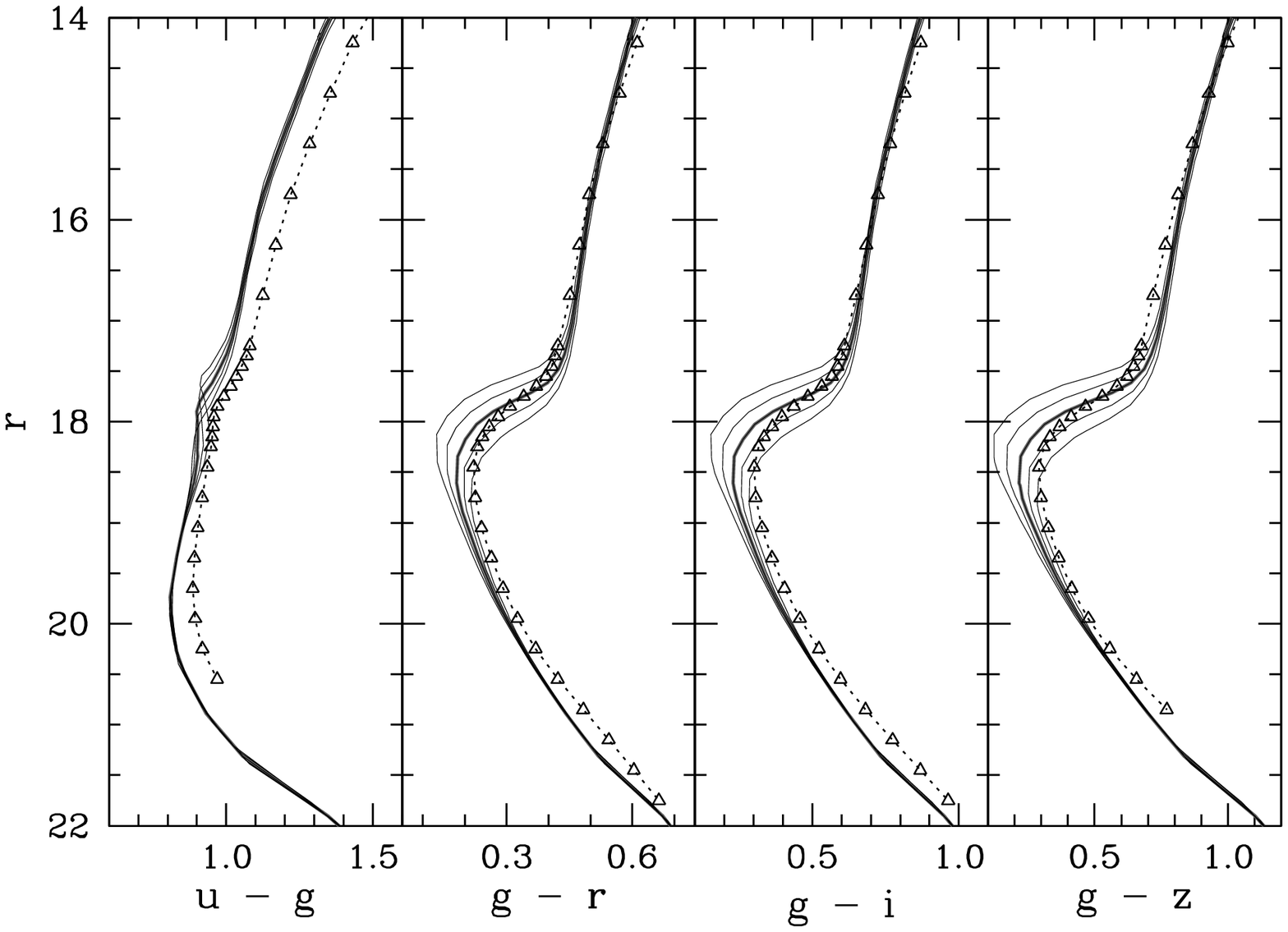}
\caption{Same as in Fig.~\ref{fig:cmdm15.pv}, but for M92.
\label{fig:cmdm92.pv}}
\end{figure*}

\begin{figure*}
\epsscale{1.0}
\plotone{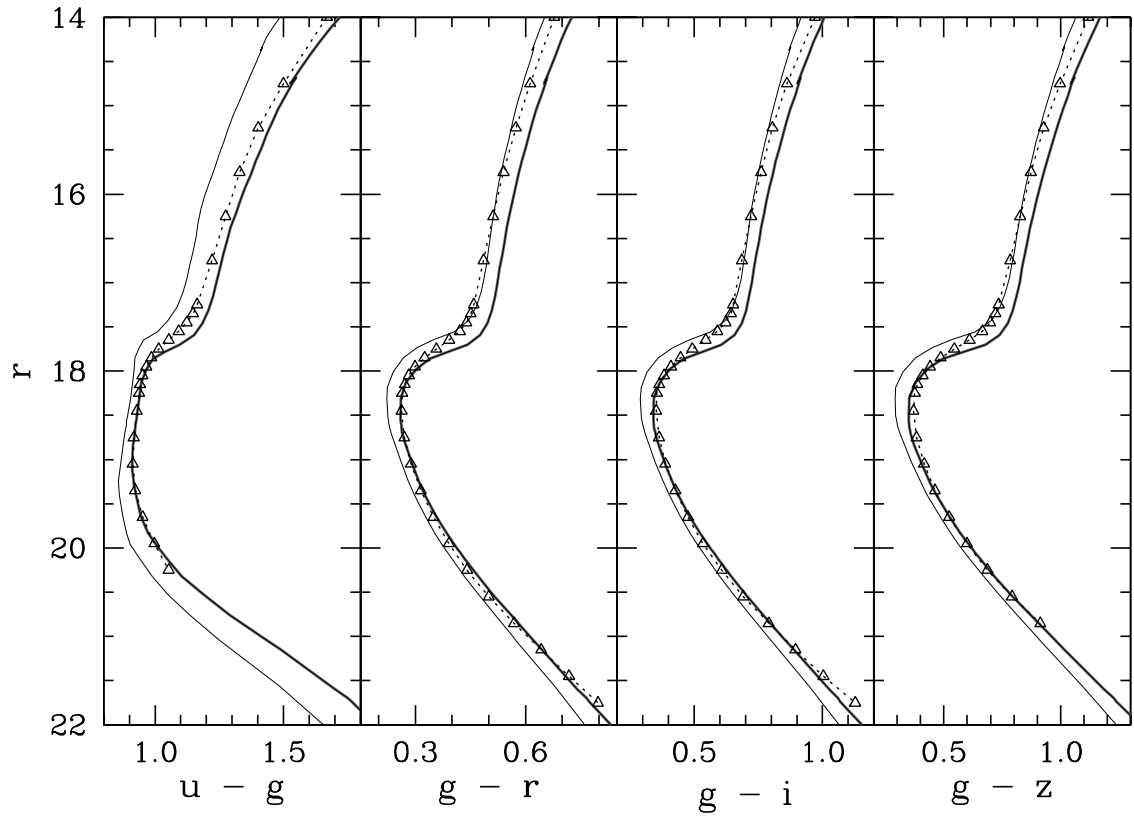}
\caption{Same as in Fig.~\ref{fig:cmdm15.pv}, but for M13.  Models are
shown for $Z = 0.0004$ (${\rm [m/H]} \approx -1.7$) and $Z = 0.0010$
(${\rm [m/H]} \approx -1.3$) at an age of 12.6~Gyr.
\label{fig:cmdm13.pv}}
\end{figure*}

\begin{figure*}
\epsscale{1.0}
\plotone{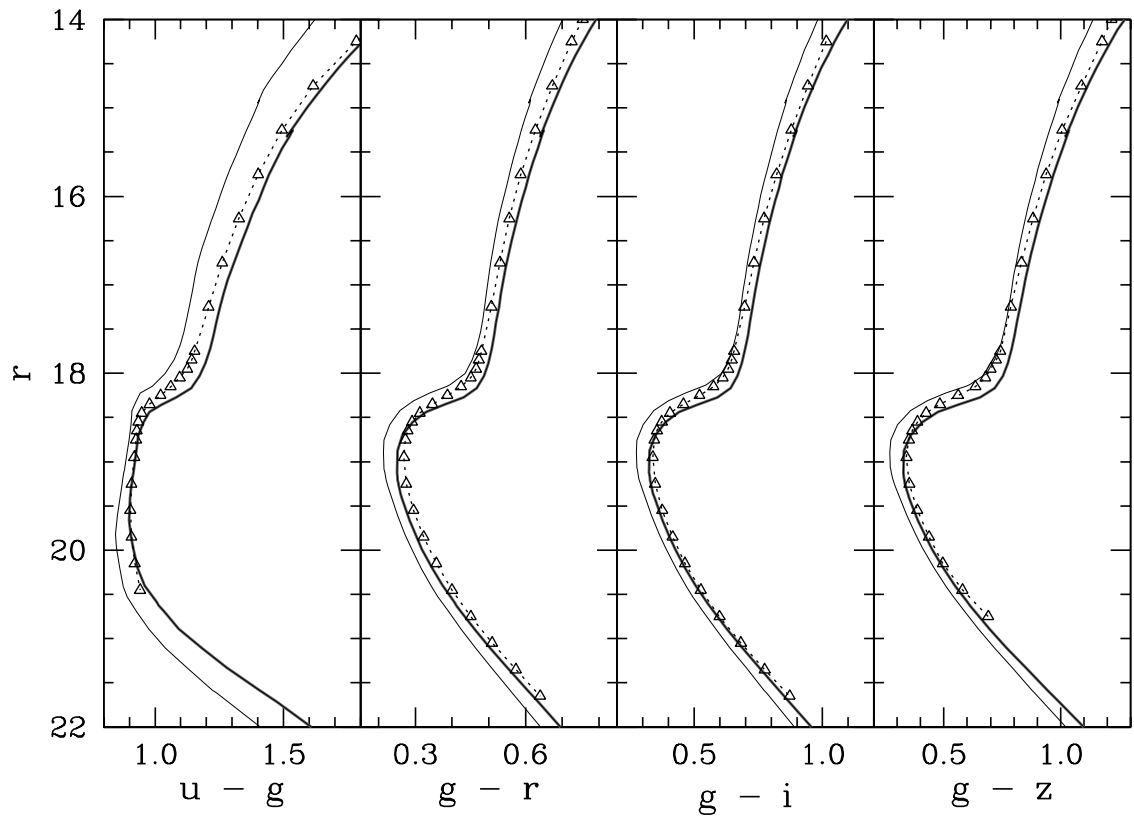}
\caption{Same as in Fig.~\ref{fig:cmdm13.pv}, but for M3.
\label{fig:cmdm3.pv}}
\end{figure*}

\begin{figure*}
\epsscale{1.0}
\plotone{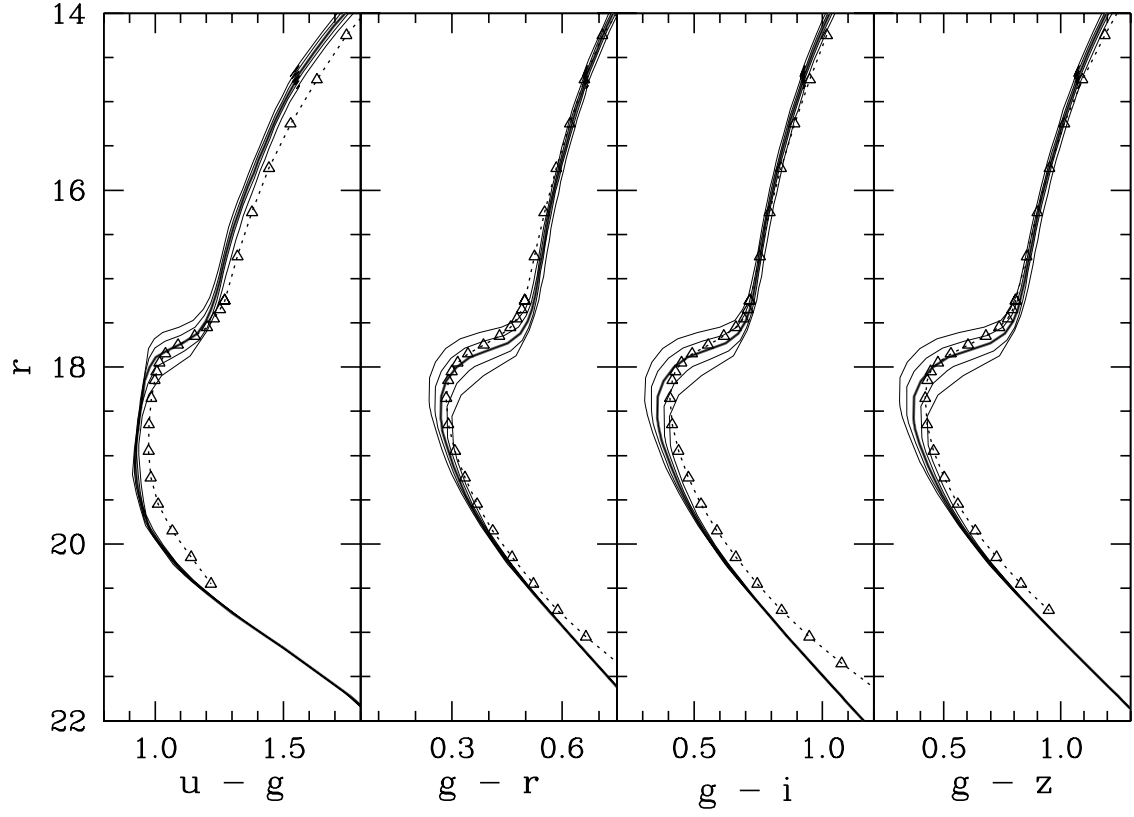}
\caption{Same as in Fig.~\ref{fig:cmdm15.pv}, but for M5.  Models are
shown for $Z = 0.0010$ (${\rm [m/H]} \approx -1.3$) at ages of 10.0,
11.2, 12.6, 14.1, and 15.9~Gyrs.\label{fig:cmdm5.pv}}
\end{figure*}

\begin{figure*}
\epsscale{1.0}
\plotone{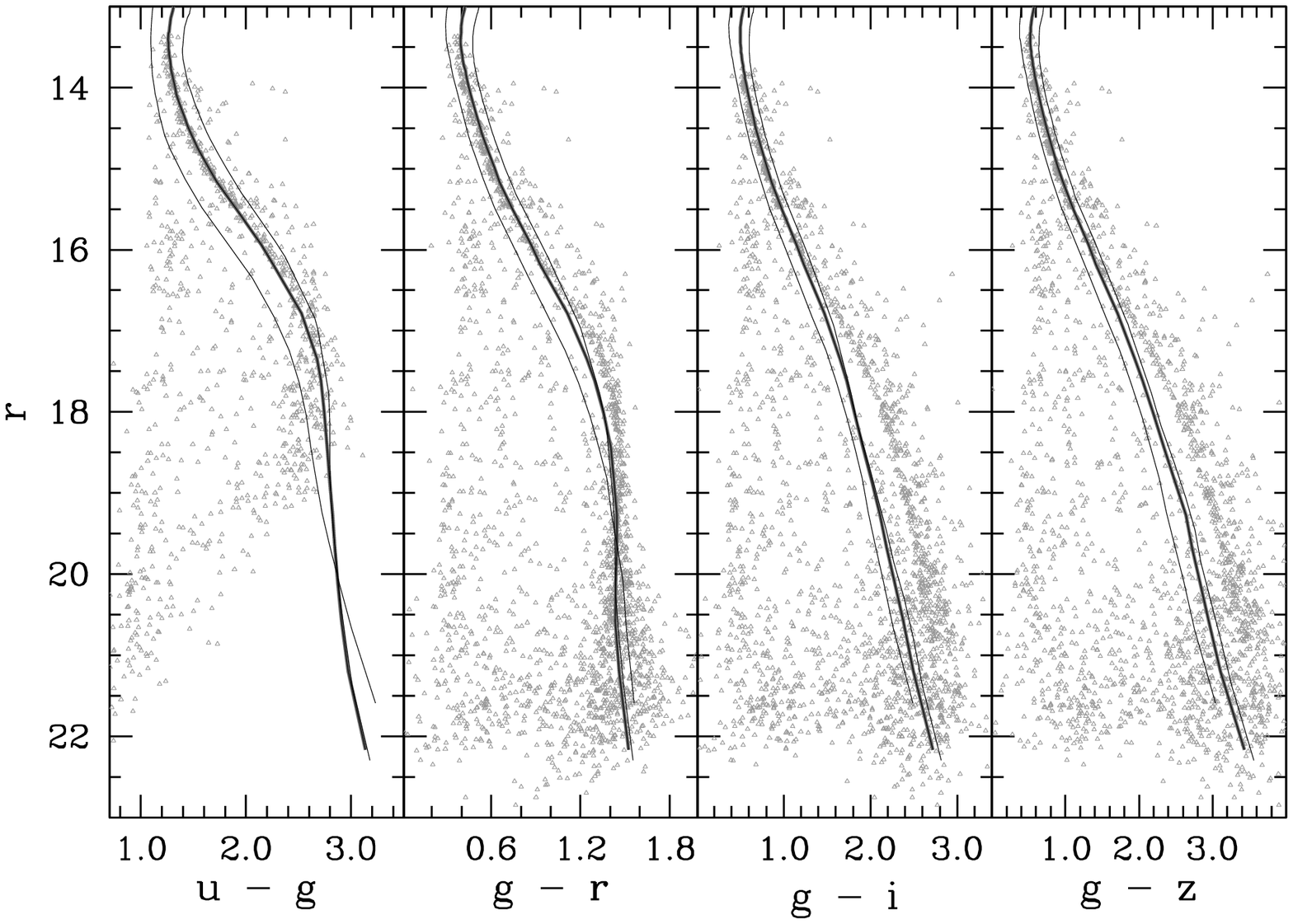}
\caption{Same as in Fig.~\ref{fig:cmdm15.pv}, but for M67.  Models are
shown for $Z = 0.0080$, $0.0190$, and $0.0300$
(${\rm [m/H]} \approx -0.4, +0.0, +0.2$) at an age of 3.5~Gyr.
\label{fig:cmdm67.pv}}
\end{figure*}

Figures~\ref{fig:cmdm15.pv} and \ref{fig:cmdm92.pv} show comparisons
between fiducial sequences for the two most metal-poor globular clusters
in our sample, M15 and M92 ({\it dotted line}), and the \citet{girardi:04}
theoretical isochrones ({\it solid line}).  Models are shown
with the heavy-element content $Z = 0.0001$ (${\rm [m/H]} \approx -2.3$)
at ages of 10.0, 11.2, 12.6, 14.1, and 15.9~Gyrs.
Comparisons in $(g - r, r)$, $(g - i, r)$, and $(g - z, r)$ show that
model colors are $\sim0.02$--$0.05$~mag bluer and redder than the fiducial
sequences for MS and RGB, respectively.  In addition, the morphology of
the model SGB does not perfectly match the observed ones.  A significant
color offset is found in $(u - g, r)$, up to as large as $\sim0.1$~mag.
While the SDSS $u$-band filters are known to have a red leak
\citep{edr}, it is probably not the reason for the discrepancy found for
stars bluer than $g - r \sim 1.2$.  We note again that our photometry
does not reach to the tips of the RGBs in all clusters because of
saturation, and that this work does not constrain the reddest part of
the RGBs for the nearest clusters.

Figures~\ref{fig:cmdm13.pv} and \ref{fig:cmdm3.pv} show comparisons
for the intermediate metallicity globular clusters M13 and M3,
respectively.  The models are shown for two different metallicities,
$Z = 0.0004$ (${\rm [m/H]} \approx -1.7$) and $Z = 0.0010$
(${\rm [m/H]} \approx -1.3$), to bracket the observed cluster abundances.
It is noted that these models assume the scaled-solar abundance
ratios, while metal-poor stars show $\alpha$-enhanced abundances
\citep[e.g.,][and references therein]{sneden:04,venn:04}.  However,
the effects of $\alpha$-enhancement can be mimicked by increasing the
total metal abundance in this low metallicity range
\citep{salaris:93,kim:02,cassisi:04}.
Nevertheless, neither isochrones simultaneously match the colors of
both the MS and RGB sequences with high precision.  We found a similar
result for M5 (${\rm [Fe/H]} = -1.26$), as shown in Figure~\ref{fig:cmdm5.pv}.

Here we neglected the effects of unresolved binaries.  \citet{an:07b}
performed extensive simulations of unresolved binaries in clusters and their
influence in the MS-fitting distances.  After the same photometric filtering as we
applied in this paper, they found that the unresolved binaries can make the MS
look brighter by $\sim0.007$~mag for a $40\%$ binary fraction\footnote{Binary
fraction is defined as the number of binaries divided by the total number of
systems.} because all of the low mass-ratio binaries cannot be detected in
the photometric filtering.  However, this tranlates into only $\sim0.001$~mag
in colors since the slope of the MS is about 5-6.  Furthermore,
the observed binary fraction of globular clusters is typcially less than
$20\%$ \citep[e.g.,][and references therein]{sollima:07,davis:08}, which makes
the influence of unresolved binaries even smaller.

Figure~\ref{fig:cmdm67.pv} shows the comparison between the
solar-metallicity open cluster M67 CMDs and 3.5~Gyr models at three
different metallicities, $Z = 0.0080$, $0.0190$, and $0.0300$
(${\rm [m/H]} \approx -0.4, +0.0, +0.2$).  We used models without
convective core overshooting, but the difference from those models
based on the overshooting assumption is small in most parts of the CMDs
in Figure~\ref{fig:cmdm67.pv}.  Near the MS turnoff, the agreement is
good between solar-metallicity models and the data.
However, the models begin to diverge from
the fiducial MS below $r \sim 16.5$~mag or $\sim 0.7 M_\odot$ in their models.
The difference becomes as large as $\sim0.5$~mag in $(g - i, r)$
and $(g - z, r)$ at the bottom of the MS.  Users of these models
should be warned about this potentially large discrepancy.

\subsection{Comparison with Fiducial Sequences in $u'g'r'i'z'$}
\label{sec:prime}

\citet{clem:08} observed four globular clusters (M3, M13, M71, M92) and
one open cluster (NGC~6791) in the $u'g'r'i'z'$ passbands with the
MegaCam wide-field imager on the Canada-France-Hawaii Telescope.  Their
data included observations of various integration times, which resulted
in highly precise CMDs extending from the tip of the RGB down to
approximately four magnitudes below the MS turnoff.

The photometry in Clem et al.\ has been calibrated to the $u'g'r'i'z'$
system defined by the \citet{smith:02} sample of standard stars, while the
SDSS photometry is on the natural $ugriz$ system of the 2.5-m survey
telescope.  Therefore, we converted their fiducial sequences in the
$u'g'r'i'z'$ system onto the SDSS 2.5-m $ugriz$ system, using the
transformation equations in \citet{tucker:06}: $u = u'$,
$g = g' + 0.060 [(g' - r') - 0.53]$, $r = r' + 0.035 [(r' - i') - 0.21]$,
$i = i' + 0.041 [(r' - i') - 0.21]$, $z = z' - 0.030 [(i' - z') - 0.09]$.
These relations were derived using stars in $0.70 \leq (u' - g') \leq 2.70$,
$0.15 \leq (g' - r') \leq 1.20$, $−0.10 \leq (r' - i') \leq 0.60$,
and $−0.20 \leq (i' - z') \leq 0.40$.

\begin{figure*}
\epsscale{1.0}
\plotone{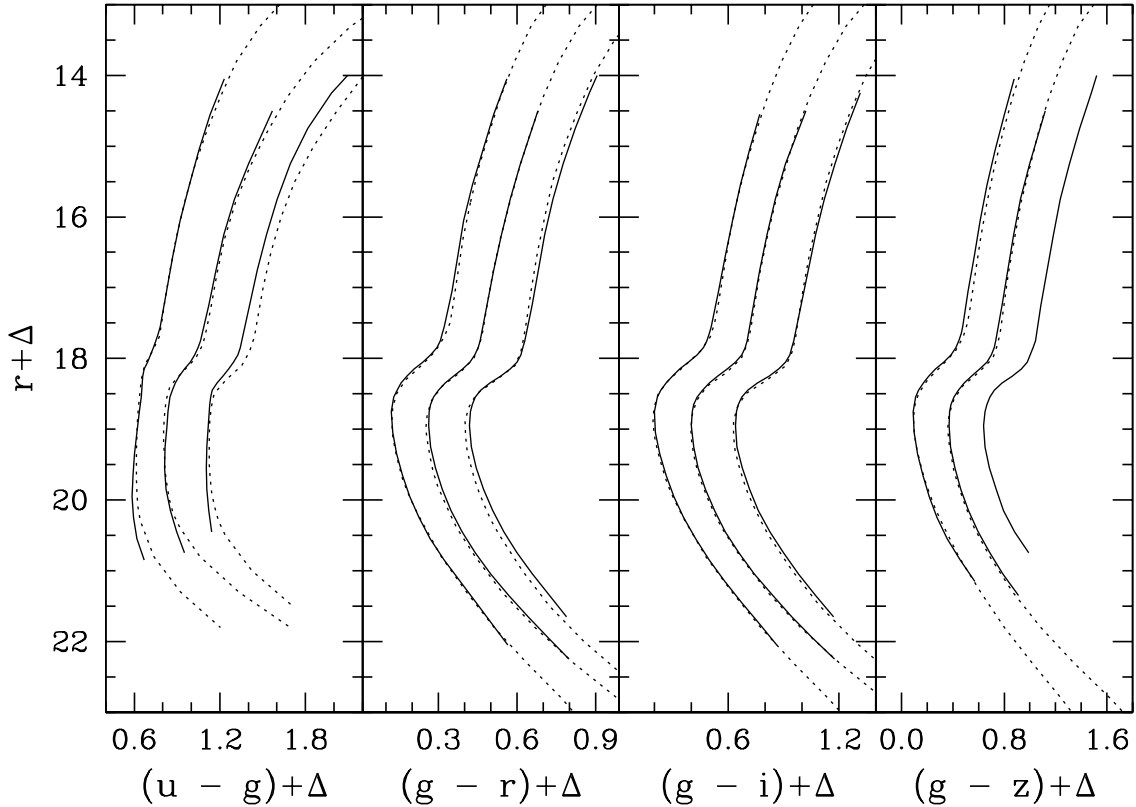}
\caption{Comparisons between fiducial sequences in this paper
({\it solid line}) and those in Clem et~al. ({\it dotted line}), after
transforming the latter from the $u'g'r'i'z'$ to the $ugriz$ system.  Fiducial
sequences are shown for M92, M13, and M3 (from {\it left} to {\it right}) with
arbitrary offsets in colors and magnitudes for clarity.  The comparison in
$g - z$ for M3 is not shown because the sequence is not available in Clem et
al.\ for that color index.\label{fig:comp.clem}}
\end{figure*}

\begin{figure*}
\epsscale{1.0}
\plotone{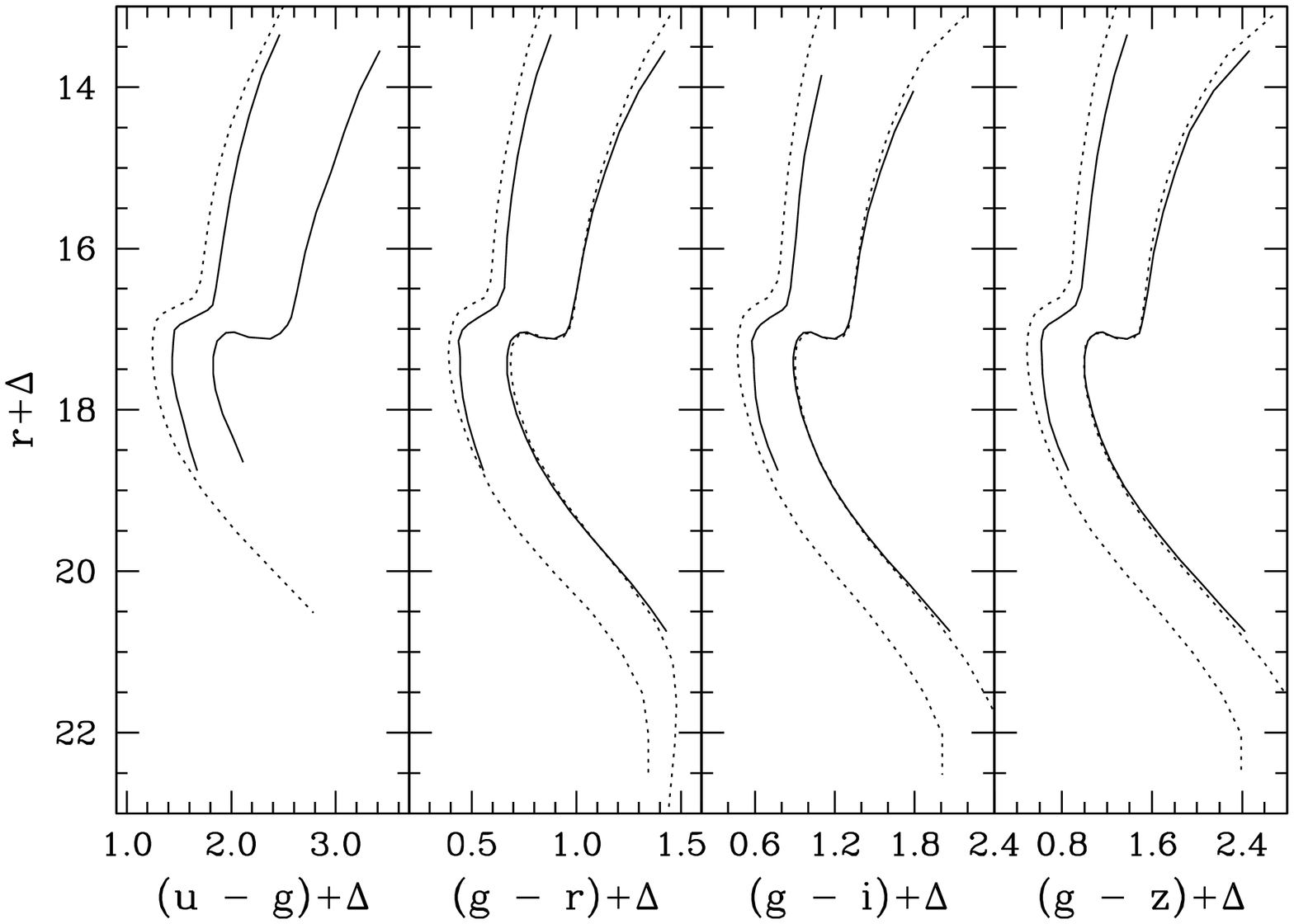}
\caption{Same as in Fig.~\ref{fig:comp.clem}, but for M71 ({\it left})
and NGC~6791 ({\it right}).  See text for a discussion on the M71
comparisons.  The comparison in $u - g$ for NGC~6791 is not shown because
the sequence is not available in Clem et al.\ for that color index.
\label{fig:comp.clem.2}}
\end{figure*}

Figures~\ref{fig:comp.clem} and \ref{fig:comp.clem.2} show comparisons
between our fiducial sequences and those in Clem et al.\ on the $ugriz$
system.  For clarity, comparisons are shown with arbitrary offsets in colors
and magnitudes for each cluster.  Note that the sequences in Clem et al.\
extend far beyond the magnitude limits of our fiducial sequences.
The comparison in $g - z$ for M3 is not shown in Figure~\ref{fig:comp.clem}
because the sequence is not available in Clem et al.\ for that color
index.  For the same reason, the comparison in $u - g$ for NGC~6791 is
not shown in Figure~\ref{fig:comp.clem.2}.

While a generally good agreement is found between the two sets of
fiducial sequences, the comparisons for M71 show particularly large
differences in all four of the color indices.  The differences in colors
are $\sim0.05$-$0.15$~mag, in the sense that our fiducial sequences are
always redder than those in Clem et al.  As we noted in
\S~\ref{sec:zeropoint}, the zero points for the M71 photometry were very
uncertain, due to the suspicious {\it Photo} magnitudes in the cluster's
flanking field.  In addition, the shapes of the fiducial sequences could
not be accurately defined due to the strong contamination from background
stars.  Caution should be given when using our DAOPHOT photometry for M71
and its fiducial sequences.

Except for M71, the differences in colors and/or magnitudes between the
two fiducial sequences are typically less than $\sim2\%$.  These differences
are smaller than those found from the comparison with theoretical
isochrones (Figs.~\ref{fig:cmdm15.pv}-\ref{fig:cmdm67.pv}).  Furthermore,
they are comparable in size to the zero-point errors in the DAOPHOT
photometry (\S~\ref{sec:error}).  Therefore, the agreement found here
not only validates the accuracy of the transformation equations between
$u'g'r'i'z'$ and $ugriz$, but also the accuracy of our fiducial
sequences derived from the single-epoch photometry.

In the case of NGC~6791, our fiducial sequences on RGB become redder
than the Clem et al.\ sequences at redder colors.  The differences at
the tip of our fiducial sequences are $\sim0.05$--$0.10$~mag.  Although different
filter responses can cause these color-dependent zero-point shifts, the
observed differences are possibly due to uncertainties in the fiducial
sequences from the sparsely populated RGB of the cluster.

\section{Conclusion}

We used the DAOPHOT/ALLFRAME suite of programs to derive photometry in
$ugriz$ filter bandpasses for 17 globular clusters and 3 open clusters
that have been observed with SDSS.  The regions close to the globular
clusters are too crowded for the standard SDSS photometric pipeline
({\it Photo}) to process, and the photometry is not available for the
most crowded regions of these clusters.   In order to exploit over 100
million stellar objects with $r < 22.5$~mag observed by SDSS, we
used the DAOPHOT crowded field photometry package to derive accurate
magnitudes and colors of stars in the Galactic clusters.  We also
derived fiducial sequences for the 20 clusters on the native SDSS
2.5-meter $ugriz$ photometric system, which can be directly applied to
the SDSS photometry without relying upon transformations from
the $u'g'r'i'z'$ system.

We showed that DAOPHOT PSF magnitudes are spatially and temporally
uniform to $\la0.5\%$ with respect to aperture photometry.  However,
comparison between the DAOPHOT and the {\it Photo} magnitudes showed
$\sim2\%$ high spatial frequency structures on a sub-field scale,
indicating an error in the {\it Photo} magnitudes.  Although the $2\%$
accuracy of {\it Photo} magnitudes already makes SDSS one of the most
successful optical surveys, our result indicates that its photometric
accuracy could be further improved in the future
\citep[e.g.,][]{ivezic:07,padmanabhan:08}.  Nevertheless, the accuracy
of the zero point in the DAOPHOT photometry is predominantly limited by
the $\sim2\%$ run-to-run zero-point variations.

From repeated flux measurements in overlapping strips/runs, we also
measured realistic photometric errors for SDSS photometry determined
by DAOPHOT.  The error distributions at the bright ends indicate errors
of $\sim1\%$ in $griz$ and $\sim2\%$ in the $u$ band, which are a factor
of two better than the $2\%$ rms photometric precision obtained with
{\it Photo} \citep{ivezic:03}.  We found slightly larger rms differences
($\sim0.025$~mag) between the {\it Photo} and the DAOPHOT magnitudes in
semi-crowded open cluster fields.

Using fiducial sequences, we performed a preliminary test of theoretical
isochrones from \citet{girardi:04}.  We found that model colors differ by
$\sim0.02$--$0.05$~mag from those of the fiducial sequences for our adopted cluster
distance and reddening values.  Furthermore, these models cannot be
simultaneously matched to the MS and RGB ridgelines of our fiducial
sequences.  In the solar-metallicity open cluster M67, model colors are
too blue by $\sim0.5$~mag at the bottom of the MS.  On the other hand, we
found a good agreement ($\la0.02$~mag in colors) with the \citet{clem:08}
empirical fiducial sequences in $u'g'r'i'z'$, after transformation to
the native $ugriz$ system using the transformation equations of
\citet{tucker:06}.  This result not only validates the accuracy of the
transformation equations between $u'g'r'i'z'$ and $ugriz$, but also the
accuracy of our fiducial sequences derived from the single-epoch photometry.

There are several projects that will benefit from our accurate cluster
photometry and fiducial sequences in $ugriz$.  The photometry is of
great value for empirical calibrations of the spectroscopic measurements
such as the SEGUE Stellar Parameter Pipeline \citep{sspp1,sspp2,sspp3},
and for deriving accurate transformations between $ugriz$ and other
photometric systems.  As templates for stellar populations, fiducial
sequences can be used to identify and characterize the dwarf companions to
the Milky Way and Andromeda galaxies.  They can be also used for tracing
the tidal structures from globular clusters \citep[e.g.,][]{odenkirchen:01}.
In addition, the distances to individual stars in SDSS can be better
determined with $ugriz$ fiducials of well-studied clusters, which is
the subject of the next paper in this series.

\acknowledgements

D.A.\ and J.A.J.\ thank Donald Terndrup, Marc Pinsonneault, and Andrew Gould
for helpful comments.  D.A.\ and J.A.J.\ acknowledge support from SSP-271.
H.L.M.\ acknowledges support from NSF grants AST-0098435 and AST-0607518.

Funding for the SDSS and SDSS-II has been provided by the Alfred P.\ Sloan
Foundation, the Participating Institutions, the National Science Foundation,
the U.S.\ Department of Energy, the National Aeronautics and Space Administration,
the Japanese Monbukagakusho, the Max Planck Society, and the Higher Education
Funding Council for England. The SDSS Web Site is http://www.sdss.org/.

The SDSS is managed by the Astrophysical Research Consortium for the Participating
Institutions. The Participating Institutions are the American Museum of Natural
History, Astrophysical Institute Potsdam, University of Basel, University of
Cambridge, Case Western Reserve University, University of Chicago, Drexel
University, Fermilab, the Institute for Advanced Study, the Japan Participation
Group, Johns Hopkins University, the Joint Institute for Nuclear Astrophysics,
the Kavli Institute for Particle Astrophysics and Cosmology, the Korean Scientist
Group, the Chinese Academy of Sciences (LAMOST), Los Alamos National Laboratory,
the Max-Planck-Institute for Astronomy (MPIA), the Max-Planck-Institute for
Astrophysics (MPA), New Mexico State University, Ohio State University, University
of Pittsburgh, University of Portsmouth, Princeton University, the United States
Naval Observatory, and the University of Washington.


\begin{thebibliography}

\bibitem[Abazajian et al.(2003)]{dr1}
Abazajian, K., et al.\ 2003, \aj, 126, 2081

\bibitem[Abazajian et al.(2004)]{dr2}
Abazajian, K., et al.\ 2004, \aj, 128, 502

\bibitem[Abazajian et al.(2005)]{dr3}
Abazajian, K., et al.\ 2005, \aj, 129, 1755

\bibitem[Adelman-McCarthy et al.(2006)]{dr4} 
Adelman-McCarthy, J.~K., et al.\ 2006, \apjs, 162, 38

\bibitem[Adelman-McCarthy et al.(2007)]{dr5}
Adelman-McCarthy, J.~K., et al.\ 2007, \apjs, 172, 634

\bibitem[Adelman-McCarthy et al.(2008)]{dr6} 
Adelman-McCarthy, J.~K., et al.\ 2008, \apjs, 175, 297

\bibitem[Allende Prieto et al.(2006)]{allendeprieto:06}
Allende Prieto, C., Beers, T.~C., Wilhelm, R., Newberg, H.~J.,
Rockosi, C.~M., Yanny, B., \& Lee, Y.~S.\ 2006, \apj, 636, 804

\bibitem[Allende Prieto et al.(2007)]{sspp3}
Allende Prieto, C., et al.\ 2007, ArXiv e-prints, 710, arXiv:0710.5780

\bibitem[An et al.(2007a)]{an:07a}
An, D., Terndrup, D.~M., \& Pinsonneault, M.~H.\ 2007a, \apj, 671, 1640

\bibitem [An et al.(2007b)]{an:07b}
An, D., Terndrup, D.~M., Pinsonneault, M.~H., Paulson, D.~B.,
Hanson, R.~B., \& Stauffer, J.~R.\ 2007b, \apj, 655, 233

\bibitem[Anthony-Twarog et al.(2006)]{anthony-twarog:06}
Anthony-Twarog, B.~J., Tanner, D., Cracraft, M., \&
Twarog, B.~A.\ 2006, \aj, 131, 461 

\bibitem[Belokurov et al.(2006)]{belokurov:06}
Belokurov, V., et al.\ 2006, \apjl, 642, L137 

\bibitem[Bessell et al.(1998)]{bessell:98}
Bessell, M.~S., Castelli, F., \& Plez, B.\ 1998, \aap, 333, 231

\bibitem[Carraro et al.(2006)]{carraro:06}
Carraro, G., Villanova, S., Demarque, P., McSwain, M.~V.,
Piotto, G., \& Bedin, L.~R.\ 2006, \apj, 643, 1151

\bibitem[Cassisi et al.(2004)]{cassisi:04}
Cassisi, S., Salaris, M., Castelli, F.,
\& Pietrinferni, A.\ 2004, \apj, 616, 498

\bibitem[Castelli et al.(1997)]{castelli:97}
Castelli, F., Gratton, R.~G., \& Kurucz, R.~L.\ 1997, \aap, 318, 841

\bibitem[Clem et al.(2008)]{clem:08}
Clem, J.~L., Vanden Berg, D.~A., \& Stetson, P.~B.\ 2008, \aj, 135, 682

\bibitem[Coleman et al.(2007)]{coleman:07}
Coleman, M.~G., Jordi, K., Rix, H.-W., Grebel, E.~K.,
\& Koch, A.\ 2007, \aj, 134, 1938

\bibitem[Davenport et al.(2007)]{davenport:07}
Davenport, J.~R.~A., Bochanski, J.~J., Covey, K.~R., Hawley, S.~L.,
West, A.~A., \& Schneider, D.~P.\ 2007, \aj, 134, 2430

\bibitem[Davis et al.(2008)]{davis:08}
Davis, D.~S., Richer, H.~B., Anderson, J., Brewer, J., Hurley, J.,
Kalirai, J.~S., Rich, R.~M., \& Stetson, P.~B.\ 2008, \aj, 135, 2155

\bibitem[de Jong et al.(2008)]{dejong:08}
de Jong, J.~T.~A., Rix, H.-W., Martin, N.~F., Zucker, D.~B.,
Dolphin, A.~E., Bell, E.~F., Belokurov, V., \& Evans, N.~W.\ 2008, \aj, 135, 1361

\bibitem[Eisenstein et al.(2006)]{eisenstein:06}
Eisenstein, D.~J., et al.\ 2006, \apjs, 167, 40

\bibitem[Fukugita et al.(1996)]{fukugita:96}
Fukugita, M., Ichikawa, T., Gunn, J.~E., Doi, M., Shimasaku, K.,
\& Schneider, D.~P.\ 1996, \aj, 111, 1748

\bibitem[Girardi et al.(2000)]{girardi:00}
Girardi, L., Bressan, A., Bertelli, G., \& Chiosi, C.\ 2000, \aaps, 141, 371

\bibitem[Girardi et al.(2004)]{girardi:04}
Girardi, L., Grebel, E.~K., Odenkirchen, M., \& Chiosi, C.\ 2004, \aap, 422, 205

\bibitem[Gratton et al.(2006)]{gratton:06}
Gratton, R., Bragaglia, A., Carretta, E., \&
Tosi, M.\ 2006, \apj, 642, 462

\bibitem[Gratton et al.(1997)]{gratton:97}
Gratton, R.~G., Fusi Pecci, F., Carretta, E., Clementini, G., Corsi, C.~E., 
\& Lattanzi, M.\ 1997, \apj, 491, 749

\bibitem[Gunn et al.(1998)]{gunn:98}
Gunn, J.~E., et al.\ 1998, \aj, 116, 3040

\bibitem[Gunn et al.(2006)]{gunn:06}
Gunn, J.~E., et al.\ 2006, \aj, 131, 2332

\bibitem[Harris(1996)]{harris:96}
Harris, W.~E.\ 1996, \aj, 112, 1487

\bibitem[Hogg et al.(2001)]{hogg:01}
Hogg, D.~W., Finkbeiner, D.~P., Schlegel, D.~J.,
\& Gunn, J.~E.\ 2001, \aj, 122, 2129

\bibitem[Holberg \& Bergeron(2006)]{holberg:06}
Holberg, J.~B., \& Bergeron, P.\ 2006, \aj, 132, 1221

\bibitem[Ivezi{\'c} et al.(2001)]{ivezic:01}
Ivezi{\'c}, {\v Z}., et al.\ 2001, \aj, 122, 2749

\bibitem[Ivezi{\'c} et al.(2003)]{ivezic:03}
Ivezi{\'c}, {\v Z}., et al.\ 2003,
Memorie della Societa Astronomica Italiana, 74, 978

\bibitem[Ivezi{\'c} et al.(2004)]{ivezic:04}
Ivezi{\'c}, {\v Z}., et al.\ 2004, Astronomische Nachrichten, 325, 583

\bibitem[Ivezi{\'c} et al.(2007)]{ivezic:07}
Ivezi{\'c}, {\v Z}., et al.\ 2007, \aj, 134, 973

\bibitem[Johnson(1957)]{johnson:57}
Johnson, H.~L.\ 1957, \apj, 126, 121

\bibitem[Juri{\'c} et al.(2008)]{juric:08}
Juri{\'c}, M., et al.\ 2008, \apj, 673, 864

\bibitem[Kaiser et al.(2002)]{kaiser:02}
Kaiser, N., et al.\ 2002, \procspie, 4836, 154

\bibitem[Kim et al.(2002)]{kim:02}
Kim, Y.-C., Demarque, P., Yi, S.~K., \& Alexander, D.~R.\ 2002,
\apjs, 143, 499

\bibitem[Kraft(1994)]{kraft:94}
Kraft, R.~P.\ 1994, \pasp, 106, 553

\bibitem[Kraft \& Ivans(2003)]{kraft:03}
Kraft, R.~P., \& Ivans, I.~I.\ 2003, \pasp, 115, 143

\bibitem[Kraft \& Ivans(2004)]{kraft:04}
Kraft, R.~P., \& Ivans, I.~I.\ 2004,
in Carnegie Observatories Astrophysics Ser.\ 4:
Origin and Evolution of the Elements, 2004,
ed. A. McWilliam \& M. Rauch (Pasadena: Carnegie Observatories),
http://www.ociw.edu/ociw/symposia/series/symposium4/proceedings.html

\bibitem[Lee et al.(2007a)]{sspp1}
Lee, Y.~S., et al.\ 2007a, ArXiv e-prints, 710, arXiv:0710.5645

\bibitem[Lee et al.(2007b)]{sspp2}
Lee, Y.~S., et al.\ 2007b, ArXiv e-prints, 710, arXiv:0710.5778

\bibitem[Lupton et al.(1999)]{lupton:99}
Lupton, R.~H., Gunn, J.~E., \& Szalay, A.~S.\ 1999, \aj, 118, 1406

\bibitem[Lupton et al.(2002)]{lupton:02}
Lupton, R.~H., Ivezic, Z., Gunn, J.~E., Knapp, G., Strauss, M.~A., 
\& Yasuda, N.\ 2002, \procspie, 4836, 350

\bibitem[Newberg et al.(2002)]{newberg:02}
Newberg, H.~J., et al.\ 2002, \apj, 569, 245

\bibitem[Odenkirchen et al.(2001)]{odenkirchen:01}
Odenkirchen, M., et al.\ 2001, \apjl, 548, L165

\bibitem[Oke \& Gunn(1983)]{oke:83}
Oke, J.~B., \& Gunn, J.~E.\ 1983, \apj, 266, 713

\bibitem[Origlia et al.(2006)]{origlia:06}
Origlia, L., Valenti, E., Rich, R.~M., \&
Ferraro, F.~R.\ 2006, \apj, 646, 499

\bibitem[Padmanabhan et al.(2008)]{padmanabhan:08}
Padmanabhan, N., et al.\ 2008, \apj, 674, 1217

\bibitem[Perryman et al.(2001)]{perryman:01}
Perryman, M.~A.~C., et al.\ 2001, \aap, 369, 339

\bibitem[Pier et al.(2003)]{pier:03}
Pier, J.~R., Munn, J.~A., Hindsley, R.~B., Hennessy, G.~S.,
Kent, S.~M., Lupton, R.~H., \& Ivezi{\'c}, {\v Z}.\ 2003, \aj, 125, 1559

\bibitem[Pinsonneault et~al.(2003)]{pinsono:03}
Pinsonneault, M. H., Terndrup, D. M., Hanson, R. B., \& Stauffer, J. R.
2003, \apj, 598, 588

\bibitem[Pinsonneault et~al.(2004)]{pinsono:04}
Pinsonneault, M. H., Terndrup, D. M., Hanson, R. B., \& Stauffer, J. R.
2004, \apj, 600, 946

\bibitem[Reid(1997)]{reid:97}
Reid, I.~N.\ 1997, \aj, 114, 161

\bibitem[Salaris et al.(1993)]{salaris:93}
Salaris, M., Chieffi, A., \& Straniero, O.\ 1993, \apj, 414, 580

\bibitem[Schechter et al.(1993)]{schechter:93}
Schechter, P.~L.,  Mateo, M., \& Saha, A.\ 1993, \pasp, 105, 1342

\bibitem[Schlegel et al.(1998)]{schlegel:98}
Schlegel, D.~J., Finkbeiner, D.~P., \& Davis, M.\ 1998, \apj, 500, 525

\bibitem[Sesar et al.(2006)]{sesar:06}
Sesar, B., et al.\ 2006, \aj, 131, 2801

\bibitem[Smith et al.(2002)]{smith:02}
Smith, J.~A., et al.\ 2002, \aj, 123, 2121

\bibitem[Smol{\v c}i{\'c} et al.(2007)]{smolcic:07}
Smol{\v c}i{\'c}, V., Zucker, D.~B., Bell, E.~F., Coleman, M.~G., Rix, H.~W.,
Schinnerer, E., Ivezi{\'c}, {\v Z}., \& Kniazev, A.\ 2007, \aj, 134, 1901

\bibitem[Sneden et al.(2004)]{sneden:04}
Sneden, C., Kraft, R.~P., Guhathakurta, P., Peterson, R.~C., 
\& Fulbright, J.~P.\ 2004, \aj, 127, 2162

\bibitem[Sollima et al.(2007)]{sollima:07}
Sollima, A., Beccari, G., Ferraro, F.~R., Fusi Pecci, F.,
\& Sarajedini, A.\ 2007, \mnras, 380, 781

\bibitem[Stetson(1987)]{stetson:87}
Stetson, P.~B.\ 1987, \pasp, 99, 191

\bibitem[Stetson(1990)]{stetson:90}
Stetson, P.~B.\ 1990, \pasp, 102, 932

\bibitem[Stetson(1994)]{stetson:94}
Stetson, P.-B.\ 1994, \pasp, 106, 250

\bibitem[Stetson et al.(2003)]{stetson:03}
Stetson, P.~B., Bruntt, H., \& Grundahl, F.\ 2003, \pasp, 115, 413

\bibitem[Stoughton et al.(2002)]{edr}
Stoughton, C., et al.\ 2002, \aj, 123, 485

\bibitem[Stubbs et al.(2004)]{stubbs:04}
Stubbs, C.~W., Sweeney, D., Tyson, J.~A., 
\& LSST 2004, Bulletin of the American Astronomical Society, 36, 1527 

\bibitem[Tucker et al.(2006)]{tucker:06}
Tucker, D.~L., et al.\ 2006, Astronomische Nachrichten, 327, 821

\bibitem[Venn et al.(2004)]{venn:04}
Venn, K.~A., Irwin, M., Shetrone, M.~D., Tout, C.~A., Hill, V., 
\& Tolstoy, E.\ 2004, \aj, 128, 1177

\bibitem[York et al.(2000)]{york:00}
York, D.~G., et al.\ 2000, \aj, 120, 1579

\end{thebibliography}
\end{document}